\documentclass[a4paper,11pt]{article}
\pdfoutput=1 

\usepackage{jheppub} 
\usepackage{amsthm,amsmath,amssymb}
\usepackage{mathrsfs}
\usepackage[citecolor=blue]{hyperref}
\usepackage{verbatim}
\usepackage{setspace}
\usepackage{array,multirow}
\usepackage[normalem]{ulem}
\usepackage{float}
\usepackage{rotating}
\usepackage{makecell}
\usepackage{mathdots}
\usepackage{color}
\usepackage[table]{xcolor}
\usepackage{adjustbox}
\usepackage{tikz}
\usepackage{pifont}
\usepackage[all]{xy}
\usepackage{bm}
\usepackage[T1]{fontenc} 
\usepackage{caption}
\usepackage{subcaption}

\def\rank{\mathrm{rank}}

\def\fkf{\mathfrak{f}}
\def\fkg{\mathfrak{g}}
\def\fkj{\mathfrak{j}}
\def\fku{\mathfrak{u}}

\def\fsu{\mathfrak{su}}
\def\fso{\mathfrak{so}}
\def\fsp{\mathfrak{sp}}

\def\CB{\mathcal{B}}
\def\CG{\mathcal{G}}
\def\CI{\mathcal{I}}
\def\CM{\mathcal{M}}
\def\CN{\mathcal{N}}
\def\CO{\mathcal{O}}
\def\CT{\mathcal{T}}

\def\bbC{\mathbb{C}}
\def\bbZ{\mathbb{Z}}

\def\tr{\mathrm{tr}}

\title{On low rank 4d $\CN=2$ SCFTs }

\author[a]{Bohan Li}
\author[a,b]{Dan Xie}
\author[a]{and Wenbin Yan}

\affiliation[a]{Yau Mathematics Science center, Tsinghua University,  Beijing, 100084,China}
\affiliation[b]{Department of Mathematics, Tsinghua University,  Beijing, 100084,China}

\emailAdd{libh19@mails.tsinghua.edu.cn}
\emailAdd{danxie@mail.tsinghua.edu.cn}
\emailAdd{wbyan@mail.tsinghua.edu.cn}

\abstract{There are two major ways of constructing 4d $\mathcal{N}=2$ superconformal field theories (SCFTs): the first one is putting a 6d $(2,0)$ theory on a punctured Riemann surface (class-S theory), and the second one is putting type IIB string theory on a 3d canonical  singularity.  As there are interests on low rank theories, we search all the possibilities from above  two constructions. 
Most of those theories are  engineered by class-S theory with irregular singularities, and we find a universal formula for 
the rank of theory so that a complete search is possible. We then compute various physical quantities of those theories, such as the central charges, flavor symmetry, associated vertex operator algebra and Higgs branch, etc.
One of interesting consequence of our results are the prediction of many new isomorphism of 2d vertex operator algebra.
}

\begin{document} 
\maketitle

\section{Introduction}
Four dimensional $\mathcal{N}=2$ superconformal field theories (SCFTs) \cite{Howe:1983wj,Seiberg:1994rs, Seiberg:1994aj} constantly provide invaluable insights into the understanding of quantum field theory in the past thirty years.  The 
studies of them also gave intimate connections to various branches of mathematics such as invariants of four manifolds, integrable systems, algebraic geometry, Hodge theory, vertex operator algebras (VOA) and etc.  One of major advancement in last ten years is that one can engineer a large class of new 4d SCFTs by  geometric methods: one class (class-S) is the compactification of 6d $(2,0)$ theories on a punctured Riemann surface \cite{Gaiotto:2009we, Gaiotto:2009hg,Xie:2012hs,Wang:2015mra,Wang:2018gvb}, 
and the other class is the compactification of type IIB string theory on a 3d canonical singularity \cite{Xie:2015rpa, Chen:2016bzh, Wang:2016yha, Chen:2017wkw}. Resulting theories are usually strongly coupled, so  conventional perturbative methods are of little usage, yet 
 geometric tools can help us understand lots of fundamental properties of these theories. One can often obtain the Seiberg-Witten geometry, central charges, the corresponding 2d VOA (and the Higgs branch from it),  3d mirror \cite{Intriligator:1996ex,Benini:2010uu, Xie:2021ewm}, relation with 5d theories \cite{Closset:2020scj,Closset:2020afy, Closset:2021lwy} and etc. 
It is such an amazing fact that one can easily compute those quantities given the strongly coupled nature of these theories.

On the other hand, there are also interest in the complete classification of 4d $\CN=2$ theories at small rank. By rank we mean the dimension of the Coulomb branch (CB). Classification of rank one and  two theories are discussed in \cite{Argyres:2015ffa, Argyres:2015gha, Argyres:2016xmc, Argyres:2016xua, Argyres:2018zay, Caorsi:2018zsq, Argyres:2022lah, Argyres:2022puv, Argyres:2022fwy,Xie:2022aad}. These classifications are mostly based on studying  constraints on CB. Their efforts lead to some new theories which have not been noticed before \cite{Argyres:2016xua, Argyres:2022lah}. It would be interesting to construct these theories by geometric methods, 
as one can use geometric tools to further study those theories, e.g. understanding their Schur sector and Higgs branch. 

Given the large landscape of the geometric engineered theories, it is not  easy to systematically scan the theory space. In fact, most of low rank theories are 
constructed using class ${\cal S}$ theory with irregular singularities, and no simple formula for the rank is known.
What makes the exhaustive search possible is the 
CB dimension formulae \ref{eq:CBdimUT} and \ref{eq:CBdimT}. Using these dimension formulae, we list all rank one, two and three geometric engineered theories. Indeed, many new theories found in \cite{Argyres:2016xua, Argyres:2022lah}
appear also in our search. Using general results of  geometric construction, one can get new information of these theories, such as the corresponding VOA and the Higgs branch, summarized in section \ref{sec:rank123}. 

Let us point out several interesting consequences of our findings: 
\begin{itemize}
\item There is a sequence of rank one AD theories with the following flavor symmetry (see also \cite{Xie:2017obm})
\begin{equation*}
A_0\subset A_1\subset A_2 \subset G_2 \subset B_3 \subset D_4 \subset F_4 \subset E_6 \subset E_7 \subset E_8,
\end{equation*}
and the corresponding VOA is the the affine Kac-Moody algebra $V_k(\mathfrak{g})$ at  level $k=-h/6-1$.
Here $h$ is the Coxeter number of the flavor group.
This sequence is different from the famous Deligne exceptional series $A_1\subset A_2\subset G_2\subset D_4\subset F_4\subset E_6\subset E_7\subset E_8$ \cite{deligne1996serie} at level $-h^\vee/6-1$, when the group is non-simply-laced. It would be interesting to further study properties of this sequence.

\item There is a sequence of theories with flavor symmetry group $SO(4N+8),~N\geq 2$ and possible enhancement, i.e.
$E_8$, $SO(20)$, $SO(24)$, and etc. These theories are 4d counterpart of 5d theory engineered by the UV limit of $SU(N)$ gauge group  with $2N+3$ flavors \cite{Yonekura:2015ksa}, and can also be engineered by compactifying 6d $A_{2N+1}$ $(2,0)$ theory on a sphere with three punctures labelled by the partition $[N^2,2],[(N+1)^2],[1^{2N+2}]$. They are very likely to have the maximal flavor symmetry at their own rank: $E_8$ for rank one theories, $SO(20)$ for rank two theories, $SO(24)$ for rank three theories and etc. These classes of theory are special in that they seem to be the theory with maximal flavor symmetry for a given rank, i.e  $E_8$ for rank one, $SO(20)$ for rank two, etc.
\item It is also easy to find many other sequence of theories, such as theories with flavor symmetry $G_2, F_4$. Some of those theories are studied in \cite{Giacomelli:2020gee}, our construction would give the VOA and the 
associated Higgs branch for these theories.
\item Many of the new theories found in \cite{Argyres:2015ffa} are constructed using discrete gauging of the known theories. In our findings, those new theories are typically constructed using 
outer automorphism twist. The relation between discrete gauging and outer automorphism twist deserves further study.
\item Since a 4d theory can be realized in many different geometric ways, this often implies the highly nontrivial isomorphism for W algebras (see section \ref{sec:VOAisom}). It would be interesting to prove those 
isomorphisms rigorously. 
\end{itemize}

This paper is organized as follows. section \ref{sec:GeoEng} reviews the geometric construction and how to compute various physical quantities such as SW geometries, central charges, etc. Section \ref{sec:rank123} summarizes 
the list of rank one, two and three theories from geometry. Section \ref{sec:VOAisom} lists  isomorphisms for W algebras derived from our constructions. Section \ref{sec:higherRank} discusses some results on theories with higher rank and summarize some rank four theories. Finally a 
conclusion is given in section \ref{sec:conclusion}.

\section{Geometric engineering of 4d $\mathcal{N}=2$ SCFTs}
\label{sec:GeoEng}
There are  two main classes of 4d $\mathcal{N}=2$ SCFTs. The first class, which is called class-S, is constructed by putting 6d $(2,0)$ theories on a Riemann surface with regular or irregular singularities \cite{Gaiotto:2009we, Gaiotto:2009hg,Xie:2012hs,Wang:2015mra,Wang:2018gvb}. The second class is engineered by putting the type IIB string theory on a 3d-canonical singularities \cite{Xie:2015rpa}. There are several advantages of these constructions:
\begin{enumerate}
\item The Coulomb branch solution (the SW geometry) is given by the spectral curve of the underlying Hitchin system for class-S construction, and is identified with the mini-versal deformation for  type IIB construction.
\item There are simple formula for the physical data like central charges $a$, $c$, $k_F$ and others.
\item The Higgs branch or more generally the Schur sector can be derived for a large portion of class-S theories, as their corresponding 2d VOA is known.
\end{enumerate}
One must keep in mind that for a generic 4d $\CN=2$ theory the above information  is not  easy to derive, hence the above two constructions are  extremely valuable for the study of 4d $\CN=2 $ SCFTs.
In the following, we will review these two constructions and how to compute various physical quantities.

\subsection{Class-S theories}
\subsubsection{Theories with regular singularities only}
One can engineer 4d $\mathcal{N}=2$ SCFTs by putting 6d $(2,0)$ theory of type $\fkj$ on a genus $g$ Riemann surface with only regular singularities \cite{Gaiotto:2009we, Gaiotto:2009hg,Nanopoulos:2009uw, Chacaltana:2012ch,  Chacaltana:2012zy}.
In the untwisted case, the regular singularities are labeled by the nilpotent orbit  $f$ of $\fkj$\footnote{We label the regular singularity by the Nahm label, its (Spaltanstein) dual is also called Hitchin label which appears in the corresponding Hitchin system.},  so an untwisted theory is specified by the data 
\begin{equation}
(\Sigma_g, \fkj, f_1, f_2, \ldots, f_n).
\end{equation}
For a twisted theory, one denotes the outer-automorphism twist of $\fkj$ by $o$, and the invariant Lie algebra obtained after the outer-automorphism by $\fkg^\vee$. Let
$\fkg$ be the Langlands dual of $\fkg^\vee$. There are two kinds of regular singularities for a twisted theory: twisted puncture $f^t$ labelled by the nilpotent orbit of $\fkg$ and untwisted puncture $f$ labelled by the nilpotent orbit of $\fkj$, therefore a twisted theory is specified by the data
\begin{equation}
(\Sigma_g, \fkj ,o,f_1^t, f_2^t, \ldots, f_n^t, f_1,\ldots, f_s).
\end{equation}
Notice that the number of twisted punctures is constrained by $o$. For example, $n$ has to be  even  if $o$ is $Z_2$.

Physical data such as CB spectrum, flavor symmetry and central charges are determined by the local data of the nilpotent orbit $f$ together with global data of $\Sigma_g$ \cite{Chacaltana:2010ks, Chacaltana:2011ze, Chacaltana:2012zy, Chacaltana:2013oka, Chacaltana:2014jba,Chacaltana:2015bna,Chacaltana:2016shw,Chacaltana:2017boe,Chacaltana:2018vhp}. 
The main feature of this class of theories is that the scaling dimensions of CB operators are almost all \textbf{integrals}, while the number of CB operators, or the rank of the untwisted theory, is given by
\begin{equation}
\label{eq:dimClassSunt}
\dim \mathcal{B}=\frac{1}{2} \left(\sum \dim (f_i^D)+(g-1)\dim(\fkj)\right).
\end{equation}
Here $f_i^D$ is the dual orbit of $f_i$. On the other hand, the rank of a twisted theory is given by 
\begin{equation}
\label{eq:dimClassSt}
\dim \mathcal{B}=\frac{1}{2}\left( \sum [dim (f_i^{t,D})+(\dim \fkj - \dim \fkg)]+ \sum \dim (f_i^D)+(g-1)\dim(\fkg^\vee)\right).
\end{equation}
For the twisted theory, the dual orbit for the twisted puncture  (which is labeled by nilpotent orbit of $\fkg$) is a nilpotent orbit of $\fkg^\vee$.  

\textit{Example}: Consider the class-S theory $(\Sigma_0,A_9,[5^2],[4^2,2],[1^{10}])$. This rank three theory has Coulomb branch operators with dimension $10$, $8$ and $6$. The flavor symmetry group is $SO(24)$ which is  maximal  in all rank three theories we found. The $a$ and $c$ central charges are $337/24$ and $101/6$.

\subsubsection{Theories with one irregular singularity}
To get theories with fractional scaling dimension CB operators, one need to add irregular singularities on the Riemann surface. Yet only spheres with one irregular and one regular singularity lead to non-trivial SCFTs in the IR. The untwisted theories are labelled by the following data \begin{equation}
 (\fkj, b,k,f),
\end{equation}
where $\fkj=ADE$ specifies the $6d$ $(2,0)$ theory, $b$ is a positive integer which are determined by $\fkj$, $k>-b$ is also an integer. $\fkj$, $k$ and $b$ together determine the form of the irregular singularity, while the regular singularity is labelled by a nilpotent orbit $f$ of $\fkj$.
The allowed value for $b,k$ are listed in table \ref{table:SWuntwisted}, where the correspondence with the three-fold singularities \cite{Wang:2015mra, Xie:2015rpa} is used.

For the twisted theory, one denote an outer-automorphism twist of $\fkj$ by $o$. Let $\fkg^\vee$ be the invariant Lie algebra after the outer-automorphism twist, and
let $\fkg$ be the Langlands dual of $\fkg^\vee$.  The twisted theories are labelled by data 
\begin{equation}
(\fkj, o, b_t, k_t, f)
\end{equation}
The allowed set for $b_t, k_t$ is summarized in table \ref{table:SWtwisted}. $f$ now labels a nilpotent orbit of $\fkg$. Details of such theories can be found in \cite{Wang:2018gvb}.

\begin{table}[!htb]\small
\begin{center}
	\begin{tabular}{ |c|c|c|c|c| }
		\hline
		$ \mathfrak{j}$& $b$  & Singularity  & Spectral curve at SCFT point & $\Delta[z]$ \\[1pt] \hline
		$A_{N-1}$&$N$ &$x_1^2+x_2^2+x_3^N+z^k=0$ &$x^{N}+z^{k}=0$ & ${N\over N+k}$ \\ \hline
		$~$  & $N-1$ & $x_1^2+x_2^2+x_3^N+x_3 z^k=0$ &$x^{N}+xz^{k}=0$ & ${N-1\over N+k-1}$ \\ \hline
		
		$D_N$  &$2N-2$ & $x_1^2+x_2^{N-1}+x_2x_3^2+z^k=0$ &$x^{2N}+x^{2}z^{k}=0$ & ${2N-2\over 2N+k-2}$ \\     \hline
		$~$  & $N$ &$x_1^2+x_2^{N-1}+x_2x_3^2+z^k x_3=0$&$x^{2N}+z^{2k}=0$ & ${N\over N+k}$  \\     \hline
		
		$E_6$&12  & $x_1^2+x_2^3+x_3^4+z^k=0$   &$x^{12}+z^{k}=0$ & ${12\over 12+k}$ \\     \hline
		$~$ &9 & $x_1^2+x_2^3+x_3^4+z^k x_3=0$   &$x^{12}+x^{3}z^{k}=0$ & ${9\over 9+k}$ \\     \hline
		$~$  &8 & $x_1^2+x_2^3+x_3^4+z^k x_2=0$    &$x^{12}+x^{4}z^{k}=0$ & ${8\over 8+k}$ \\     \hline
		
		$E_7$& 18  & $x_1^2+x_2^3+x_2x_3^3+z^k=0$   &$x^{18}+z^{k}=0$ & ${18\over 18+k}$ \\     \hline
		$~$&14   & $x_1^2+x_2^3+x_2x_3^3+z^kx_3=0$    &$x^{18}+x^{4}z^{k}=0$ & ${14\over 14+k}$ \\     \hline

		$E_8$ &30   & $x_1^2+x_2^3+x_3^5+z^k=0$  &$x^{30}+z^{k}=0$ & ${30\over 30+k}$ \\     \hline
		$~$  &24  & $x_1^2+x_2^3+x_3^5+z^k x_3=0$ &$x^{30}+x^{6}z^{k}=0$ & ${24\over 24+k}$  \\     \hline
		$~$  & 20  & $x_1^2+x_2^3+x_3^5+z^k x_2=0$  &$x^{30}+x^{10}z^{k}=0$ & ${20\over 20+k}$ \\     \hline
	\end{tabular}
\end{center}
\caption{\label{table:SWuntwisted}Three-fold isolated quasi-homogenous singularities of cDV type corresponding to the $(\fkj,b,k)$  irregular punctures of the regular-semisimple type in \cite{Wang:2015mra}. These 3d singularity is very useful in extracting the CB spectrum \cite{Xie:2015rpa}. }
\label{table:sing}
\end{table} 

\begin{table}[!htb]\tiny
	\begin{center}
		\begin{tabular}{|c|c|c|c|c|c|}
			\hline
			$\mathfrak{j}/o$ & $\fkg$ & $b_t$ & SW geometry at SCFT point & Spectral curve at SCFT point & $\Delta[z]$ \\[1pt] \hline
			$A_{2N}/\mathbb{Z}_2$ & $C^{(1)}_N$ & $4N+2$ &$x_1^2+x_2^2+x^{2N+1}+z^{k'+{1\over2}}=0$ & $x^{2N+1}+z^{k'+{1\over2}}=0$ & ${4N+2\over 4N+2k'+3}$  \\[3pt] \hline
			~&~& $2N$ & $x_1^2+x_2^2+x^{2N+1}+xz^{k'}=0$ & $x^{2N+1}+xz^{k'}=0$ & ${2N\over k'+2N}$ \\[3pt] \hline
			$A_{2N-1}/\mathbb{Z}_2$& $B_N$ & $4N-2$ & $x_1^2+x_2^2+x^{2N}+xz^{k'+{1\over2}}=0$ & $x^{2N}+x z^{k'+{1\over2}}=0$ & ${4N-2\over 4N+2k'-1}$ \\[3pt] \hline
			~ & ~& $2N$ &$x_1^2+x_2^2+x^{2N}+z^{k'}=0$ & $x^{2N}+z^{k'}=0$ & ${2N\over 2N+k'}$  \\[3pt] \hline
			$D_{N+1}/\mathbb{Z}_{2}$& $C^{(2)}_N$ & $2N+2$ & $x_1^2+x_2^{N}+x_2x_3^2+x_3z^{k'+{1\over2}}=0$ & $x^{2N+2}+z^{2k'+1}=0$ & ${2N+2\over 2k' +2N+3}$ \\[3pt] \hline
			~ & ~ & $2N$ &$x_1^2+x_2^{N}+x_2x_3^2+z^{k'}=0$ & $x^{2N+2}+x^{2}z^{k'}=0$ & ${2N\over k'+2N}$  \\[3pt] \hline
			$D_4/\mathbb{Z}_3$ & $G_2$ & $12$ &$x_1^2+x_2^{3}+x_2x_3^2+x_3z^{k'\pm {1\over3}}=0$ & $x^{8}+z^{2k'\pm \frac{2}{3}}=0$ & ${12\over 12+3k'\pm1}$  \\[3pt] \hline
			~& ~ & $6$ &$x_1^2+x_2^{3}+x_2x_3^2+z^{k'}=0$ & $x^{8}+x^{2}z^{k'}=0$ & ${6\over 6+k'}$  \\[3pt] \hline
			$E_6/\mathbb{Z}_2$& $F_4$ & $18$ &$x_1^2+x_2^{3}+x_3^4+x_3z^{k'+{1\over2}}=0$ & $x^{12}+x^{3}z^{k'+{1\over2}}=0$ & ${18\over 18+2k'+1}$  \\[3pt] \hline
			~& ~ & $12$ &$x_1^2+x_2^{3}+x_3^4+z^{k'}=0$ & $x^{12}+z^{k'}=0$ & ${12\over 12+k'}$  \\ [3pt]\hline
			~ & ~ & $8$ &$x_1^2+x_2^{3}+x_3^4+x_2z^{k'}=0$ & $x^{12}+x^{4}z^{k'}=0$ & ${8\over 8+k'}$  \\[3pt] \hline
		\end{tabular}
		\caption{Seiberg-Witten geometry of twisted theories at the SCFT point. Relations between $k_{t}$ and $k'$ which appears in SW geometry are: $k_t=2k'+1$ if $z^{k'+\frac{1}{2}}$ appears in SW geometry, $k_t = 3k' \pm 1$ if $z^{k'\pm\frac{1}{3}}$ in SW geometry, and finally $k_t=k'$ if $z^{k'}$ in SW geometry.}
		\label{table:SWtwisted}
	\end{center}
\end{table}

\textbf{SW curve and CB spectrum}: The SW curve can be identified with the spectral curve of the underlying Hitchin system, and the detailed scaling dimensions can be found from the underlying Newton polygon \cite{Xie:2017aqx}.
For an untwisted theory specified by $(\fkj, b, k, f)$, its rank  can be described by the following formula \cite{Xie:2019pre}
\begin{equation}
\label{eq:CBdimUT}
\boxed{
	\text{dim}\mathcal{B}=\frac{1}{2}\left[(h_{\fkj}\frac{k}{b}+h_{\fkj}-1)\rank(\fkj)-f_{0}-\dim\mathcal{O}_{prin}+\dim\mathcal{O}_{f}^{Hitchin}\right],
	}
\end{equation}
where $f_0$ is the number of mass deformations from the irregular singularity computed in \cite{Xie:2017aqx, Wang:2018gvb}, $h_\fkj$ is the Coxeter number of $\fkj$ and $\CO_f^{Hitchin}$ is the Spaltenstein dual of the nilpotent orbit $f$. Similarly for a twisted theory specified by 
$(\fkj, o,\fkg, b_t, k_t, f)$, its rank is  \cite{Xie:2019pre}
\begin{equation}
\label{eq:CBdimT}
\boxed{
	\text{dim}\mathcal{B}=\frac{1}{2}\left[(h_{\theta}\frac{k_{t}}{b_{t}}+h_{\theta}-1)\rank(\mathfrak{g})-f_{0}-\dim\mathcal{O}_{prin}+\dim\mathcal{O}_{f}^{Hitchin}\right],
	}
\end{equation}
where $h_{\theta}$ take value in Table \ref{table:htheta1} instead of the Coxeter number\footnote{$h_\theta$ is called the twisted Coxeter number in \cite{oblomkov2016geometric}.}.

\begin{table}[htb]
	\begin{center}
		\begin{tabular}{ |c|c| c|c|c|c| }
			\hline
			$ \mathfrak{j}/o $ ~&$A_{2N}/Z_2$ &$A_{2N-1}/Z_2$ & $D_{N+1}/Z_2$  &$E_6/Z_2$&$D_4/Z_3$ \\ \hline
			$h_{\theta}$   &$4N+2$ &$4N-2$& $2N+2$  & $18$&$12$\\     \hline
		\end{tabular}
	\end{center}
	\caption{The $h_{\theta}$ for different twisted AD theories.}
	\label{table:htheta1}
\end{table}

{\textit Example 1:} $(A_7, 8, -6,[2^2,1^4])$ theory. In this example, $\fkg=A_7$, $b=8$, $k=-6$ and $f=[2^2,1^4]$. Then $f_0=1$, $\dim\CO_{prin}=56$ and $\CO_f^{Hitchin}=[4,2^2]$ with dimension $52$. According to \ref{eq:CBdimUT}, its rank is
\begin{equation}
\frac{1}{2}\left[(8\times \frac{-6}{8}+8-1)\times 7 - 1 - 56+52\right]=1.
\end{equation}

\begin{figure}[htb]
	\centering
	\begin{minipage}{0.4\textwidth}
		\includegraphics[width=1\textwidth]{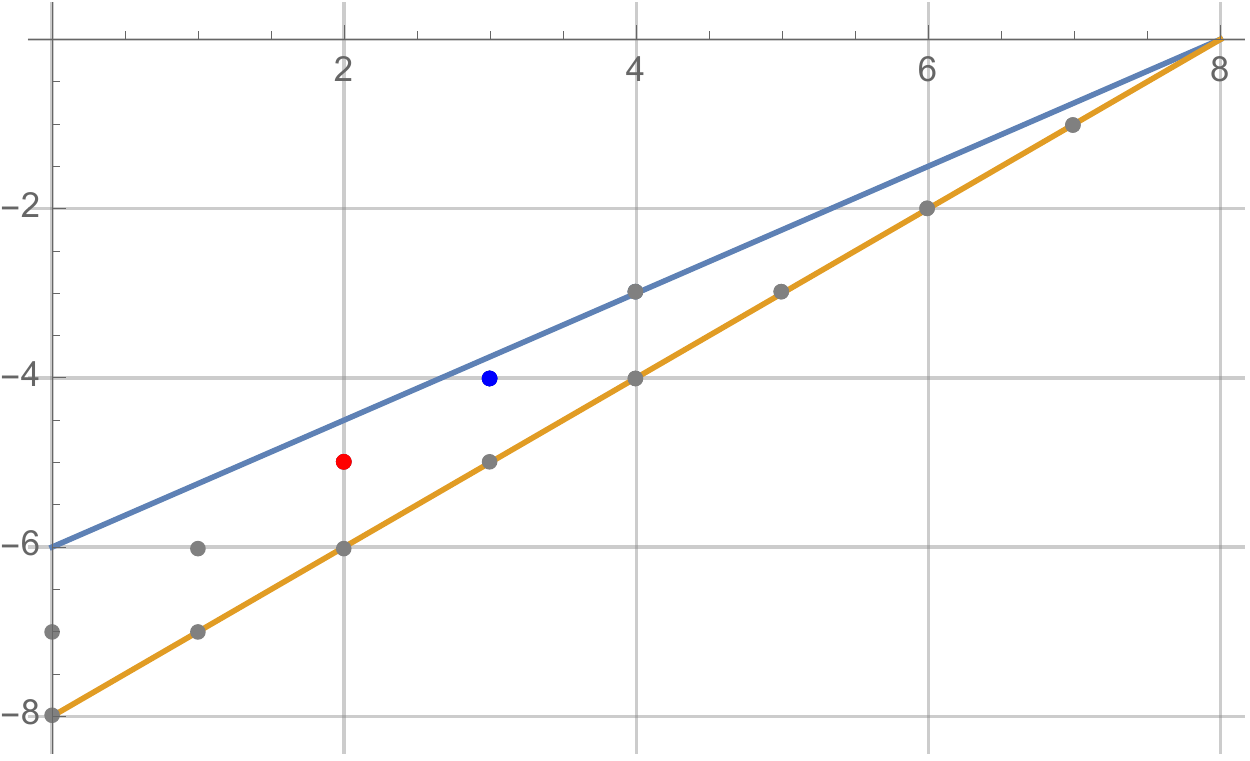}
	\end{minipage}%
	\caption{Newton Polygons for the theory $(A_7,8,-6,[2^2,1^4])$. Grey dots represents terms beyond the corresponding pole order. The red dot corresponds to the CB operator, while the blue dot is the mass parameter. } 
	\label{fig:neA7}
\end{figure}

The SW curve of this theory looks like
$x^8+z^{-6} = \sum_{i=1}^7\phi_{d_i}(z)x^{8-d_i}$, 
where $\{d_i\}=\{2,3,\cdots,8\}$ are degrees of Casimir of $\fsu(8)$. For $f=[2^2,1^4]$ the pole order is $\{p_2,p_3,\cdots, p_8\}=\{1,2,3,4,5,5,6\}$. The SW curve can be visualized by a Newton polygon (figure  \ref{fig:neA7}) where each term $x^az^b$ in the curve is represented by a point $(a,b)$ inside the polygon \cite{Xie:2017aqx}. In this example there are only two points in the polygon representing the term $u x^2z^{-5}+ m x^3z^{-4}$, so the SW curve is
\begin{equation}
x^8+z^{-6} =  ux^2z^{-5}+ m x^3z^{-4},
\end{equation}
where $u$ has conformal dimension $2$ and $m$ has conformal dimension $1$. In the end, the theory has one dimension $2$ CB operator and one mass parameter.

\textit{Example 2}: $(\mathfrak{e}_{6}, 12, -7,A_{4})$ theory. In this example, $\fkg=\mathfrak{e}_{6}$, $b=12$, $k=-7$ and $f=A_{4}$. Then $f_0=0$, $\dim\CO_{prin}=72$ and $\CO_f^{Hitchin}=A_{3}$ with dimension $52$. Using \ref{eq:CBdimUT}, we get its rank being
\begin{equation}
	\frac{1}{2}\left[(12\times \frac{-7}{12}+12-1)\times 6 - 0 - 72+52\right]=2.
\end{equation}

\begin{figure}[H]
	\centering
	\begin{minipage}{0.4\textwidth}
		\includegraphics[width=1\textwidth]{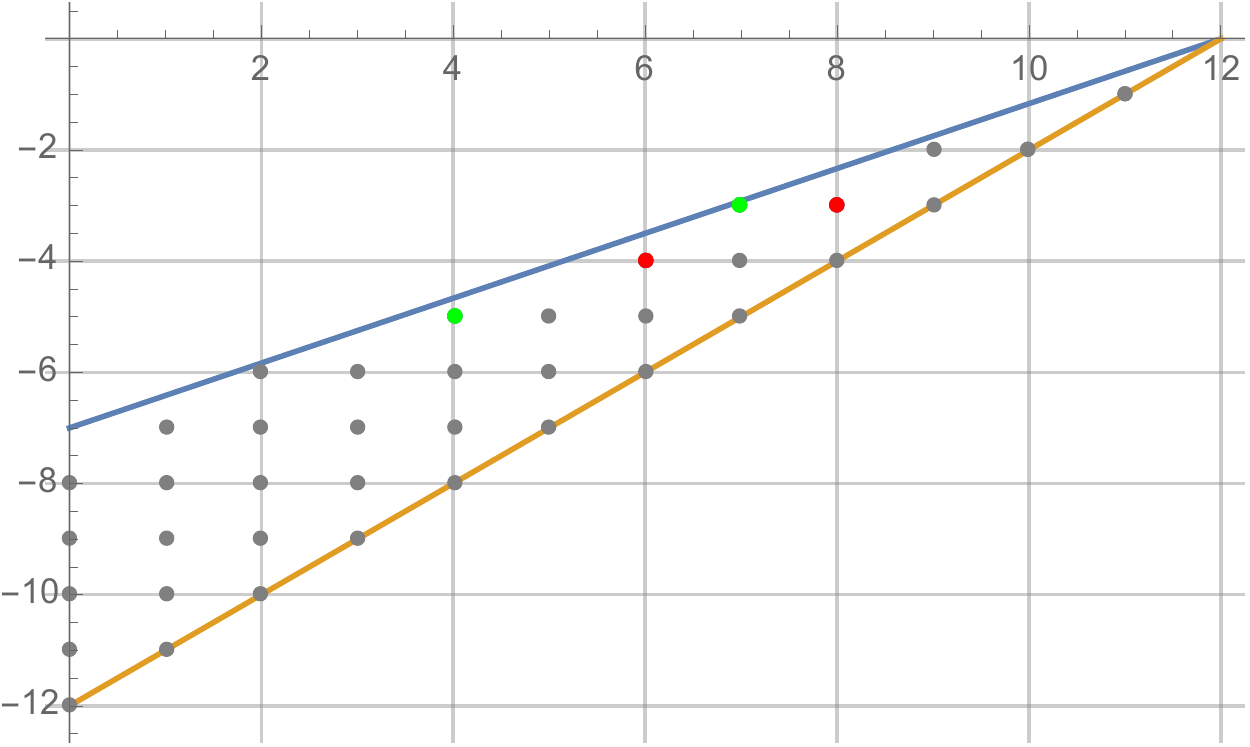}
	\end{minipage}
	\caption{Newton Polygons for $(\mathfrak{e}_{6},12,-7,A_{4})$ theory. Grey dots represents terms beyond the corresponding pole order.  Red dots correspond to  CB operators, while  green dots are  mass parameters. } 
	\label{fig:e6}
\end{figure}

For $f=A_4$ there is an extra degree $4$ differential $\phi_4(z)$ in the SW curve, so the set of degrees $\{d_i\}=\{2,4,5,6,8,9,12\}$. The pole order structure after imposing constraints is $\{p_2, p_4,p_5,p_{6},p_{8},p_{9},p_{12}\}=\{1,3, 3,4,5,5,7\}$ \cite{Chacaltana:2014jba}. Its Newton polygon is shown in figure  \ref{fig:e6}, and the SW curve is
\begin{equation}
	x^{12}+z^{-7} =  u_{1}x^{6}z^{-4}+u_{2} x^{8}z^{-3}+c_{1}x^{7}z^{-3}+c_{2}x^{4}z^{-5},
\end{equation}
where two CB operators $u_{1}, u_{2}$ have conformal dimension $\frac{8}{5},\frac{6}{5}$ respectively. The coupling $c_{1}$ and $c_2$ have conformal dimension $\frac{1}{5}$  and  $\frac{4}{5}$ respectively.

\textit{Example 3} $(\mathfrak{e}_{6},\mathbb{Z}_{2},\mathfrak{f}_{4},8,-7,0)$ theory. In this example, $\mathfrak{j}=\mathfrak{e}_{6}$, $o=\mathbb{Z}_{2}$, $\fkg=\mathfrak{f}_{4}$, $b_{t}=8$, $k_{t}=-7$ and $f=0$. Then $f_0=1$, $\dim\CO_{prin}=48$ and $\CO_f^{Hitchin}=F_{4}$ with dimension $48$. According to \ref{eq:CBdimT}, its rank is
\begin{equation}
	\frac{1}{2}\left[(18\times \frac{-7}{8}+18-1)\times 4 - 1 - 48+48\right]=2.
\end{equation}

\begin{figure}[H]
	\centering
	\begin{minipage}{0.4\textwidth}
		\includegraphics[width=1\textwidth]{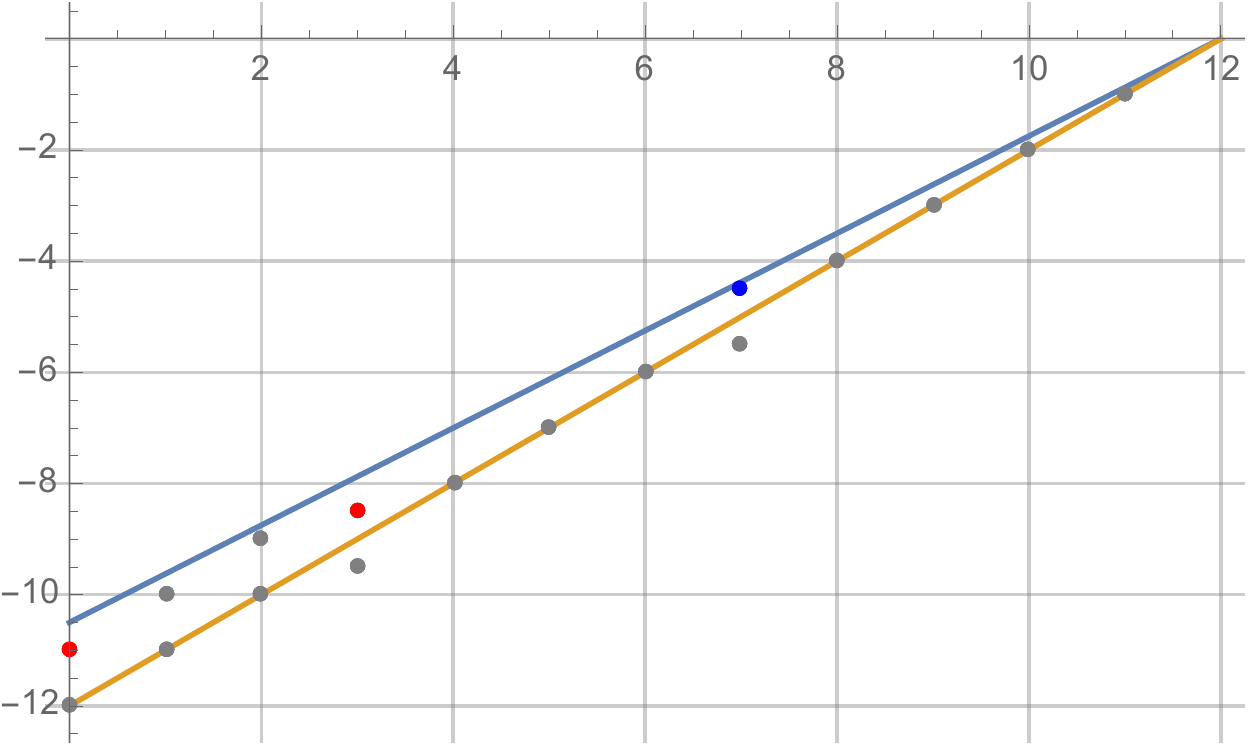}
	\end{minipage}
	\caption{Newton Polygons for $(\mathfrak{e}_{6},\mathbb{Z}_{2},\mathfrak{f}_{4},8,-7,0)$ theory. Grey dots represents terms beyond the corresponding pole order. Red dots correspond to  CB operators, while the blue dot represents the mass parameter.} 
	\label{fig:f4}
\end{figure}

For twisted punctures, $p_{5}$ are $p_{9}$ are half-integer. For $f=0$  the pole order structure is $\{p_2,p_5,p_{6},p_{8},p_{9},p_{12}\}=\{1,\frac{9}{2},5,7,\frac{17}{2},11\}$.  The  Newton polygon is drawn in figure  \ref{fig:f4}, and the SW curve is,
\begin{equation}
x^{12}+x^{4}z^{-7}=u_{1}z^{-11}+u_{2}x^{3}z^{-\frac{17}{2}}+mx^{7}z^{-\frac{9}{2}}
\end{equation}
All in all, the theory has two CB operators with conformal dimension $5$ and $4$ and one mass parameter. The $(a,c)$ central charges of this theory are $(\frac{14}{3},\frac{16}{3})$, and the flavor symmetry is $(\mathfrak{f}_4)_{5}\times \mathfrak{u}(1)$\footnote{Our normalization of flavor anomaly is half of the one in \cite{Argyres:2022lah}}. Using this data we identify this theory as one of the theories in table 6 of \cite{Argyres:2022lah}, and we provide its SW curve as well.

\textbf{Remark}:  Our dimension formulae \ref{eq:CBdimUT} and \ref{eq:CBdimT} actually computes the dimension of the Hitchin base, then identify it as the rank of the theory.  Another way of computing the rank of the theory is to count the number of 
 graded Coulomb branch operators by using the pole order of each individual differential in the SW curve. In rare occasions,
the number of graded Coulomb branch operators is greater than the dimension of the Hitchin base. That is because the dimension of the Hitchin base is the sum of the virtual dimension of each individual differential. However, the virtual dimension of each individual differential might be negative, therefore the dimension of the Hitchin base is smaller than the number of the graded Coulomb branch operators in such cases.

\textbf{Central charges $a_{4d}$ and $c_{4d}$}:  The $a_{4d}$ and $c_{4d}$ central charges are computed by using the following relations. The first one relates  central charges with the CB spectrum \cite{Shapere:2008zf,Xie:2013jc},
\begin{equation}
\boxed{
2a_{4d}-c_{4d} = \frac{1}{4}\sum_{i}(2[u_i]-1),
}
\end{equation}
where $[u_i]$ is the dimension of the CB operator $u_i$ and $i$ runs over all CB operators. The other two relations require data of the corresponding VOA \cite{Beem:2013sza,Xie:2019zlb}
\begin{equation}
\label{eq:centralChargesVOA}
\boxed{\begin{split}
&c_{4d}=-\frac{1}{12}c_{2d},\\
&a_{4d}-c_{4d}=-\frac{1}{48}\CG.
\end{split}}
\end{equation}
Here $c_{2d}$ is the central charge of the corresponding VOA, $\CG$ the growth function of its vacuum module. Any two of these relations can be used to work out $a_{4d}$ and $c_{4d}$ while the third one provides a consistency check. Here we will mainly discuss computing central charges using relations \ref{eq:centralChargesVOA}, that is, computing the central charge and growth of the corresponding VOA.

For an untwisted (twisted) theory labelled by $(\fkj,b,k,f)$ ($(\fkj,o,\fkg,b_t,k_t,f)$ for twisted), if the irregular singularity has no mass deformation ($f_0=0$), the corresponding VOA is the W-algebra $W_{\tilde{k}}(\fkj,f)$ with $\tilde{k} = -h^\vee_{\fkj} +\frac{b}{k+b}$ ($W_{\tilde{k}}(\fkg,f)$ with $\tilde{k} = -h^\vee_{\fkg} +\frac{1}{m}\frac{b_t}{k_t+b_t}$ for twisted, where $m$ is the lacety number of $\fkg$). The central charge $c_{W_{\tilde{k}}(\fkg,f)}$ is given by \cite{Kac_2003}
\begin{equation}
	\begin{aligned}
		c_{W_{\tilde{k}}(\mathfrak{g},f)}
		=&\text{dim}\mathfrak{g}^{0}-\frac{1}{2}\text{dim}\mathfrak{g}^{\frac{1}{2}}-\frac{12}{\tilde{k}+h^{\vee}_{\mathfrak{g}}}\left|\rho-(\tilde{k}+h^{\vee}_{\mathfrak{g}})\frac{h}{2}\right|^{2}.
	\end{aligned}
\end{equation}
Here the $\mathfrak{sl}_{2} $ triple $(e,f,h)$ is fixed by the Dynkin grading. The Weyl vector $\rho$ is  sum of all fundamental weights.  $\text{dim}\mathfrak{g}^{0}$ and $\text{dim}\mathfrak{g}^{\frac{1}{2}}$ are the number of positive roots with grading $0$ and $\frac{1}{2}$. One way to compute them is by looking at the decomposition of the adjoint representation of $\mathfrak{g}$ under the subalgebra $\Lambda(\mathfrak{su}(2))\oplus\mathfrak{f}_{\Lambda}$, where $\Lambda$ is the embedding map $\Lambda:\mathfrak{su}(2)\rightarrow
\mathfrak{g}$,
\begin{equation}
	\mathfrak{g}=\bigoplus _{j\in\frac{1}{2}\mathbb{Z}_{\geq 0}}V_{j}\otimes \mathcal{R}_{j}
\end{equation}
where $V_{j}$ is the spin-$j$ representation and $R_{j}$ in some representation of $\mathfrak{f}_{\Lambda}$, then
\begin{equation}
	\text{dim}\mathfrak{g}^{0}=\sum_{j\in\mathbb{Z}_{\geq 0}}\text{dim}\, \mathcal{R}_{j} \qquad \text{dim}\mathfrak{g}^{\frac{1}{2}}=\sum_{j\in\frac{1}{2}+\mathbb{Z}_{\geq 0}}\text{dim}\, \mathcal{R}_{j}.
\end{equation}
If the irregular singularity has $f_0\neq 0$ mass parameters, the corresponding VOA should be an extension of $ W_{\tilde{k}}(\fkg,f)\oplus M(1)^{\oplus f_0}$, where we denote the Heisenberg algebra by $M(1)$, therefore the $2d$ central charge equals to $c_{W_{\tilde{k}}(\fkg,f)}+f_0$.  Combining these two situations, we see that the $4d$ central charge $c_{4d}$ is
\begin{equation}
\boxed{
	c_{4d}=-\frac{1}{12}\left(\text{dim}\mathfrak{g}^{0}-\frac{1}{2}\text{dim}\mathfrak{g}^{\frac{1}{2}}-\frac{12}{\tilde{k}+h^{\vee}_{\mathfrak{g}}}\left|\rho-(\tilde{k}+h^{\vee}_{\mathfrak{g}})\frac{h}{2}\right|^{2}\right)-\frac{f_{0}}{12},
	}
\end{equation}
with $\tilde{k} = -h^\vee_{\fkg} +\frac{b}{k+b}$ for untwisted theories and $\tilde{k} = -h^\vee_{\fkg} +\frac{1}{m}\frac{b_t}{k_t+b_t}$ for twisted theories. The data relevant for $h$ can be found from the weighted Dynkin diagram
associated with the nilpotent orbit $f$.

On the other hand, computing the growth function of the vacuum module of a VOA requires more work. For a W-algebra $W_{\tilde{k}}(\fkg,f)$ at boundary admissible level $\tilde{k}=-h^\vee+\frac{h^\vee}{u}$, the growth function of its vacuum module can be found in math literature (e.g. theorem 2.16 of \cite{kac2008rat}).
If the VOA is non-admissible or an extension of $W_{\tilde{k}}(\fkg,f)\oplus M(1)^{\oplus f_0}$, the growth function of its vacuum module is conjectured to have the following universal form \cite{Xie:2019vzr}
\begin{equation}
\boxed{
\CG=\mathrm{rank}\fkg+f_0-\frac{d(n)}{u}+\dim\CO_{\mathrm{prin}}-\dim\CO_f.
}
\label{growth}
\end{equation}
Here $n$ and $u$ is related to the level $\tilde{k}$ by
\begin{equation}
\frac{n}{u}=\tilde{k}+h^\vee,\quad \mathrm{gcd}(n,u)=1.
\end{equation}
The number $d(n)$ depends only on $\fkg$ and $n$. To determine $d(n)$ one first consider the case $f=trivial$ so that the VOA is an extension of $V_{\tilde{k}}(\fkg)\oplus u(1)^{\oplus f_0}$ with $2d$ central charge
\begin{equation}
\frac{\tilde{k}\dim\fkg}{\tilde{k}+h^\vee}+f_0.
\end{equation}
Together with the CB spectrum of the corresponding 4d theory, one can determine the growth function $\CG$ of its vacuum module, hence determine $d(n)$. Results of $d(n)$ are listed in tables of appendix \ref{sec:dnlists}. Here we also include the data for irregular singularities with mass deformation which were not discussed in \cite{Xie:2019vzr}.

$\textbf{Example}$: Consider the $A_{3}/\mathbb{Z}_{2}$ twisted theory  $\mathcal{T}\equiv (A_{3},\mathbb{Z}_{2},B_{2},4,-1,[3,1^{2}])$. Its CB spectrum is $\{4/3,4/3\}$, so
\begin{equation}
	2a_{4d}-c_{4d}=\frac{1}{4}\times2\left(2\times\frac{4}{3}-1\right)=\frac{5}{6}.
\end{equation} 
The irregular singularity has a mass deformation and the corresponding VOA is an extension of $W_{-3+\frac{2}{3}}(B_{2},[3,1^{2}])\oplus M(1)$. Then we have
\begin{equation}
	c_{2d}=-\frac{1}{12},
\end{equation}
then central charges of this theory should be
\begin{equation}
	a_{4d}=\frac{11}{12}, \qquad c_{4d}=1.
\end{equation}
For a consistency check, notice that the growth function of the vacuum module is
\begin{equation}
	\mathcal{G}=2\times(2\times 2+1)+1-\frac{d(2)}{3}-6=4,
\end{equation}
and indeed $a_{4d}-c_{4d} = -\CG/48=-1/12$. From central charges and CB spectrum, one conjecture that the VOA $V_{\CT}$ is isomophsic to two copies of $V_{-\frac{4}{3}}(\mathfrak{su}(2))$,
\begin{equation}
	W_{-3+\frac{2}{3}}(B_{2},[3,1^{2}])\oplus M(1)\subset V_{\mathcal{T}}\cong V_{-\frac{4}{3}}(\mathfrak{su}(2))\oplus V_{-\frac{4}{3}}(\mathfrak{su}(2))
\end{equation}
where $V_{-\frac{4}{3}}(\mathfrak{su}(2))$ means that the simple affine vertex algebra at level $k=-\frac{4}{3}$.

\textbf{Flavor central charge}: The flavor symmetry of our theories are determined by both the regular singularity $f$ and the number of mass deformations $f_0$ of the irregular singularity. Given $f$ and an embedding $\Lambda:\fsu(2)\rightarrow\fkg$,  Let $\fkf_\Lambda$ be the centralizer of $\Lambda(\fsu(2))$ in $\fkg$, then the naive flavor symmetry algebra of the theory is  $\fkf_\Lambda\oplus \fku(1)^{\oplus f_0}$. Notice that the full flavor symmetry may get enhanced to a larger algebra with $\fkf_\Lambda\oplus \fku(1)^{\oplus f_0}$ as subalgebra.

Usually $\fkf_\Lambda$  decomposes into several simple factors 
\begin{equation}
\fkf_\Lambda=\oplus_{i}\fkf_i,
\end{equation}
and we are interested in the 4d flavor anomaly $k_{4d,i}$ for each simple factor $\fkf_i$. Using the 4d/VOA correspondence, this means that there are some simple factors of affine vertex algebra as the subalgebra of the $W$-algebra corresponding the theory in consideration,
\begin{equation}
\oplus_i V_{k_{i}}(\mathfrak{g}_{i})\oplus M(1)^{\oplus f_0}\subset W_{\tilde{k}}(\mathfrak{g},f)\oplus M(1)^{\oplus f_0}\subset V(\CT).
\end{equation}
One can compute the level of each affine vertex subalgebra $V_{k_i}(\mathfrak{g}_i)$ in the following way (see e.g.  \cite{deBoer:1993iz,Chacaltana:2012zy}). Let $T^a$ and $T^b$ be generators of $\fkf_i$ such that $\tr_{\fkf_i}T^aT^b=h^\vee_{\fkf_i}\delta^{ab}$. Denote by $n$ the natural embedding $n:\fkf_\Lambda\rightarrow \fkg$. The level $k_i$ of the affine vertex subalgebra $V_{k_i}(\mathfrak{g}_i)$ is then 
\begin{equation}
\label{eq:falvorAnomaly}
k_i(\mathfrak{f}_i)\delta^{ab}=\frac{\tilde{k}_{2d}}{h^{\vee}_{\mathfrak{g}}}\text{tr}_{\mathfrak{g}}n(T^{a})n(T^{b})+\sum_{j}2j\text{tr}_{\mathcal{R}_{j}}T^{a}T^{b},
\end{equation}
where $R_j$'s are $\fkf_\Lambda$ representations appear in the decomposition
\begin{equation}
	\mathfrak{g}=\bigoplus _{j\in\frac{1}{2}\mathbb{Z}_{\geq 0}}V_{j}\otimes \mathcal{R}_{j}.
\end{equation}
Once we have 2d level $k_i$, the 4d flavor anomaly is $k_{4d, i}= - k_i$ by the 4d/VOA correspondence.

{\bf Example: } Consider the theory $(\mathfrak{e}_6,9,-4, (A_{2},\mathbb{Z}_{2}))$, one has
$\mathfrak{f}_{\Lambda}=\mathfrak{su}(2)$ and $\tilde{k}=-12+\frac{9}{5}$. 
The decomposition of the adjoint representation of $\mathfrak{e}_{6}$ is
\begin{equation}
	\mathfrak{e}_{6}\rightarrow (\bm{1};\bm{3})+(\bm{3};\bm{1})+(\bm{4};\bm{2})+(\bm{5};\bm{1})+(\bm{6};\bm{2})+(\bm{7};\bm{1})+(\bm{9};\bm{1})+(\bm{10};\bm{2})+(\bm{11};\bm{1}),
\end{equation}
where the first entry in the bracket is the dimension of the $\fsu(2)$ representation $V_{j}$, and the second entry the dimension of the representation of $\mathfrak{f}_{\Lambda}$.
Normalizing the Dynkin index of the representation of $\mathfrak{su}(2)$ as the following
\begin{equation}
2\text{tr}_{\text{adj}}T^{a}T^{b}=4 , \qquad 2\text{tr}_{\bm{2}}T^{a}T^{b}=1,
\end{equation}
the flavor anomaly of $\fkf_\Lambda$ is then
\begin{equation}
k_{4d}=-k_{2d}=\frac{17}{10}.
\end{equation}

We need to emphasize that equation \ref{eq:falvorAnomaly} does not work for $\fku(1)$ factor in $\fkf_\Lambda$. However, there is a way to compute the  level of $\fsu(1)$ in $\fkf_\Lambda$ in terms of  "pyramids" \cite{elashvili2005classification,brundan2007good}. We refer to \cite{Arakawa:2021ogm} for the details.

\subsection{Three dimensional canonical singularities}
One can also construct 4d $\mathcal{N}=2$ SCFTs  by putting type IIB string theory on a 3d canonical singularity \cite{Xie:2015rpa, Chen:2016bzh, Wang:2016yha, Chen:2017wkw}.  Two most interesting classes of 
canonical singularities are isolated quasi-homogeneous hypersurface singularity defined by a polynomial $f$ and a complete intersection singularity defined by two polynomials $(f_1, f_2)$.  

\textbf{Hypersurface singularity}: Given a hypersurface singularity $f(z_1, z_2, z_3, z_4)$, one can compute the CB spectrum as follows \cite{Xie:2015rpa}: The quasi-homogeneity of  $f$ means that there is a $C^\ast$ action
\begin{equation}
f(\lambda^{q_i} z_i)=\lambda f,~~\sum q_i<1,\quad q_i>0,
\end{equation}
One can also define a Jacobian algebra $J_f$
\begin{equation}
J_f={\bbC[z_1, z_2, z_3, z_4]\over \{ {\partial f\over \partial z_1},\ldots, {\partial f\over \partial z_4}\}}.
\end{equation}
Let $\{\phi_\alpha\}$  be a monomial basis of $J_f$, the SW geometry is then
 \begin{equation}
 F=f+\sum \lambda_\alpha \phi_\alpha.
 \end{equation}
 The scaling dimension for $\lambda_\alpha$ can be worked out
 \begin{equation}
 \Delta(\lambda_\alpha)={(1-Q(\phi_\alpha) )\over \sum_i q_i}.
  \end{equation}
Here $Q(\phi_\alpha)$ is the weight for the monomial $\phi_\alpha$.  For the SW curve, only $\lambda_\alpha$ with scaling dimension greater or equal to zero would be preserved as physical deformation parameters. 

The dimension of the charge lattice $\mu=2r+f$ (Here $r$ is the rank of the theory, and $f$ the number of mass parameters) is given by the dimension of $J_f$ (also called the Milnor number), which are related to weights of $z_i$'s by the following formula
\begin{equation}
\mu=(1-\frac{1}{q_1})(1-\frac{1}{q_2})(1-\frac{1}{q_3})(1-\frac{1}{q_4}).
\end{equation}
Since one can find the number of mass parameters using the scaling formula, one can easily find the rank of the theory.   The central charge can be computed as follows
\begin{equation}
a={R(A)\over 4}+{R(B)\over6}+{5 r\over 24},~~c={R(B)\over 3}+{r\over 6},
\label{central}
\end{equation} 
where $R(A)$ and $R(B)$ are determined by the CB spectrum
 \begin{equation}
 R(A)=\sum_i([u_i]-1), \quad R(B)={1\over 4} \mu [u]_{max}.  
 \end{equation}
Here $ [u]_{max}$ is the maximal scaling dimension, $\mu$ is the dimension of $J_f$ (Milnor number) defined before.

\textbf{Complete intersection singularity}:  Given a complete intersection singularity $f=(f_1, f_2)$,  and a $C^\ast$ action
\begin{equation}
f_1(\lambda^{w_i}z_i)=\lambda f_1(z_i),~~~f_2(\lambda^{w_i}z_i)=\lambda^d f_2(z_i),
\end{equation} 
there is a distinguished $(3,0)$ form $\Omega$ satisfying
\begin{equation}
\Omega \wedge df_1\wedge df_2=dz_1\wedge dz_2\wedge \ldots \wedge dz_5,
\end{equation}
which has weight $\sum w_i-1-d$ under the $C^*$ action. To define a sensible 4d SCFT, we require this weight to be positive, which leads to a constraint on $w_i$'s
\begin{equation}
\sum w_i>1+d.
\end{equation}
The SW solution is described by the mini-versal deformation of the singularity \cite{Chen:2016bzh, Wang:2016yha}
\begin{equation}
F(\lambda, z_i)=f(z_i)+\sum_{\alpha=1}^\mu \lambda_{\alpha} \phi_{\alpha},
\label{SW}
\end{equation}
where $\phi_{\alpha}$ is the monomial basis of the Jacobi module of $f$, and $\mu$ is the Milnor number of $J_f$. The coefficient $\lambda_{\alpha}$ is identified 
with the parameters on Coulomb branch. The scaling dimension of $\lambda_{\alpha}$ is determined by the requirement that $\Omega$  have dimension one, because the integration of $\Omega$ over
the middle homology cycle of the Milnor fibration gives the mass of BPS particle. At the end the dimension of $\phi_\alpha$ is computed by the following formulae
\begin{align}
&\phi_{\alpha}=[\psi_{\alpha},0]:~~[\lambda_{\alpha}]={1-Q_{\alpha}\over \sum w_i-1-d}, \nonumber\\
&\phi_{\alpha}=[0,\psi_{\alpha}]:~~[\lambda_{\alpha}]={d-Q_{\alpha}\over \sum w_i-1-d}.
\label{spectrum}
\end{align}
Here $Q_{\alpha}$ is the $C^*$ weight of the monomial $\psi_{\alpha}$.  The dimension of charge lattice $\mu$ can also be found from the 
weights \cite{Chen:2016bzh, Wang:2016yha}.

\section{Summary of theories with  rank one, two and three}
\label{sec:rank123}

Now we can scan the theory space reviewed in the last section for  SCFTs with small rank. Theories from spheres with one irregular singularity and one regular singularity with rank one, two and three are listed in table \ref{table:rank1sum},  \ref{table:rank2sum1}, 
and  \ref{table:rank3sum1} respectively. For each theory, we list its central charges, Coulomb branch spectrum, flavor symmetry and the corresponding VOA. The complete lists with constructions are summarized in appendix \ref{sec:completList1}, \ref{sec:completList2} and \ref{sec:completList3}. The predictions of VOA isomorphisms are list in section \ref{sec:VOAisom}. We also scan the theory engineered by compactifying $A_{n}$ type theory on  sphere with three or more regular singularities, and results are shown in table \ref{table:rank2other} and \ref{table:rank3int}. The corresponding VOA of theories with regular singularities only is constructed rigorously by mathematicians \cite{Arakawa:2018egx}.

We also search the rank one, two, three theories from the list of hypersurface and complete intersection singularities. They do not provide new possibilities, but we also include 
the construction in our tables. The class is labeled by two Dynkin diagrams $(G, G^{'})$ so that the 3-fold singularity is $f_G(x,y)+f_{G^{'}}(w,z)=0$, and $f_{G=ADE}$ is 
the corresponding ADE singularity.

All theories in table \ref{table:rank1sum},  \ref{table:rank2sum1}, 
and  \ref{table:rank3sum1} satisfy the following bound
\begin{equation}
\label{eq:boundSCFTs}
\frac{12\left(\sum_i(u_i-1)^2\right)}{u_{max}(u_{max}-1)} \geq 2 r +f,
\end{equation}
where $\{u_i\}$ is the spectrum of CB operators, $u_{max}$ is the maximal dimension of the CB operators, $r$ is the rank of the theory and $f$ is the rank of the flavor symmetry. Theories saturating this bound is marked with $*$ in the tables.
Those theories seem special to us.

Notice that some  theories could be the decoupled sum of two lower rank theories. They are summarized in table \ref{table:rank2sum1decomp} and \ref{table:rank3sum1decomp}. Below we provide detailed analysis for one example of this situation.

{\bf Example:} Consider  the theory with CB spectrum $\{\frac{8}{5},\frac{6}{5},\frac{6}{5}\}$ in table \ref{table:rank3sum1decomp}. It is expected to be comprised of two decoupled theories with CB spectrum $\{\frac{8}{5},\frac{6}{5}\}$ and $\{\frac{6}{5}\}$ respectively. Using the corresponding VOA, one can denote these theories by  $\CT[W_{-7+\frac{7}{5}}(B_4,[5,2^2])]$, $\CT[V_{-2+\frac{2}{5}}(\fsu(2))]$ and $\CT[\mathrm{Vir}(2,5)]$ repectively. Since all of these VOAs are admissible, one can compute its Schur index using the formula in \cite{Xie:2019zlb} and check if it is the product of the indices of the two component theories. 

For nilpotent orbit $[5,2^2]$, the flavor symmetry is $SU(2)$ and the adjoint representation of $SO(9)$ decomposes as
\begin{equation}
\mathrm{adj}_{SO(9)}=V_0\otimes U_1\oplus 2V_1\oplus V_{\frac{3}{2}}\otimes U_{\frac{1}{2}}\oplus V_{\frac{5}{2}}\otimes U_{\frac{1}{2}} \oplus V_3,
\end{equation}
where $V_j$ is the spin $j$ representation of $\fsu(2)$ and $U_j$ is the spin $j$ representation of the flavor $SU(2)$. The Schur index (the normalized vacuum character of $W_{-7+\frac{7}{5}}(B_4,[5,2^2])$ ) is then
\begin{equation}
\CI_{\CT[W_{-7+\frac{7}{5}}(B_4,[5,2^2])]}
=\mathrm{PE}\left[\frac{(q-q^5)\chi_{1}(z)+(q^2-q^4)}{(1-q)(1-q^5)}\right],
\end{equation}
where $\mathrm{PE}[f(x,y,z,\cdots)]=\exp\left(\sum_{n=1}^\infty\frac{1}{n}f(x^n,y^n,z^n,\cdots)\right)$ being the plethystic exponential and $\chi_1(z)$ is the character of the adjoint representation of $SU(2)$. On the other hand, the indices of $\CT[V_{-2+\frac{2}{5}}(\fsu(2))]$ and $\CT[\mathrm{Vir}(2,5)]$ are
\begin{equation}
\CI_{\CT[V_{-2+\frac{2}{5}}(\fsu(2))]} = \mathrm{PE}\left[\frac{(q-q^5)\chi_{1}(z)}{(1-q)(1-q^5)}\right],\quad
\CI_{\CT[\mathrm{Vir}(2,5)]} = \mathrm{PE}\left[\frac{q^2-q^4}{(1-q)(1-q^5)}\right].
\end{equation}
It is clear that
\begin{equation}
\begin{split}
\CI_{\CT[W_{-7+\frac{7}{5}}(B_4,[5,2^2])]}
=&\CI_{\CT[V_{-2+\frac{2}{5}}(\fsu(2))]} \CI_{\CT[\mathrm{Vir}(2,5)]}.
\end{split}
\end{equation}
This also leads to a possible isomorphism of VOAs
\begin{equation}
W_{-7+\frac{7}{5}}(B_4,[5,2^2]) \cong V_{-2+\frac{2}{5}}(\fsu(2)) \oplus \mathrm{Vir}(2,5).
\end{equation}


 \begin{table}[H]\scriptsize
	\centering
	\begin{tabular}{|c|c|c|c|c|c|}
		\hline
		$(a,c)$ & $\Delta_{\mathrm{Coulomb}}$ & $\mathfrak{f}$ & VOA  & $\CM_H$/Asso. Var. & Singularity\\
		\hline
		
		\hline
		$*(\frac{43}{120},\frac{11}{30})$ & $\frac{6}{5}$ & $-$ & $W_{-\frac{8}{5}}(\mathfrak{su}(2),[2])\cong\text{Vir}(2,5)$ & $-$ & $(A_1, A_2)$ \\
		\hline
		
		\hline
		$*(\frac{11}{24},\frac{1}{2})$ & $\frac{4}{3}$ & $\mathfrak{su}(2)_{\frac{4}{3}}$ & $V_{-\frac{4}{3}}(\mathfrak{su}(2))$ & $[2]$ & $(A_1, A_3)$\\
		\hline
		
		\hline
		$*(\frac{7}{12},\frac{2}{3})$ & $\frac{3}{2}$ & $\mathfrak{su}(3)_{\frac{3}{2}}$ & $ V_{-\frac{3}{2}}(\mathfrak{su}(3))$ & $[2,1]$ &$(A_1, D_4)$ \\ 
		\hline
		
		\hline
		$*(\frac{23}{24},\frac{7}{6})$ & $2$ & $\mathfrak{so}(8)_{2}$ & $ V_{-2}(\mathfrak{so}(8))$ & $[2^2,1^4]$ & \\
		\hline
		$(\frac{23}{24},\frac{7}{6})$ & $2$ & $\mathfrak{so}(7)_{2}$ & $V_{-2}(\mathfrak{so}(7))$ & $[3,1^4]$ &  \\
		\hline
		$(\frac{23}{24},\frac{7}{6})$ & $2$ & $(\mathfrak{g}_{2})_{2}$ & $V_{-2}(\mathfrak{g}_2)$ & $G_2(a_1)$ & \\
		
		\hline
		$(\frac{3}{4},\frac{3}{4})$ & $2$ & $\mathfrak{su}(2)_{\frac{3}{2}}$ & $W_{-6+2}(D_{4},[3,2^{2},1])$ & $\overline{\CO}_{[3^2,1^2]}\cap S_{[3,2^2,1]}$ & \\
		\hline
		
		\hline

		$*(\frac{41}{24},\frac{13}{6})$ & $3$ & $(\mathfrak{e}_{6})_{3}$ & $V_{-3}(\mathfrak{e}_{6})$ & $A_1$ & \\
		\hline
		
		\hline
		$(\frac{41}{24},\frac{13}{6})$ & $3$ & $(\mathfrak{f}_{4})_{3}$ & $V_{-3}(\mathfrak{f}_{4})$ & $\tilde{A}_1$& \\
		\hline
		$(\frac{17}{12},\frac{19}{12})$  & $3$ & $\mathfrak{sp}(2)_{2}\times\mathfrak{u}(1)_{f_{0}}$ & $W_{-4+\frac{3}{2}}(C_{3},[2,1^{4}]) \oplus M(1)$ & ~& \\
		\hline
		
		\hline
		
		$*(\frac{59}{24},\frac{19}{6})$ & $4$ & $(\mathfrak{e}_{7})_{4}$ & $V_{-4}(\mathfrak{e}_{7})$ & $A_1$&\\
		\hline
		$(\frac{25}{12},\frac{29}{12})$ & $4$ & $\mathfrak{sp}(3)_{\frac{5}{2}}\times\mathfrak{u}(1)_{f_{0}}$ & $W_{-18+7}(\mathfrak{e}_{7},4A_{1}) \oplus M(1)$  & ~&\\
		\hline
		
		\hline
		
		$*(\frac{95}{24},\frac{31}{6})$ & $6$ & $(\mathfrak{e}_{8})_{6}$ & $V_{-6}(\mathfrak{e}_{8})$ & $A_1$&\\
		\hline

	\end{tabular}
	\caption{\label{table:rank1sum}List of rank one AD theories with central charges $(a,c)$, CB spectrum $\Delta_{\mathrm{Coulomb}}$, flavor symmetry algebra $\fkf$, one corresponding VOA and Higgs branch. Theories with $*$ saturate the bound \ref{eq:boundSCFTs}. $\fku(1)$ with a subscript $f_0$ means the $\fku(1)$ algebra from the irregular singularity. $W\oplus M(1)^{\oplus f_0}$ means the actual VOA is possibly an extension of $W\oplus M(1)^{\oplus f_0}$. In the last coulomb, a single nilpotent orbit $f$ means that the corresponding Higgs branch or associated variety is the closure of the nilpotent orbit $\overline{\CO}_f$, while $S_{f'}$ means the Slodowy slice of $f'$.}
\end{table}

 \begin{table}[H]\tiny
	\centering
	\begin{tabular}{|c|c|c|c|c|c|}
		\hline
		$(a,c)$ & $\Delta_{\mathrm{Coulomb}}$ & $\mathfrak{f}$ & VOA & $\CM_H$/Asso. Var. &Singularity \\
		\hline
		
		\hline
		
		
		$*(\frac{67}{84},\frac{17}{21})$ & $\frac{10}{7}$, $\frac{8}{7}$ & $-$ & $W_{-2+\frac{2}{7}}(\mathfrak{su}(2),[2])$ & $-$ &$(A_1,A_4)$\\
		\hline
		
		\hline
		$*(\frac{11}{12},\frac{23}{24})$ & $\frac{3}{2}$, $\frac{5}{4}$ & $\mathfrak{u}(1)_{f_{0}}$ & $W_{-5+\frac{5}{4}}(\mathfrak{su}(5),[3,2])$ & ~& $(A_1, A_5)$\\
		\hline
		
		\hline
		$(\frac{7}{6},\frac{4}{3})$ & $\frac{3}{2}$, $\frac{3}{2}$ & $\mathfrak{su}(3)_{3}$ & $W_{-30+\frac{30}{4}}(\mathfrak{e}_{8},D_{4}(a_{1})+A_{2})$ &  $\overline{\CO}_{A_4+2A_1}\cap S_{D_{4}(a_{1})+A_{2}}$ &\\
		\hline
		
		\hline
		
		$*(\frac{19}{20},1)$ & $\frac{8}{5}$, $\frac{6}{5}$ & $\mathfrak{su}(2)_{\frac{8}{5}}$ & {$V_{-2+\frac{2}{5}}(\mathfrak{su}(2))$} & $[2]$ &$(A_1, D_5)$\\
		\hline
		
		\hline
		$*(\frac{13}{12},\frac{7}{6})$ & $\frac{5}{3}$, $\frac{4}{3}$ & $\mathfrak{su}(2)_{\frac{5}{3}}\times\mathfrak{u}(1)_{f_{0}}$ & $W_{-30+\frac{30}{9}}(\mathfrak{e}_{8},E_{7}(a_{3}))$ & $\overline{\CO}_{E_8(a_6)}\cap S_{E_{7}(a_{3})}$ & $(A_1, D_6)$ \\
		\hline
		
		\hline
		
		
		$(\frac{7}{4},2)$ & $2$, $2$ & $\mathfrak{u}(1)^{5}_{f_{0}}$ & $W_{-8+1}(D_{5},[7,3])\oplus M(1)^{\oplus 5}$ & ~& $(\sum_{i=1}^5 x_i^2,\sum_{i=1}^5 i  x_i^2)$\\
		\hline
		
		
		$(\frac{7}{4},2)$ & $2$, $2$ & $\mathfrak{su}(2)_{2}\times\mathfrak{su}(2)_{2}\times\mathfrak{su}(2)_{6}$ & $W_{-18+6}(\mathfrak{e}_{7},A_{2}+2A_{1})$ & $\overline{\CO}_{D_4(a_1)}\cap S_{A_{2}+2A_{1}} $ &\\
		\hline
		$(\frac{7}{4},2)$ & $2$, $2$ & $\mathfrak{su}(2)_{8}\times\mathfrak{su}(2)_{2}$ & $W_{-30+6}(\mathfrak{e}_{8},D_{5}(a_{1})+A_{1})$ & $ \overline{\CO}_{E_8(a_7)}\cap S_{D_{5}(a_{1})+A_{1}}$  &\\
		\hline
		$(\frac{7}{4},2)$ & $2$, $2$ & $\mathfrak{su}(2)_{10}$ & $W_{-30+6}(\mathfrak{e}_{8},A_{4}+A_{2}+A_{1})$ & $ \overline{\CO}_{E_8(a_7)}\cap S_{A_{4}+A_{2}+A_{1}}$ &\\
		\hline

		$(\frac{19}{12},\frac{5}{3})$ & $2$, $2$ & $\mathfrak{sp}(2)_{2}$ & {$V_{-3+1}(\mathfrak{so}(5))$} & $[3,1^2]$ &\\
		\hline
		
		\hline
		
		$(\frac{163}{120},\frac{17}{12})$ & $\frac{12}{5}$, $\frac{6}{5}$ & $\mathfrak{su}(2)_{\frac{17}{10}}$ & $W_{-12+\frac{9}{5}}(\mathfrak{e}_{6},A_{5})$ & $\overline{\CO}_{E_6(a_3)} \cap S_{A_5} $  &\\
		\hline
		
		\hline
		$*(\frac{7}{4},2)$ & $\frac{5}{2}$, $\frac{3}{2}$ & $\mathfrak{su}(5)_{\frac{5}{2}}$ & {$V_{-5+\frac{5}{2}}(\mathfrak{su}(5))$} & $[2^2,1]$ & $D_2(SU(5))$ \\
		
		\hline
		$(\frac{19}{12},\frac{5}{3})$ & $\frac{5}{2}$, $\frac{3}{2}$ & $\mathfrak{su}(2)_{\frac{7}{4}}\times\mathfrak{u}(1)_{f_{0}}$ & $W_{-4+\frac{3}{4}}(C_{3},[4,1^{2}])\oplus M(1)$ & ~ &\\
		\hline
		
		\hline
		$(\frac{13}{8},\frac{7}{4})$ & $\frac{8}{3}$, $\frac{4}{3}$ & $\mathfrak{so}(3)_{\frac{4}{3}}\times\mathfrak{su}(2)_{\frac{11}{6}}$ & $W_{-5+\frac{5}{3}}(C_{4},[2^{3},1^{2}])$ & $\overline{\CO}_{[3^2,2]} \cap S_{[2^3,1^2]} $  &\\
		\hline
		
		\hline
		%
		%
		%
		%
		$(\frac{47}{24},\frac{13}{6})$ & $3$, $\frac{3}{2}$ & $\mathfrak{su}(3)_{3}\times\mathfrak{su}(2)_{2}$ & $W_{-12+\frac{9}{2}}(\mathfrak{e}_{6},3A_{1})$ & $\overline{\CO}_{A_2+2A_1} \cap S_{3A_1} $  &\\
		\hline
		
		\hline
		$*(\frac{29}{12},\frac{17}{6})$ & $3$, $2$ & $\mathfrak{su}(6)_{3}\times\mathfrak{u}(1)_{f_{0}}$ & $V_{-6+3}(\mathfrak{su}(6))\oplus M(1)$ & ~ &\\
		
		\hline
		$(2,2)$ & $3$, $2$ & $\mathfrak{su}(2)_{4}$ & $W_{-4+1}(\mathfrak{g}_{2},A_{1})$ & $ \overline{\CO}_{G_2(a_1)} \cap S_{A_1}$  &\\
		\hline
		
		\hline
		$(\frac{61}{24},\frac{17}{6})$ & $3$, $\frac{5}{2}$ & $\mathfrak{sp}(3)_{\frac{5}{2}}\times\mathfrak{u}(1)_{f_{0}}$ & $V_{-4+\frac{3}{2}}(C_{3})\oplus M(1)$ & ~ &\\
		\hline
		
		\hline
		$(2,\frac{13}{6})$ & $\frac{10}{3}$, $\frac{4}{3}$ & $\mathfrak{sp}(2)_{\frac{13}{6}}$ & $W_{-10+\frac{10}{3}}(D_{6},[3,2^{4},1])$ & $\overline{\CO}_{[3^4]}\cap S_{[3,2^4,1]}$  & \\
		\hline
		
		\hline
		$*(\frac{37}{12},\frac{11}{3})$ & $4$,  $2$ & $\mathfrak{spin}(12)_{4}$ & {$V_{-10+6}(D_6)$ }& $[2^4,1^4]$ &\\
		
		\hline
		$(\frac{71}{24},\frac{41}{12})$ & $4$, $2$ & $\mathfrak{su}(3)_{4}\times\mathfrak{su}(2)_{\frac{5}{2}}\times\mathfrak{u}(1)^{2}_{f_{0}}$ & $W_{-12+4}(\mathfrak{e}_{6},3A_{1}) \oplus M(1)^{\oplus 2}$ & ~ &\\
		\hline
		$(\frac{71}{24},\frac{41}{12})$ & $4$, $2$ & $(\mathfrak{g}_{2})_{4}\times\mathfrak{su}(2)_{\frac{5}{2}}$ & $W_{-30+10}(\mathfrak{e}_{8},A_{2}+3A_{1})$ & $\overline{\CO}_{D_4(a_1)+A_1}\cap S_{A_2+3A_1}$  &\\
		
		\hline
		$(\frac{17}{6},\frac{19}{6})$ & $4$,  $2$ & $\mathfrak{sp}(2)_{\frac{5}{2}}\times\mathfrak{u}(1)^{2}_{f_{0}}$ & $W_{-10+3}(D_{6},[3,2^{4},1])\oplus M(1)^{\oplus 2}$ & ~ &\\
		\hline
		
		\hline
		$(\frac{83}{24},\frac{47}{12})$ & $4$, $3$ & $\mathfrak{sp}(3)_{3}\times\mathfrak{su}(2)_{\frac{5}{2}}\times\mathfrak{u}(1)_{f_{0}}$ & $W_{-18+7}(\mathfrak{e}_{7},(3A_{1})')\oplus M(1)$ & ~ &\\
		\hline
		
		\hline
		$*(\frac{53}{12},\frac{16}{3})$ &  $5$, $3$ & $\mathfrak{spin}(12)_{5}\times\mathfrak{u}(1)^{2}_{f_{0}}$ & $V_{-10+5}(D_{6} )\oplus M(1)^{\oplus 2}$ &~ &\\
		
		\hline
		$(\frac{95}{24},\frac{53}{12})$ & $5$, $3$ & $\mathfrak{sp}(4)_{3}\times\mathfrak{u}(1)_{f_{0}}$ & $W_{-6+\frac{5}{2}}(C_{5},[2,1^{8}]) \oplus M(1)$ &~ &\\
		\hline
		
		\hline

		$(\frac{14}{3},\frac{16}{3})$ & $5$, $4$ & $(\mathfrak{f}_{4})_{5}\times\mathfrak{u}(1)_{f_{0}}$ & $V_{-9+4}(\mathfrak{f}_{4}) \oplus M(1)$ & ~ &\\
		\hline
		
		\hline

		$*(\frac{101}{12},\frac{31}{3})$ & $10$, $4$ & $(\mathfrak{e}_{8})_{10}$ & {$V_{-30+20}(\mathfrak{e}_8)$} & $2A_1$  &\\
		\hline

	\end{tabular}
	\caption{\label{table:rank2sum1}List of rank two AD theories with central charges $(a,c)$, CB spectrum $\Delta_{\mathrm{Coulomb}}$, flavor symmetry algebra $\fkf$, one corresponding VOA and Higgs branch. Theories with $*$ saturate the bound \ref{eq:boundSCFTs}. $\fku(1)$ with a subscript $f_0$ means the $\fku(1)$ algebra from the irregular singularity. $W\oplus M(1)^{\oplus f_0}$ means the actual VOA is possibly an extension of $W\oplus M(1)^{\oplus f_0}$. In the last coulomb, a single nilpotent orbit $f$ means that the corresponding Higgs branch or associated variety is the closure of the nilpotent orbit $\overline{\CO}_f$, while $S_{f'}$ means the Slodowy slice of $f'$.}

\end{table}

\begin{table}[H]
\centering
		\begin{tabular}{|c|c|c|c|}
		\hline
		$(a,c)$ & $\Delta_{\mathrm{Coulomb}}$ & $\mathfrak{f}$ & VOA  \\
		\hline
		
		\hline
		$*(\frac{43}{60},\frac{11}{15})=2(\frac{43}{120},\frac{11}{30})$ & $\frac{6}{5}$, $\frac{6}{5}$ & $-$ & $W_{-12+\frac{6}{5}}(\mathfrak{e}_{6},E_{6}(a_{1}))$\\
		\hline
		
		\hline
		$*(\frac{11}{12},1)=2(\frac{11}{24},\frac{1}{2})$ & $\frac{4}{3}$, $\frac{4}{3}$ & $\mathfrak{su}(2)_{\frac{4}{3}}\times\mathfrak{su}(2)_{\frac{4}{3}}$ & $W_{-9+\frac{5}{3}}(B_{5},[7,1^{4}])$\\
		\hline
		
		\hline
		$*(\frac{7}{6},\frac{4}{3})=2(\frac{7}{12},\frac{2}{3})$ & $\frac{3}{2}$, $\frac{3}{2}$ & $\mathfrak{su}(3)_{\frac{3}{2}}\times\mathfrak{su}(3)_{\frac{3}{2}}$ & $W_{-12+\frac{9}{2}}(\mathfrak{e}_{6},A_{2})$\\
		\hline
		
		\hline
		$(\frac{23}{12},\frac{7}{3})=2(\frac{23}{24},\frac{7}{6})$ & $2$, $2$ & $(\mathfrak{g}_{2})_{2}\times(\mathfrak{g}_{2})_{2}$ & $W_{-30+10}(\mathfrak{e}_{8},2A_{2})$\\
		\hline

	\end{tabular}
	\caption{\label{table:rank2sum1decomp}Rank 2 theories which are likely to be two decouple rank 1 theories.}
\end{table}

 \begin{table}[H]\tiny
	\centering
	\begin{tabular}{|c|c|c|c|c|c|}
		\hline
		$(a,c)$ & $\Delta_{\mathrm{Coulomb}}$ & $\mathfrak{f}$ & VOA & $\CM_H$/Asso. Var. & Singularity\\
		\hline
		
		\hline
		$(\frac{3}{2},\frac{13}{8})$ & $\frac{5}{4},\frac{3}{2},\frac{3}{2}$ & $\mathfrak{su}(2)_{\frac{3}{2}}\times\mathfrak{u}(1)_{f_{0}}$ & $W_{-18+\frac{9}{4}}(\mathfrak{e}_{7},D_{5}+A_{1})+M(1)$ & ~ &\\
		\hline
		
		\hline

		$(\frac{7}{4},2)$ & $\frac{3}{2},\frac{3}{2},\frac{3}{2}$ & $\mathfrak{su}(2)^{3}_{\frac{3}{2}}\times\mathfrak{u}(1)_{f_{0}}$ & $W_{-18+\frac{9}{2}}(\mathfrak{e}_{7},D_{4}(a_{1}))+M(1)$ &~ &\\
		\hline
		
		\hline
		
		$*(\frac{91}{72},\frac{23}{18})$ & $\frac{14}{9},\frac{4}{3},\frac{10}{9}$ & $-$ & $W_{-2+\frac{2}{9}}(\mathfrak{su}(2),[2])$ & $-$ & $(A_1,A_6)$ \\
		\hline
		
		\hline

		$*(\frac{167}{120},\frac{43}{30})$ & $\frac{8}{5},\frac{7}{5},\frac{6}{5}$ & $\mathfrak{u}(1)_{f_{0}}$ & $W_{-4+\frac{4}{5}}(\mathfrak{su}(4),[3,1])$ & ~& $(A_1, A_7)$\\
		\hline
		
		\hline
		
		$(\frac{75}{56},\frac{19}{14})$ & $\frac{12}{7},\frac{9}{7},\frac{8}{7}$ & $-$ & $W_{-3+\frac{3}{7}}(\mathfrak{su}(3),[3])$ & $-$ &$(A_1,E_6)$\\
		\hline
		
		\hline
		$*(\frac{81}{56},\frac{3}{2})$ & $\frac{12}{7},\frac{10}{7},\frac{8}{7}$ & $\mathfrak{su}(2)_{\frac{12}{7}}$ & {$V_{-2+\frac{2}{7}}(\mathfrak{su}(2))$} & $[2]$ & $(A_1, D_7)$\\
		\hline
		
		\hline
		$*(\frac{19}{12},\frac{5}{3})$ & $\frac{7}{4},\frac{3}{2},\frac{5}{4}$ & $\mathfrak{su}(2)_{\frac{7}{4}}\times\mathfrak{u}(1)_{f_{0}}$ & $W_{-5+\frac{5}{4}}(\mathfrak{su}(5),[3,1^{2}])$ & ~ & $(A_1, D_8)$\\
		\hline
		
		\hline
		
		$*(\frac{3}{2},\frac{31}{20})$ & $\frac{9}{5},\frac{7}{5},\frac{6}{5}$ & $\mathfrak{u}(1)_{\frac{9}{5}}$ & $W_{-3+\frac{3}{5}}(\mathfrak{su}(3),[2,1])$ & $\overline{\CO}_{[3]} \cap S_{[2,1]} $  & $(A_1,E_7)$\\
		\hline
		
		\hline
		$*(\frac{15}{8},2)$ & $2,\frac{3}{2},\frac{3}{2}$ & $\mathfrak{u}(1)_{2}\times\mathfrak{u}(1)^{2}_{f_{0}}$ & $W_{-3+\frac{1}{2}}(\mathfrak{su}(3),[2,1])+M(1)^{\oplus 2}$ & ~ &\\
		\hline
		
		\hline
		
		$(\frac{61}{24},\frac{17}{6})$ & $2,2,2$ & $\mathfrak{su}(2)_{2}\times\mathfrak{u}(1)_{2}\times\mathfrak{u}(1)^{4}_{f_{0}}$ & $W_{-5+1}(\mathfrak{su}(5),[3,1^{2}])+M(1)^{\oplus 4}$ & ~ &\\
		\hline
		$(\frac{61}{24},\frac{17}{6})$ & $2,2,2$ & $\mathfrak{su}(2)_{2}\times\mathfrak{su}(2)_{1}$ & $W_{-30+6}(\mathfrak{e}_{8},A_{4}+A_{2})$ & $\overline{\CO}_{E_8(a_7)} \cap S_{A_4+A_2}$  &\\
		\hline
		
		$(\frac{19}{8},\frac{5}{2})$ & $2,2,2$ & $\mathfrak{su}(2)_{2}\times\mathfrak{su}(2)_{2}\times\mathfrak{su}(2)_{2}$ & $W_{-6+2}(D_{4},[2^{2},1^{4}])$ & $\overline{\CO}_{[3^2,1^2]} \cap S_{[2^{2},1^{4}]}$  &\\
		\hline

		$(\frac{53}{24},\frac{13}{6})$ & $2,2,2$ & $\mathfrak{u}(1)$ & $W_{-10+2}(D_{6},[5,3^{2},1])$ & $\overline{\CO}_{[5^2,1^2]} \cap S_{[5,3^2,1]}$  &\\
		\hline

		%
		\hline
		
		$*(\frac{15}{8},2)$ & $\frac{9}{4},\frac{3}{2},\frac{5}{4}$ & $\mathfrak{su}(3)_{\frac{9}{4}}$ & {$V_{-3+\frac{3}{4}}(\mathfrak{su}(3))$} & $[3]$ & $D_4 (SU(3))$\\
		\hline
		
		\hline
		
		$*(\frac{25}{12},\frac{9}{4})$ & $\frac{7}{3},\frac{5}{3},\frac{4}{3}$ & $\mathfrak{su}(3)_{\frac{7}{3}}\times\mathfrak{u}(1)_{\frac{5}{3}}$ & $W_{-5+\frac{5}{3}}(\mathfrak{su}(5),[2,1^{3}])$ & $\overline{\CO}_{[3,2]} \cap S_{[2,1^3]} $  &\\
		\hline
		
		%
		\hline
		$(\frac{103}{60},\frac{107}{60})$ & $\frac{12}{5},\frac{6}{5},\frac{6}{5}$ & $\mathfrak{su}(2)_{\frac{17}{10}}$ & $W_{-18+\frac{14}{5}}(\mathfrak{e}_{7},D_{6}(a_{2}))$ & $\overline{\CO}_{E_7(a_5)}\cap S_{D_6(a_2)}$ & \\
		\hline
		
		\hline
		
		$(\frac{85}{42},\frac{25}{12})$ & $\ \frac{18}{7}, \frac{12}{7},\frac{8}{7}$ & $\mathfrak{su}(2)_{\frac{25}{14}}$ & $W_{-12+\frac{12}{7}}(\mathfrak{e}_{6},A_{5})$ & $\overline{\CO}_{E_6(a_3)}\cap S_{A_5}$  &\\
		\hline
		
		\hline
		
		
		%
		%
		%
		%
		%
		%
		%
		%
		$(\frac{55}{24},\frac{5}{2})$ & $\frac{8}{3},\frac{5}{3},\frac{4}{3}$ & $\mathfrak{su}(4)_{\frac{8}{3}}$ & {$V_{-4+\frac{4}{3}}(\mathfrak{su}(4))$} & $[3,1]$ & $D_3(SU(4))$ \\
		\hline
		
		\hline
		$(\frac{21}{8},\frac{17}{6})$ & $\frac{8}{3},\frac{7}{3},\frac{4}{3}$ & $\mathfrak{sp}(2)_{\frac{7}{3}}\times\mathfrak{u}(1)_{\frac{4}{3}}$ & $W_{-8+\frac{8}{3}}(D_{5},[2^{4},1^{2}])$ & $\overline{\CO}_{[3^3,1]} \cap S_{[2^{4},1^{2}]} $   &\\
		\hline
		
		\hline
		
		$(\frac{32}{15},\frac{133}{60})$ & $\frac{14}{5},\frac{8}{5},\frac{6}{5}$ & $\mathfrak{su}(2)_{\frac{19}{10}}$ & $W_{-8+\frac{8}{5}}(D_{5},[5,2^{2},1])$ & $\overline{\CO}_{[5^2]}\cap S_{[5,2^{2},1]}$  &\\
		\hline
		
		\hline
		
		$(\frac{8}{3},\frac{17}{6})$ & $3,2,\frac{3}{2}$ & $\mathfrak{su}(2)_{2}\times\mathfrak{u}(1)^{2}_{f_{0}}$ & $W_{-10+\frac{3}{2}}(D_{6},[7,2^{2},1])+M(1)^{\oplus 2}$ & ~ &\\
		\hline
		
		\hline
		$*(\frac{25}{8},\frac{7}{2})$ & $3,2,2$ & $\mathfrak{su}(4)_{3}\times\mathfrak{u}(1)^{3}_{f_{0}}$ & $V_{-4+1}(\mathfrak{su}(4))+M(1)^{\oplus 3}$ & ~ &\\
		\hline
		
		\hline
		
		$(\frac{73}{24},\frac{10}{3})$ & $3,\frac{5}{2},\frac{3}{2}$ & $\mathfrak{sp}(2)_{\frac{5}{2}}\times\mathfrak{u}(1)_{\frac{3}{2}}\times\mathfrak{u}(1)_{f_{0}}$ & $W_{-8+\frac{5}{2}}(D_{5},[2^{4},1^{2}])+M(1)$ & ~ &\\
		\hline
		
		\hline
		
		$(\frac{27}{8},\frac{7}{2})$ & $3,3,2$ & $(\mathfrak{g}_{2})_{3}$ & {$V_{-4+1}(\mathfrak{g}_{2})$} & $G_2(a_1)$  &\\
		\hline
		
		\hline
		
		$(\frac{29}{8},4)$ & $3,\frac{7}{2},\frac{3}{2}$ & $\mathfrak{so}(7)_{\frac{7}{2}}\times\mathfrak{u}(1)_{f_{0}}$ & $V_{-5+\frac{3}{2}}(B_{3})+M(1)$ & ~ &\\
		\hline
		
		\hline
		
		$(\frac{49}{12},\frac{53}{12})$ & $3,\frac{9}{2},\frac{3}{2}$ & $\mathfrak{su}(2)_{\frac{9}{2}}\times\mathfrak{su}(2)_{\frac{11}{2}}\times\mathfrak{u}(1)_{f_{0}}$ & $W_{-18+\frac{9}{2}}(\mathfrak{e}_{7},2A_{2}+A_{1})+M(1)$ & ~ &\\
		\hline
		
		\hline
		
		$(\frac{77}{24},\frac{7}{2})$ & $\frac{10}{3},\frac{8}{3},\frac{4}{3}$ & $\mathfrak{spin}(7)_{\frac{10}{3}}$ & {$V_{-5+\frac{5}{3}}(\mathfrak{so}(7))$ }& $[3^2,1]$ &  $D_3(SO(7))$\\
		\hline
		$(\frac{77}{24},\frac{7}{2})$ & $\frac{10}{3},\frac{8}{3},\frac{4}{3}$ & $\mathfrak{sp}(3)_{\frac{8}{3}}$ & {$V_{-4+\frac{4}{3}}(C_{3})$} & $[3^2]$ &  $D_3(Sp(3))$\\
		\hline
		
		$(\frac{19}{6},\frac{41}{12})$ & $\frac{10}{3},\frac{8}{3},\frac{4}{3}$ & $\mathfrak{sp}(2)_{6}\times\mathfrak{su}(2)_{\frac{13}{6}}$ & $W_{-10+\frac{10}{3}}(D_{6},[3,2^{2},1^{5}])$ & $\overline{\CO}_{[3^4]}\cap S_{[3,2^2,1^5]}$  &\\
		\hline

		\hline
		
		$*(\frac{7}{2},4)$ & $\frac{7}{2},\frac{5}{2},\frac{3}{2}$ & $\mathfrak{su}(7)_{\frac{7}{2}}$ & {$V_{-7+\frac{7}{2}}(\mathfrak{su}(7))$} & $[2^3,1]$  & $D_2(SU(7))$ \\
		\hline
		
		\hline
		
		$(\frac{71}{24},\frac{46}{15})$ & $\frac{18}{5},\frac{12}{5},\frac{6}{5}$ & $\mathfrak{su}(2)_{\frac{23}{5}}$ & $W_{-18+\frac{14}{5}}(\mathfrak{e}_{7},A_{5}+A_{1})$ & $\overline{\CO}_{E_7(a_5)} \cap S_{A_5+A_1}$  & \\
		\hline
		
		\hline
		
		$(\frac{83}{24},\frac{11}{3})$ & $4,2,2$ & $\mathfrak{sp}(2)_{\frac{5}{2}}\times\mathfrak{su}(2)_{4}$ & $W_{-30+8}(\mathfrak{e}_{8},A_{3}+2A_{1})$ & $\overline{\CO}_{D_4(a_1)+A_2} \cap S_{A_{3}+2A_{1}} $ &\\
		\hline
		
		\hline
		
		$(\frac{83}{24},\frac{11}{3})$ & $4,\frac{8}{3},\frac{4}{3}$ & $\mathfrak{su}(2)_{5}\times\mathfrak{su}(2)_{4}$ & $W_{-18+\frac{14}{3}}(\mathfrak{e}_{7},2A_{2}+A_{1})$ & $\overline{\CO}_{A_3+A_2+A_1}\cap S_{2A_2+A_1}$  &\\
		\hline
		
		\hline
		$*(\frac{107}{24},\frac{31}{6})$ & $4,3,2$ & $\mathfrak{su}(8)_{4}\times\mathfrak{u}(1)_{f_{0}}$ & $V_{-8+4}(\mathfrak{su}(8))+M(1)$ & ~ &\\
		
		\hline
		
		$(\frac{101}{24},\frac{14}{3})$ & $4,3,2$ & $\mathfrak{sp}(3)_{3}\times\mathfrak{u}(1)^{2}_{f_{0}}$ & $W_{-10+3}(D_{6},[2^{6}])+M(1)^{\oplus 2}$ & ~ &\\		
		
		\hline
		$(\frac{25}{6},\frac{55}{12})$ & $4,3,2$ & $\mathfrak{sp}(2)_{3}\times\mathfrak{su}(2)_{\frac{5}{2}}\times\mathfrak{u}(1)^{2}_{f_{0}}$ & $W_{-10+3}(D_{6},[3,2^{2},1^{5}])+M(1)^{\oplus 2}$ & ~ &\\
		\hline
		
		$(\frac{15}{4},\frac{15}{4})$ & $4,3,2$ & $\mathfrak{su}(2)_{\frac{15}{2}}$ & $W_{-9+2}(\mathfrak{f}_{4},A_{2}+\tilde{A}_{1})$ & $\overline{\CO}_{F_4(a_3)} \cap S_{A_2+\tilde{A}_1}$  &\\
		\hline

		\hline
		
		$(\frac{39}{8},\frac{11}{2})$ & $4,4,2$ & $\mathfrak{so}(7)_{4}\times\mathfrak{u}(1)_{6}\times\mathfrak{u}(1)^{2}_{f_{0}}$ & $W_{-12+4}(\mathfrak{e}_{6},2A_{1})+M(1)^{\oplus 2}$ & ~ &\\
		\hline
		$(\frac{39}{8},\frac{11}{2})$ & $4,4,2$ & $\mathfrak{so}(7)_{4}\times\mathfrak{su}(2)_{12}$ & $W_{-30+10}(\mathfrak{e}_{8},A_{2}+2A_{1})$ & $\overline{\CO}_{D_4(a_1)+A_1}\cap S_{A_{2}+2A_{1}}$  &\\
		\hline
		
		\hline
		$(\frac{47}{12},\frac{17}{4})$ & $\frac{14}{3},\frac{8}{3},\frac{4}{3}$ & $\mathfrak{sp}(3)_{\frac{17}{6}}$ & $W_{-5+\frac{5}{3}}(C_{4},[2,1^{6}])$ & $\overline{CO}_{[3^2,2]} \cap S_{[2,1^{6}]} $ &\\
		\hline
		
		\hline
		$(\frac{17}{3},\frac{19}{3})$ & $5,\frac{7}{2},3$ & $\mathfrak{sp}(5)_{\frac{7}{2}}\times\mathfrak{u}(1)_{f_{0}}$ & $V_{-6+\frac{5}{2}}(C_{5})+M(1)$ & ~ &\\
		\hline
		
		\hline
		
		$(\frac{143}{24},\frac{20}{3})$ & $5,4,3$ & $\mathfrak{so}(9)_{5}\times\mathfrak{su}(2)_{3}\times\mathfrak{u}(1)_{f_{0}}$ & $W_{-18+7}(\mathfrak{e}_{7},2A_{1})+M(1)$ & ~ &\\
		\hline
		
		\hline
		
		$(\frac{19}{4},\frac{21}{4})$ & $6,2,2$ & $\mathfrak{sp}(3)_{\frac{7}{2}}$ & $W_{-18+6}(\mathfrak{e}_{7},4A_{1})$ & $ \overline{\CO}_{D_4(a_1)}\cap S_{4A_1}$  &\\
		\hline
		
		\hline
		
		$*(\frac{45}{8},\frac{13}{2})$ & $6,3,2$ & $(\mathfrak{e}_{6})_{6}$ & {$V_{-12+6}(\mathfrak{e}_{6})$ }& $A_2$ &\\
		\hline
		
		\hline
		
		$(\frac{133}{24},\frac{35}{6})$ & $6,4,2$ & $\mathfrak{sp}(2)_{7}$ & $W_{-30+8}(\mathfrak{e}_{8},2A_{2}+2A_{1})$ & $\overline{\CO}_{D_4(a_1)+A_2} \cap S_{2A_{2}+2A_{1}}$   &\\
		\hline
		
		\hline
		
		$(\frac{43}{6},\frac{97}{12})$ & $6,6,2$ & $(\mathfrak{f}_{4})_{6}\times\mathfrak{su}(2)_{\frac{7}{2}}$ & $W_{-30+12}(\mathfrak{e}_{8},3A_{1})$ & $\overline{\CO}_{2A_2}\cap S_{3A_1}$ &\\
		\hline
		
		\hline
		
		$*(\frac{75}{8},11)$ & $9,5,3$ & $(\mathfrak{e}_{7})_{9}\times\mathfrak{u}(1)_{f_{0}}$ & $V_{-18+9}(\mathfrak{e}_{7})+M(1)$ & ~ &\\
		\hline
		$(\frac{75}{8},11)$ & $9,5,3$ & $(\mathfrak{e}_{7})_{9}$ & $W_{-30+15}(\mathfrak{e}_{8},A_{1})$ & $\overline{\CO}_{A_2+A_1}\cap S_{A_1}$  &\\
		\hline

	\end{tabular}
	\caption{\label{table:rank3sum1}List of rank three AD theories with central charges $(a,c)$, CB spectrum $\Delta_{\mathrm{Coulomb}}$, flavor symmetry algebra $\fkf$, one corresponding VOA and Higgs branch. Theories with $*$ saturate the bound \ref{eq:boundSCFTs}. $\fku(1)$ with a subscript $f_0$ means the $\fku(1)$ algebra from the irregular singularity. $W\oplus M(1)^{\oplus f_0}$ means the actual VOA is possibly an extension of $W\oplus M(1)^{\oplus f_0}$. In the last coulomb, a single nilpotent orbit $f$ means that the corresponding Higgs branch or associated variety is the closure of the nilpotent orbit $\overline{\CO}_f$, while $S_{f'}$ means the Slodowy slice of $f'$.}
	
\end{table}

\begin{table}[H]\tiny
	\begin{tabular}{|c|c|c|c|}
 		\hline
 		$(a,c)$ & $\Delta_{\mathrm{Coulomb}}$ & $\mathfrak{f}$ & VOA \\
 		\hline
		
 		\hline
		 $*(\frac{43}{40},\frac{11}{10})=3(\frac{43}{120},\frac{11}{30})$ & $\frac{6}{5},\frac{6}{5},\frac{6}{5}$ & $-$ & $W_{-18+\frac{6}{5}}(\mathfrak{e}_{7},E_{7}(a_{1}))$\\
 		\hline
		
		\hline
		 $(\frac{157}{120},\frac{41}{30})={\color{red}(\frac{19}{20},1)}+{\color{blue} (\frac{43}{120},\frac{11}{30})}$ & ${\color{red}\frac{8}{5},\frac{6}{5}},{\color{blue}\frac{6}{5}}$ & ${\color{red}\mathfrak{su}(2)_{\frac{8}{5}}}$ & $W_{-7+\frac{7}{5}}(B_{4},[5,2^{2}])$\\
 		\hline
		
 		\hline
		$*(\frac{11}{8},\frac{3}{2})=3(\frac{11}{24},\frac{1}{2})$ & $\frac{4}{3},\frac{4}{3},\frac{4}{3}$ & $\mathfrak{su}(2)^{3}_{\frac{4}{3}}$ & $W_{-18+\frac{14}{3}}(\mathfrak{e}_{7},D_{4}(a_{1}))$\\
 		\hline
		
		\hline
		$(\frac{37}{24},\frac{5}{3})={\color{red}(\frac{13}{12},\frac{7}{6})}+{\color{blue}(\frac{11}{24},\frac{1}{2})}$ & ${\color{red}\frac{5}{3},\frac{4}{3}},{\color{blue}\frac{4}{3}}$ & ${\color{red}\mathfrak{su}(2)_{\frac{5}{3}}\times\mathfrak{u}(1)_{f_0}}\times{\color{blue}\mathfrak{su}(2)_{\frac{4}{3}}}$ & $W_{-18+\frac{14}{6}}(\mathfrak{e}_{7},D_{5})+M(1)$\\
 		\hline
		
		\hline
		$(\frac{25}{12},\frac{9}{4})={\color{red}(\frac{13}{8},\frac{7}{4})}+{\color{blue}(\frac{11}{24},\frac{1}{2})}$ & ${\color{red}\frac{8}{3},\frac{4}{3}},{\color{blue}\frac{4}{3}}$ & ${\color{red}\mathfrak{su}(2)_{\frac{11}{6}}\times\mathfrak{su}(2)_{\frac{8}{3}}}\times{\color{blue}\mathfrak{su}(2)_{\frac{4}{3}}}$ & $W_{-18+\frac{14}{3}}(\mathfrak{e}_{7},(A_{3}+A_{1})')$\\
 		\hline
		
		 \hline
 		$(\frac{61}{24},\frac{17}{6})={\color{red}(\frac{47}{24},\frac{13}{6})}+{\color{blue}(\frac{7}{12},\frac{2}{3})}$ & ${\color{red} 3,\frac{3}{2}},{\color{blue}\frac{3}{2}}$ & ${\color{red}\fsu(2)_3\times\fsu(2)_2}\times{\color{blue}\fsu(2)_{\frac{3}{2}}\times \mathfrak{u}_{f_0}}$ & $W_{-18+\frac{9}{2}}(\mathfrak{e}_{7},(A_{3}+A_{1})')+M(1)$\\
 		\hline

 	\end{tabular}
 	\caption{\label{table:rank3sum1decomp}Rank 3 theories that are likely to be several decoupled lower rank theories. Data with the same color come from the same theory.}

\end{table}

 \begin{table}[H]\scriptsize
	\centering
	\begin{tabular}{|c|c|c|c|}
		\hline
		$(a,c)$ & $\Delta_{\mathrm{Coulomb}}$ & $\mathfrak{f}$ & Fixtures \\
		\hline
		
		\hline
		
		$(\frac{7}{4},2)$ & $2,2$ & $\fsu(2)^5$ & $[1^2],[1^2],[1^2],[1^2],[1^2]$ \\
		\hline
		
		\hline
		$*(\frac{37}{12},\frac{11}{3})$ & $4,2$ &  $\fso(12)$  & $[3,1],[2^2],[2^2],[1^4]$  \\
		\hline
		
		$(3,\frac{7}{2})$ & $4,2$ &  $\fso(8)\times \fsu(2)$  & $[2^2],[2^2],[2^2],[2,1^2]$  \\
		\hline
		
		$(\frac{15}{4},\frac{9}{2})$ & $4,3$ & $\fsu(8)\times \fsu(2)$ & $[2,1^2],[1^4],[1^4]$  \\
		\hline
		
		\hline
				$*(\frac{53}{12},\frac{16}{3})$ & $5,3$ & $\fso(14)\times \mathfrak{u}(1)$ & $[2^2,1],[2^2,1],[1^5]$ \\
		\hline
				
		$(\frac{61}{12}, \frac{37}{6})$ & $5,4$ & $\fsu(10)$ & $[3,2],[1^5],[1^5]$ \\
		\hline

		\hline
		$(5,6)$ & $6,3$ & $\mathfrak{e}_6 \times \fsu(2)$ & $[2^3],[2^3],[2^2,1^2]$ \\
		\hline
		
		$(\frac{23}{4},7)$ & $6,4$ & $\fso(16) \times \fsu(2)$  & $[3^2],[2^2,1^2],[1^6]$  \\
		\hline

		\hline
		
		$(7,\frac{17}{2})$ & $8,4$ & $\mathfrak{e}_7 \times \fsu(2)$ & $[4^2],[2^4],[2^3,1^2]$  \\
		\hline

		$(\frac{101}{12},\frac{31}{3})$ & $8,6$ & $\fso(20)$ & $[4^2],[3^2,2],[1^8]$  \\
		\hline
		
		\hline
		$*(\frac{101}{12},\frac{31}{3})$ & $10,4$ & $\mathfrak{e}_8$ & $[5^2],[3^3,1],[2^5]$  \\
		\hline

		\hline
		$(11,\frac{27}{2})$ & $12,6$ & $\mathfrak{e}_8\times \fsu(2)$ & $[6^2],[4^3],[2^5,1^2]$ \\
		\hline
		
	\end{tabular}
	\caption{\label{table:rank2other}Rank two theories constructed from sphere with three or more regular singularities of A type. We list central charges, Coulomb branch spectrum, flavor symmetry and type of regular singularities. Theories with $*$ saturate the bound \ref{eq:boundSCFTs}. }
	
\end{table}

 \begin{table}[H]\scriptsize
	\centering
	\begin{tabular}{|c|c|c|c|}
		\hline
		$(a,c)$ & $\Delta_{\mathrm{Coulomb}}$ & $\mathfrak{f}$  & Fixtures \\
		\hline
		
		\hline
		
		$*(\frac{25}{8},\frac{7}{2})$ & $3, 2, 2 $ & $\fsu(4) \times \mathfrak{u}(1)^3$ & $[2,1],[2,1],[2,1],[2,1],[1^3]$ \\
		\hline
		
		\hline
		$*(\frac{89}{24},\frac{25}{6})$ & $4, 2, 2 $ & $\fso(8) \times \mathfrak{u}(1)$ & $[3,1],[3,1],[2^2],[2^2],[2^2]$ \\
		\hline

		$*(\frac{107}{24},\frac{31}{6})$ & $4, 3, 2 $ & $\fsu(8) \times \mathfrak{u}(1)$ & $[3,1],[3,1],[1^4],[1^4]$ \\
		\hline
		
		$(\frac{35}{8},5)$ & $4, 3, 2 $ & $\fsu(6) \times \fsu(2) \times \mathfrak{u}(1)$ & $[3,1],[2^2],[2,1^2],[1^4]$ \\
		\hline
		
		$(\frac{103}{24},\frac{29}{6}$ & $4, 3, 2 $ & $\fsu(4) \times \fsu(2)^2 \times \mathfrak{u}(1)$ & $[2^2],[2^2],[2,1^2],[2,1^2]$ \\
		\hline
		
		$(\frac{39}{8},\frac{11}{2}$ & $4, 4, 2 $ & $\fsu(4) \times \fsu(2)^3$ & $[2^2],[2^2],[2^2],[1^4]$ \\
		\hline
		
		$(\frac{45}{8},\frac{13}{2})$ & $4, 4, 3 $ & $\fsu(4)^3$ & $[1^4],[1^4],[1^4]$ \\
		\hline
		
		\hline
		$(\frac{51}{8},\frac{15}{2})$ & $5,4,3$ & $\fsu(10)\times \fsu(2)$ & $[3,1^2], [1^5],[1^5]$ \\
		\hline
		
		$(\frac{25}{4},\frac{29}{4})$ & $5,4,3$ & $\fsu(7)\times \fsu(3)\times \mathfrak{u}(1)$ & $[2^2,1],[2,1^3],[1^5]$ \\
		\hline

		\hline
		$*(\frac{45}{8},\frac{13}{2})$ & $6,3,2$ & $\mathfrak{e}_6$ & $[5,1],[3^2],[2^3],[2^3]$ \\
		\hline
		
		$*(\frac{51}{8},\frac{15}{2})$ & $6,4,2$ & $\fso(16)$ & $[5,1],[3^2],[3^2],[1^6]$ \\
		\hline
		
		$(\frac{145}{24},\frac{41}{6})$ & $6,4,2$ & $\fso(10)$ & $[4,2],[3^2],[3^2],[2^3]$ \\
		\hline
		
		$(6,\frac{27}{4})$ & $6,4,2$ & $\fso(8)\times \mathfrak{u}(1)$ & $[3^2],[3^2],[3^2],[3,2,1]$ \\
		\hline
		
		$(\frac{163}{24},\frac{47}{6})$ & $6,4,3$ & $\fso(10)\times \fsu(2)^2\times \mathfrak{u}(1)$ & $[2^3], [2^2,1^2],[2^2,1^2]$ \\
		\hline

		$(\frac{59}{8},\frac{17}{2})$ & $6,5,3$ & $\fsu(6)\times \fsu(4)$ & $[2^3],[2^3],[2,1^4]$ \\
		\hline

		$(\frac{15}{2},\frac{35}{4})$ & $6,5,3$ & $\fsu(9)\times \mathfrak{u}(1)$ & $[3,2,1],[2^3],[1^6]$ \\
		\hline
		
		$(\frac{199}{24},\frac{59}{6})$ & $6,5,4$ & $\fsu(12)$ & $[4,2],[1^6],[1^6]$ \\
		\hline
		
		$(\frac{65}{8},\frac{19}{2})$ & $6,5,4$ & $\fsu(8)\times \fsu(4)$ & $[3^2],[2,1^4],[1^6]$ \\
		\hline		
		
		\hline
		$*(\frac{179}{24},\frac{26}{3})$ & $7,4,3$ & $\mathfrak{e}_6\times \mathfrak{u}(1)^2$ & $[3^2,1],[2^3,1],[2^3,1]$ \\
		\hline
		
		$*(\frac{199}{24},\frac{59}{6})$ & $7,5,3$ & $\fso(18)\times \mathfrak{u}(1)$ & $[3^2,1],[3^2,1],[1^7]$ \\
		\hline
		
		$(\frac{75}{8},11)$ & $7,6,4$ & $\fsu(10)\times \mathfrak{u}(1)$ & $[4,3],[2^3,1],[1^7]$ \\
		\hline
		
		\hline
		$(\frac{223}{24},\frac{65}{6})$ & $8,5,4$ & $\mathfrak{e}_6\times \fsu(2)^2 \times \mathfrak{u}(1)$ & $[4^2],[2^3,1^2],[2^3,1^2]$ \\
		\hline
		
		$(\frac{73}{8},\frac{21}{2})$ & $8,6,3$ & $\fso(12)\times \mathfrak{u}(1)$ & $[3^2,2],[3^2,2],[2^4]$ \\
		\hline
		
		$(\frac{254}{24},\frac{73}{6})$ & $8,6,4$ & $\fso(20)\times \fsu(2)$ & $[4^2],[3^2,1^2],[1^8]$ \\
		\hline
		
		$(\frac{79}{8},\frac{23}{2})$ & $8,6,4$ & $\fso(12)\times \fsu(4)$ & $[4^2],[2^4],[2^2,1^4]$ \\
		\hline
		
		\hline
		$(\frac{39}{4},\frac{45}{4})$ & $9,6,3$ & $\mathfrak{e}_6\times \mathfrak{u}(1)$ & $[3^3],[3^3],[3^2,2,1]$ \\
		\hline
		
		$*(\frac{75}{8},11)$ & $9,5,3$ & $\mathfrak{e}_7\times \mathfrak{u}(1)$ & $[4^2,1],[3^3],[2^4,1]$ \\
		\hline
		
		$(\frac{105}{8},\frac{31}{2})$ & $9,8,6$ & $\fsu(12)$ & $[5,4],[3^3],[1^9]$ \\
		\hline
		
		\hline
		$(\frac{269}{24},\frac{79}{6})$ & $10,6,4$ & $\mathfrak{e}_7\times \fsu(2)\times \mathfrak{u}(1)$ & $[5^2],[3^3,1],[2^4,1^2]$ \\
		\hline
		
		$(\frac{293}{24},\frac{85}{6})$ & $10,8,4$ & $\fso(14)\times \fsu(2)$ & $[5^2],[3^2,2^2],[2^5]$ \\
		\hline
		
		$(\frac{337}{24},\frac{101}{6})$ & $10,8,6$ & $\fso(24)$ & $[5^2],[4^2,2],[1^{10}]$ \\
		\hline
		
		\hline
		$(\frac{27}{2},\frac{63}{4})$ & $12, 8, 4$ & $\mathfrak{e}_7\times \mathfrak{u}(1)$ & $[6^2],[3^4],[3^3,2,1]$ \\
		\hline
		
		$(\frac{119}{8},\frac{35}{2})$ & $12, 8, 6$ & $\mathfrak{e}_7\times \fsu(4)$ & $[6^2],[4^3],[2^4,1^4]$ \\
		\hline
		
		\hline
		$(\frac{147}{8},\frac{43}{2})$ & $14, 12, 6$ & $\fso(18)$ & $[7^2],[5^2,4],[2^7]$ \\
		\hline
		
		\hline
		$(21,\frac{99}{4})$ & $18, 12, 6$ & $\mathfrak{e}_8\times \mathfrak{u}(1)$ & $[9^2],[6^3],[3^5,2,1]$ \\
		\hline

	\end{tabular}
	\caption{\label{table:rank3int}Rank three theories  constructed from sphere with three or more regular singularities of  A type. We list central charges, Coulomb branch spectrum, flavor symmetry and type of regular singularities. Theories with $*$ saturate the bound \ref{eq:boundSCFTs}. }
\end{table}

\section{Implications on isomorphism of W-algebras}
\label{sec:VOAisom}
In our construction, it is common that one can engineer the same 4d $\mathcal{N}=2$ SCFTs by  quite different 
data.  Since two 4d theories are very likely to be the same if they share the same CB spectrum, central charges and the rank of flavor symmetry (there might be enhancement of the flavor symmetry), using observables such as CB spectrum, central charges and flavor symmetry, one can identify theories from different construction. Because each data $(\fkj, b, k, f)$ or $(\fkj,o,\fkg, b_t, k_t, f)$ leads to a corresponding VOA, the duality of 4d theories implies possible isomorphisms of W-algebras. 

In table \ref{table:isomRank1}, \ref{table:isomRank2}, \ref{table:isomRank3} and \ref{table:isomRank3more} we list different W-algebras which correspond to the same 4d theory. In each table we list the CB spectrum, central charges and flavor symmetry of the 4d theory (aka the affine vertex subalgebra in VOA) in the first three columns. Isomorphic VOAs are listed in the subsequent columns. Note that in the table when we write $W\oplus M(1)^{\oplus n}$, we  mean that the actual W-algebra might  be an extension of $W\oplus M(1)^{\oplus f_0}$\footnote{This happens when there are $f_0$ mass deformations from the irregular singularity}. W-algebras marked with $\ast$ means the naive flavor symmetry may enhance to the flavor symmetry listed in the third column. Finally, theories with only one corresponding VOA found are not listed here.
%
%

In some cases, one finds that the W-algebra is isomorphic to an affine vertex algebra. This phenomenon is called collapsing level \cite{adamovic2018conformal, adamovic2020application}. One may also read off possible collapsing levels of W-algebras from our tables. For example,
\begin{equation}
	\begin{aligned}
		&W_{-7+\frac{7}{3}}(\mathfrak{su}(7),[3,1^{4}])\cong V_{-\frac{8}{3}}(\mathfrak{su}(4)), \qquad W_{-7+\frac{7}{4}}(\mathfrak{su}(7),[4,1^{3}])\cong V_{-\frac{9}{4}}(\mathfrak{su}(3)),\\
		&W_{-9+\frac{9}{2}}(\mathfrak{su}(9),[2,1^{7}])\cong V_{-\frac{7}{2}}(\mathfrak{su}(7)), \qquad W_{-9+\frac{9}{7}}(\mathfrak{su}(9),[7,1^{2}])\cong V_{-\frac{12}{7}}(\mathfrak{su}(2)).\\
	\end{aligned}
\end{equation}
These results are examples of theorem 8.7 of \cite{Arakawa:2021ogm}.
%
One also finds other possible collapsing levels of W-algebras whose nilpotent orbits are not labelled by the partition of form $[s,1^{q}]$ as above. For example,
\begin{equation}
	W_{-8+\frac{8}{3}}(\mathfrak{su}(8),[2^{4}])\cong V_{-\frac{8}{3}}(\mathfrak{su}(4)), \qquad W_{-9+\frac{9}{4}}(\mathfrak{su}(9),[3^{3}])\cong V_{-\frac{9}{4}}(\mathfrak{su}(3)).
\end{equation}
These are examples of collapsing levels proved recently in \cite{Adamovic:2022nsr}. There are more examples of collapsing levels from our tables, some of which even involve non-admissible W-algebras.  

In \cite{Kaidi:2021tgr}, the authors construct a new rank 2 Argyres-Douglas theory, the Schur index of their proposed theory is,
\begin{equation}
	\mathcal{I}_{\text{AD}(C_{2})}=\text{PE}\left[\frac{1}{1-q}\sum_{i\in 3\mathbb{N}}(\chi^{C_{2}}_{\text{adj}}q^{i+1}+\chi^{C_{2}}_{\bm4}q^{i+\frac{3}{2}}-\chi^{C_{2}}_{\bm4}q^{i+\frac{5}{2}}-\chi^{C_{2}}_{\text{adj}}q^{i+3})\right]  
\end{equation}
This  theory  has the CB spectrum $(\frac{4}{3},\frac{10}{3})$, central charge $(a,c)=(2,\frac{13}{6})$ and flavor symmetry group $Sp(4)_{\frac{13}{3}}$. Using our results, one corresponding VOA of this theory is $W_{-10+\frac{10}{3}}(\mathfrak{so}(12),[3,2^{4},1])$. Since this W-algebra is admissible, its normalised character can be written down easily,
\begin{equation}
	\mathcal{I}_{W_{-10+\frac{10}{3}}(\fso(12),[3,2^{4},1])}(q,z)=\mathrm{PE}\left[\frac{q\chi^{C_{2}}_{\text{adj}}(z)+q^{\frac{3}{2}}\chi^{C_{2}}_{\bm {4}}(z)-q^{\frac{5}{2}}\chi^{C_{2}}_{\bm {4}}(z)-q^{3}\chi^{C_{2}}_{\text{adj}}(z)}{(1-q)(1-q^{3})}\right],
\end{equation}
which matches $\mathcal{I}_{AD(C_{2})}$ exactly. From this vacuum module of W-algebra, we can 
recognize this character not equal to the vacuum character of the affine vertex algebra $V_{-\frac{13}{6}}(\fsp(4))$
because of the extra terms $q^{\frac{3}{2}}\chi^{C_{2}}_{\bm {4}}(z)-q^{\frac{5}{2}}\chi^{C_{2}}_{\bm {4}}(z)$. The Higgs branch (associated variety of $W_{-10+\frac{10}{3}}(\fso(12),[3,2^{4},1])$) is $\overline{\CO}_{[3^4]} \cap S_{[3,2^4,1]}$.

This example might suggest that the VOA for $Z_2$ twisted $A_{2n}$ theory with the regular puncture being full type might not be the simple 
Kac-Moody algebra as predicted in \cite{Xie:2019zlb}, and there should be a finite extension. There are some strange features of $A_{2n}/Z_2$ twisted theories, 
and this fact might also be one of it.

{\linespread{2.0}
	\begin{table}[H]\tiny
			\caption{The VOA isomorphism from rank one  theories.}
		\centering
		\resizebox{\linewidth}{!}{
			\begin{tabular}{|c|c|c|c|c|c|}\hline
				{$\Delta_{\text{Coulomb}}$}&{$(a,c)$}&$\text{AKM subalgebra} $& {$W_{\tilde{k}}(\mathfrak{g},f)$}&{$W_{\tilde{k}}(\mathfrak{g},f)$}&{$W_{\tilde{k}}(\mathfrak{g},f)$}\\
				\hline
				$\frac{6}{5}$&\makecell{$(\frac{43}{120},\frac{11}{30})$}&\makecell{$-$}&\makecell{$W_{-3+\frac{3}{5}}(\mathfrak{su}(3),[3])$\\$W_{-6+\frac{4}{5}}(D_{4},[7,1])$\\$W_{-12+\frac{12}{5}}(\mathfrak{e}_{6},A_{4}+A_{1})$\\$W_{-18+\frac{14}{15}}(\mathfrak{e}_{7},E_{7})$} &\makecell{$W_{-7+\frac{7}{5}}(\mathfrak{su}(7),[5,2])$\\$W_{-10+\frac{6}{5}}(D_{6},[9,3])$\\$W_{-12+\frac{9}{10}}(\mathfrak{e}_{6},E_{6})$\\$W_{-30+\frac{24}{25}}(\mathfrak{e}_{8},E_{8})$}&\makecell{$W_{-8+\frac{8}{5}}(\mathfrak{su}(8),[5,3])$\\$W_{-6+\frac{6}{5}}(D_{4},[5,3])$\\$W_{-18+\frac{18}{5}}(\mathfrak{e}_{7},A_{4}+A_{2})$\\$ $}\\
				\hline
				{$\frac{4}{3}$}&\makecell{$(\frac{11}{24},\frac{1}{2})$}&\makecell{$\mathfrak{su}(2)_{-\frac{4}{3}}$}&\makecell{$W_{-4+\frac{4}{3}}(\mathfrak{su}(4),[2^{2}])$\\$W_{-7+\frac{7}{3}}(\mathfrak{su}(7),[3,2^{2}])$\\$W_{-2+\frac{1}{3}}(\mathfrak{su}(2),[2])+M(1)(*)$\\$W_{-7+\frac{7}{3}}(B_{4},[3^{2},1^{3}])$\\$W_{-5+\frac{5}{3}}(C_{4},[3^{2},1^{2}])$\\$W_{-10+\frac{10}{3}}(D_{6},[3^{3},1^{3}])$\\$W_{-8+\frac{5}{6}}(D_{5},[9,1])+M(1)(*)$\\$W_{-18+\frac{14}{3}}(\mathfrak{e}_{7},A_{3}+A_{2})$\\$W_{-30+\frac{30}{9}}(\mathfrak{e}_{8},D_{7}(a_{1}))(*)$}&\makecell{$W_{-4+\frac{4}{6}}(\mathfrak{su}(4),[4])+M(1)(*)$\\$W_{-8+\frac{8}{3}}(\mathfrak{su}(8),[3^{2},1^{2}])$\\$ W_{-5+\frac{5}{3}}(B_{3},[3,2^{2}])$\\$W_{-9+\frac{5}{3}}(B_{5},[7,2^{2}])$\\$W_{-6+\frac{4}{3}}(D_{4},[4^{2}])$\\$W_{-6+\frac{4}{3}}(D_{4},[5,1^{3}])$\\$W_{-8+\frac{8}{3}}(D_{5},[3^{2},2^{2}])$\\$W_{-18+\frac{14}{9}}(\mathfrak{e}_{7},E_{6})$\\$W_{-30+\frac{20}{3}}(\mathfrak{e}_{8},A_{4}+A_{2})$}&\makecell{$W_{-5+\frac{5}{3}}(\mathfrak{su}(5),[3,1^{2}])$\\$W_{-8+\frac{8}{6}}(\mathfrak{su}(8),[6,2])+M(1)(*)$\\$W_{-9+\frac{5}{3}}(B_{5},[5^{2},1])(*)$\\$W_{-4+\frac{4}{3}}(C_{3},[2^{3}])$\\$W_{-4+\frac{4}{3}}(D_{3},[3,1^{3}])$\\$W_{-4+\frac{4}{6}}(D_{3},(5,1))+M(1)(*)$\\$W_{-12+\frac{8}{9}}(\mathfrak{e}_{6},E_{6})(*)$\\$W_{-18+\frac{14}{6}}(\mathfrak{e}_{7},E_{7}(a_{4}))+M(1)(*)$\\$W_{-9+\frac{4}{3}}(\mathfrak{f}_{4},F_{4}(a_{2}))+M(1)(*)$} \\
				\hline
				{$\frac{3}{2}$}&\makecell{$(\frac{5}{8},\frac{3}{4})$}&\makecell{$\mathfrak{su}(2)_{-\frac{3}{2}}\times\mathfrak{u}(1)^{2}$}&\makecell{$W_{-4+\frac{3}{2}}(\mathfrak{su}(4),[1^{4}])+M(1)$\\$W_{-8+\frac{5}{2}}(D_{5},[3^{2},2^{2}])+M(1)$} &\makecell{$W_{-7+\frac{6}{4}}(\mathfrak{su}(7),[4,3])+M(1)^{\oplus 2}(*)$\\$W_{-10+\frac{6}{4}}(D_{6},[6^{2}])+M(1)^{\oplus 2}(*)$}&\makecell{$W_{-4+\frac{3}{2}}(D_{3},[2^{2},1^{2}])+M(1)$\\$$}\\ 
				\hline
				{$\frac{3}{2}$}&\makecell{$(\frac{7}{12},\frac{2}{3})$}&\makecell{$\mathfrak{su}(3)_{-\frac{3}{2}}$}&\makecell{$W_{-3+\frac{3}{2}}(\mathfrak{su}(3),[1^{3}])$\\$W_{-6+\frac{6}{4}}(\mathfrak{su}(6),[4,1^{2}])+M(1)(*)$\\$W_{-9+\frac{9}{2}}(\mathfrak{su}(9),[2^{3},1^{3}])$\\$W_{-8+\frac{5}{4}}[D_{5},(7,1^{3}])+M(1)(*)$\\$W_{-12+\frac{9}{4}}(\mathfrak{e}_{6},D_{4})$\\$W_{-30+\frac{30}{4}}(\mathfrak{e}_{8},A_{4}+A_{1})$}&\makecell{$W_{-3+\frac{3}{6}}(\mathfrak{su}(3),[3])+M(1)^{\oplus2}(*)$\\$W_{-6+\frac{6}{4}}(\mathfrak{su}(6),[3^{2}])+M(1)(*)$\\$W_{-9+\frac{9}{6}}(\mathfrak{su}(9),[6,3])+M(1)^{\oplus2}(*)$\\$W_{-4+\frac{3}{2}}(C_{3},[2^{2},1^{2}]+M(1)(*) $\\$W_{-6+\frac{6}{8}}D_{4},[7,1])+M(1)^{\oplus2}(*)$\\$W_{-18+\frac{14}{4}}(\mathfrak{e}_{7},D_{5}(a_{1}))+M(1)(*)$}&\makecell{$W_{-5+\frac{5}{2}}(\mathfrak{su}(5),[2,1^{3}])$\\$W_{-7+\frac{7}{2}}(\mathfrak{su}(7),[2^{2},1^{4}])$\\$W_{-2+\frac{1}{2}}(\mathfrak{su}(2),[1^{2}])+M(1)(*)$\\$W_{-6+\frac{5}{2}}(C_{5},[2^{4},1^{2}]+M(1)(*)$\\$W_{-12+\frac{9}{2}}(\mathfrak{e}_{6},A_{2}+A_{1})$\\$W_{-18+\frac{14}{4}}(\mathfrak{e}_{7},A_{4}+A_{2})+M(1)(*)$} \\
				\hline
				{$2$}&\makecell{$(\frac{23}{24},\frac{7}{6})$}&\makecell{$\mathfrak{so}(8)_{-2}$}&\makecell{$W_{-3+\frac{3}{3}}(\mathfrak{su}(3),[1^{3}])+M(1)^{\oplus2}(*)$\\$W_{-6+\frac{6}{2}}(\mathfrak{su}(6),[2,1^{4}])+M(1)(*)$\\$W_{-8+\frac{8}{2}}(\mathfrak{su}(8),[2^{2},1^{4}])+M(1)(*)$\\$W_{-9+\frac{9}{3}}(\mathfrak{su}(9),[3,2^{3}])+M(1)^{\oplus2}(*)$\\$W_{-4+\frac{3}{3}}(D_{3},[3,1^{3}])+M(1)^{\oplus3}(*)$\\$W_{-10+\frac{6}{1}}(D_{6},[2^{2},1^{8}])$\\$W_{-30+\frac{30}{3}}(\mathfrak{e}_{8},D_{4}(a_{1}))$}&\makecell{$W_{-4+\frac{4}{2}}(\mathfrak{su}(4),[1^{4}])+M(1)(*)$\\$W_{-6+\frac{6}{3}}(\mathfrak{su}(6),[3,1^{3}])+M(1)^{\oplus2}(*)$\\$W_{-8+\frac{8}{4}}(\mathfrak{su}(8),[4,2^{2}])+M(1)^{\oplus3}(*)$\\$W_{-7+2}(B_{4},[3^{2},1^{3}])+M(1)^{\oplus2}(*)$\\$W_{-6+\frac{4}{1}}(D_{4},[1^{8}])$\\$W^{-10+\frac{6}{2}}(D_{6},[3^{3},1^{3}])+M(1)^{\oplus2}(*)$\\$W_{-18+\frac{14}{2}}(\mathfrak{e}_{7},A_{2}+A_{1})+M(1)(*)$}&\makecell{$W_{-4+\frac{4}{4}}(\mathfrak{su}(4),[2^{2}])+M(1)^{\oplus3}(*)$\\$W_{-6+\frac{6}{3}}(\mathfrak{su}(6),[2^{3}])+M(1)^{\oplus2}(*)$\\$W_{-9+\frac{9}{3}}(\mathfrak{su}(9),[3^{2},1^{3}])+M(1)^{\oplus2}(*)$\\$W_{-9+5}(B_{5},(3,1^{8}))$\\$W_{-6+\frac{4}{4}}(D_{4},[5,3])+M(1)^{\oplus4}(*)$\\$W_{-4+\frac{4}{2}}(D_{3},[1^{6}])+M(1)(*)$\\$W_{-9+4}(\mathfrak{f}_{4},\tilde{A}_{1})+M(1)(*)$} \\
				\hline
				{$2$}&\makecell{$(\frac{23}{24},\frac{7}{6})$}&\makecell{$(\mathfrak{g}_{2})_{-2}$}&\makecell{$W_{-4+2}(\mathfrak{g}_{2},0)$\\$W_{-18+6}(\mathfrak{e}_{7},2A_{2})$} &\makecell{$W_{-9+3}(\mathfrak{f}_{4},\tilde{A}_{2})$\\$W_{-30+6}(\mathfrak{e}_{8},E_{6}(a_{3}))$}&\makecell{$W_{-18+6}(\mathfrak{e}_{7},A_{2}+3A_{1})$\\$$}\\ 
				\hline
				{$2$}&\makecell{$(\frac{23}{24},\frac{7}{6})$}&\makecell{$\mathfrak{su}(3)_{-2}$}&\makecell{$W_{-30+6}(\mathfrak{e}_{8},D_{4}+A_{2})$} &\makecell{$W_{-9+3}(\mathfrak{f}_{4},A_{2})$}&\makecell{$$}\\ 
				\hline
				{$2$}&\makecell{$(\frac{23}{24},\frac{7}{6})$}&\makecell{$\mathfrak{so}(7)_{-2}$}&\makecell{$W_{-5+3}(B_{3},[1^{7}])$\\$W_{-18+6}(\mathfrak{e}_{7},A_{3})$} &\makecell{$W_{-30+12}(\mathfrak{e}_{8},A_{2}+2A_{1})$\\$ $}&\makecell{$W_{-12+6}(\mathfrak{e}_{6},2A_{1})$\\$ $}\\ 
				\hline
				{$2$}&\makecell{$(\frac{3}{4},\frac{3}{4})$}&\makecell{$\mathfrak{su}(2)_{-\frac{3}{2}}$}&\makecell{$W_{-6+2}(D_{4},[3,2^{2},1])$\\$W_{-6+2}(C_{5},[3^{2},2,1^{2}])$} &\makecell{$W_{-30+8}(\mathfrak{e}_{8},A_{3}+A_{2}+A_{1})$\\$ $}&\makecell{$W_{-3+1}(B_{2},[2^{2},1])$\\$ $}\\ 
				\hline
				{$2$}&\makecell{$(1,\frac{5}{4})$}&\makecell{$\mathfrak{su}(3)_{-2}\times\mathfrak{u}(1)^{3}$}&\makecell{$W_{-5+\frac{4}{2}}(\mathfrak{su}(5),[2,1^{3}])+M(1)^{\oplus 2}$} &\makecell{$W_{-7+\frac{6}{3}}(\mathfrak{su}(7),[3,2^{2}])+M(1)^{\oplus 3}(*)$}&\makecell{$ $}\\ 
				\hline
				{$2$}&\makecell{$(1,\frac{5}{4})$}&\makecell{$\mathfrak{so}(7)_{-2}\times\mathfrak{su}(2)_{-\frac{1}{2}}$}&\makecell{$W_{-30+10}(\mathfrak{e}_{8},A_{3}+A_{1})$} &\makecell{$W_{-9+5}(B_{5},[2^{2},1^{7}])$}&\makecell{$ $}\\ 
				\hline
				{$3$}&\makecell{$(\frac{41}{24},\frac{13}{6})$}&\makecell{$(\mathfrak{e}_{6})_{-3}$}&\makecell{$W_{-30+\frac{30}{2}}(\mathfrak{e}_{8},A_{2})$} &\makecell{$W_{-8+\frac{5}{1}}(D_{5},[1^{10}])+M(1)(*)$}&\makecell{$W_{-6+\frac{6}{2}}(D_{4},[1^{8}])+M(1)^{\oplus2}(*)$}\\
				\hline
				{$3$}&\makecell{$(\frac{17}{12},\frac{19}{12})$}&\makecell{$\mathfrak{sp}(2)_{-2}\times\mathfrak{u}(1)$}&\makecell{$W_{-4+\frac{3}{2}}(C_{3},[2,1^{4}])+M(1)$} &\makecell{$W_{-6+\frac{5}{2}}(C_{5},[2^{3},1^{4}])+M(1)$}&\makecell{$$}\\
				\hline
				{$4$}&\makecell{$(\frac{59}{24},\frac{19}{6})$}&\makecell{$(\mathfrak{e}_{7})_{-4}$}&\makecell{$W_{-12+8}(\mathfrak{e}_{6},0)+M(1)(*)$} &\makecell{$W_{-30+20}(\mathfrak{e}_{8},A_{1})$}&\makecell{$ $}\\ 
				\hline
				{$4$}&\makecell{$(\frac{25}{12},\frac{29}{12})$}&\makecell{$\mathfrak{sp}(3)_{-\frac{5}{2}}\times\mathfrak{u}(1)$}&\makecell{$W_{-18+7}(\mathfrak{e}_{7},4A_{1})+M(1)$} &\makecell{$W_{-9+4}(\mathfrak{f}_{4},A_{1})+M(1)$}&\makecell{$ $}\\ 
				\hline
		\end{tabular}}
		\label{table:isomRank1}
\end{table}}

{\linespread{2.5}
	\begin{table}[H]\tiny
			\caption{The VOA isomorphism from rank two  theories.}
		\centering
		\resizebox{\linewidth}{!}{
			\begin{tabular}{|c|c|c|c|c|c|}\hline
				{$\Delta_{\text{Coulomb}}$}&{$(a,c)$}&\makecell{$\text{AKM subalgebra}$}& {$W^{\tilde{k}}(\mathfrak{g},f)$}&{$W^{\tilde{k}}(\mathfrak{g},f)$}&{$W^{\tilde{k}}(\mathfrak{g},f)$}\\
				\hline
				\makecell{$\frac{6}{5},\frac{6}{5}$}&\makecell{$(\frac{43}{60},\frac{11}{15})$}&\makecell{$-$}&\makecell{$W_{-12+\frac{12}{10}}(\mathfrak{e}_{6},E_{6}(a_{1}))$\\$W_{-4+\frac{4}{5}}(C_{3},[4,2])$}&\makecell{$W_{-12+\frac{9}{5}}(\mathfrak{e}_{6},E_{6}(a_{3}))$\\$W_{-5+\frac{3}{5}(B_{3},[7])} $}&\makecell{$W_{-7+\frac{7}{5}}(B_{4},[5,3,1])$\\$$}\\
				\hline
				\makecell{$\frac{4}{3},\frac{4}{3}$}&\makecell{$(\frac{111}{12},1$}&\makecell{$\mathfrak{su}(2)_{-\frac{4}{3}}\times\mathfrak{su}(2)_{-\frac{4}{3}}$\\}&\makecell{$W_{-8+\frac{5}{3}}(D_{5},[5,3,1^{2}])+M(1)(*)$\\$W_{-18+\frac{14}{3}}(\mathfrak{e}_{7},D_{4}(a_{1})+A_{1})$\\$W_{-3+\frac{2}{3}}(B_{2},[3,1^{2}])+M(1)(*)$\\$W_{-5+\frac{5}{3}}(C_{4},[2^{4}])$}&\makecell{$W_{-8+\frac{8}{3}}(D_{5},[3^{2},1^{4}])$\\$W_{-18+\frac{14}{6}}(\mathfrak{e}_{7},D_{5}+A_{1})+M(1)(*)$\\$W_{-7+\frac{2}{3}}(B_{4},[9])+M(1)^{\oplus 2}(*)$\\$W_{-9+\frac{4}{3}}(\mathfrak{f}_{4},B_{3})+M(1)(*)$}&\makecell{$W_{-8+\frac{8}{6}}(D_{5},[5^{2}])+M(1)(*)$\\$W_{-5+\frac{5}{3}}(B_{3},[3,1^{4}])$\\$W_{-9+\frac{5}{3}}(B_{5},[7,1^{4}])$}\\
				\hline
				\makecell{$\frac{10}{7},\frac{8}{7}$}&\makecell{$(\frac{67}{84},\frac{17}{21})$}&\makecell{$-$}&\makecell{$W_{-2+\frac{2}{7}}(\mathfrak{su}(2),[2])$
					\\$W_{-10+\frac{6}{7}}(D_{6},[11,1])$\\$W^{-30+\frac{30}{7}}(\mathfrak{e}_{8},A_{6}+A_{1})$\\$W_{-9+\frac{9}{7}}(\mathfrak{f}_{4},F_{4}(a_{2}))$}&\makecell{$W_{-5+\frac{5}{7}}(\mathfrak{su}(5),[5])$\\$W^{-8+\frac{8}{7}}(D_{5},[7,3])$\\$W_{-30+\frac{20}{21}}(\mathfrak{e}_{8},E_{8})$\\$ $}&\makecell{$W_{-9+\frac{9}{7}}(\mathfrak{su}(9),[7,2])$\\$W_{-10+\frac{10}{7}}(D_{6},[7,5])$\\$W_{-3+\frac{3}{7}}(B_{2},[5])$\\$ $}\\
				\hline
				\makecell{$\frac{3}{2},\frac{5}{4}$}&\makecell{$(\frac{11}{12},\frac{23}{24})$}&\makecell{$\mathfrak{u}(1)$}&\makecell{$W_{-2+\frac{2}{8}}(\mathfrak{su}(2),[2])+M(1)$\\$W_{-7+\frac{7}{4}}(\mathfrak{su}(7),[4,2,1])$\\$W_{-18+\frac{14}{8}}(\mathfrak{e}_{7},E_{7}(a_{3}))+M(1)$\\$W_{-6+\frac{5}{4}}(C_{5},[4^{2},2]+M(1)$}&\makecell{$W_{-5+\frac{5}{4}}(\mathfrak{su}(5),[3,2])$\\$W_{-9+\frac{9}{4}}(\mathfrak{su}(9),[4,3,2])$\\$W_{-30+\frac{30}{8}}(\mathfrak{e}_{8},E_{6}(a_{1})+A_{1})$\\$W_{-3+\frac{3}{4}}(\mathfrak{su}(3),[2,1]) $}&\makecell{$W_{-6+\frac{6}{8}}(\mathfrak{su}(6),[6])+M(1)$\\$W_{-12+\frac{9}{4}}(\mathfrak{e}_{6},A_{4}+A_{1})$\\$W_{-4+\frac{3}{4}}(C_{3},[4,2]+M(1)$\\$ $}\\
				\hline
				\makecell{$\frac{3}{2},\frac{3}{2}$}&\makecell{$(\frac{17}{6},\frac{4}{3})$}&\makecell{$\mathfrak{su}(3)_{-\frac{3}{2}}\times\mathfrak{su}(3)_{-\frac{3}{2}}$}&\makecell{$ W_{-8+\frac{5}{2}}(D_{5},[3^{2},1^{4}])+M(1)(*)$}&\makecell{$W_{-6+\frac{6}{4}}(D_{4},[3^{2},1^{2}])+M(1)^{\oplus 2}(*)$}&\makecell{$W_{-12+\frac{9}{2}}(\mathfrak{e}_{6},A_{2})$}\\
				\hline
				\makecell{$\frac{8}{5},\frac{6}{5}$}&\makecell{$(\frac{19}{20},1)$}&\makecell{$\mathfrak{su}(2)_{-\frac{8}{5}}$}&\makecell{$W_{-2+\frac{2}{5}}(\mathfrak{su}(2),[1^{2}])$\\$ W_{-4+\frac{3}{5}}(\mathfrak{su}(4),[4])+M(1)(*)$\\$W_{-6+\frac{6}{5}}(D_{4},[4^{2}])$\\$W_{-12+\frac{12}{5}}(\mathfrak{e}_{6},A_{4})$}&\makecell{$W_{-7+\frac{7}{5}}(\mathfrak{su}(7),[5,1^{2}])$\\$W_{-4+\frac{3}{5}}(D_{3},[5,1])+M(1)(*)$\\$W_{-6+\frac{6}{5}}(D_{4},[5,1^{3}])$\\$W_{-3+\frac{2}{5}}(B_{2},[5])+M(1)(*)$}&\makecell{$W_{-8+\frac{8}{5}}(\mathfrak{su}(8),[4^{2}])$\\$W_{-10+\frac{6}{5}}(D_{6},[9,1^{3}])$\\$W_{-8+\frac{8}{10}}(D_{5},[9,1])+M(1)(*)$\\$ $}\\
				\hline
				\makecell{$\frac{5}{3},\frac{4}{3}$}&\makecell{$(\frac{13}{12},\frac{7}{6})$}&\makecell{$\mathfrak{su}(2)_{-\frac{5}{3}}\times\mathfrak{u}(1)$}&\makecell{$W_{-2+\frac{2}{6}}(\mathfrak{su}(2),[1^{2}])+M(1)$\\$\mathcal{T}[W_{-7+\frac{7}{3}}(\mathfrak{su}(7),[3,2,1^{2}])]$\\$W_{-3+\frac{2}{3}}(\mathfrak{su}(3),[2,1])+M(1)(*)$\\$W_{-10+\frac{10}{3}}(D_{6},[3^{2},2^{2},1^{2}])$\\$W_{-18+\frac{14}{6}}(\mathfrak{e}_{7},A_{6})+M(1)$\\$W_{-7+\frac{7}{3}}(B_{4},[3,2^{2},1^{2}])$}&\makecell{$W_{-4+\frac{4}{3}}(\mathfrak{su}(4),[2,1^{2}]) $\\$W_{-8+\frac{8}{3}}(\mathfrak{su}(8),[3,2^{2},1])$\\$W_{-5+\frac{4}{6}}(\mathfrak{su}(5),[5])+M(1)^{\oplus 2}(*)$\\$W_{-10+\frac{10}{12}}(D_{6},[11,1])+M(1)^{\oplus 2}(*)$\\$W_{-30+\frac{30}{9}}(\mathfrak{e}_{8},E_{7}(a_{3}))$\\$W_{-4+\frac{4}{3}}(C_{3},[2^{2},1^{2}])$}&\makecell{$W_{-5+\frac{5}{3}}(\mathfrak{su}(5),[2^{2},1])$\\$W_{-8+\frac{8}{6}}(\mathfrak{su}(8),[6,1^{2}])+M(1)$\\$W_{-4+\frac{4}{3}}(D_{3},[2^{2},1^{2}])$\\$W_{-18+\frac{14}{6}}(\mathfrak{e}_{7},D_{6}(a_{1}))+M(1)$\\$W_{-30+\frac{20}{3}}(\mathfrak{e}_{8},A_{4}+2A_{1})$\\$W_{-9+\frac{4}{3}}(\mathfrak{f}_{4},C_{3})+M(1)$}\\
				\hline
				\makecell{$2,2$}&\makecell{$(\frac{7}{4},2)$}&\makecell{$\mathfrak{su}(2)_{-2}\times\mathfrak{su}(2)_{-2}\times\mathfrak{u}(1)^{3}$}&\makecell{$W_{-4+\frac{4}{4}}(\mathfrak{su}(4),[2,1^{2}])+M(1)_{\oplus 3}(*)$\\$W_{-6+\frac{6}{3}}(\mathfrak{su}(6),[2^{2},1^{2}])+M(1)_{\oplus 2}$\\$W_{-8+\frac{8}{4}}(\mathfrak{su}(8),[3^{2},2])+M(1)^{\oplus 3}(*)$\\$W_{-6+\frac{4}{4}}(D_{4},[4^{2}])+M(1)^{\oplus 4}(*)$}&\makecell{$W_{-5+\frac{5}{5}}(\mathfrak{su}(5),[3,2])+M(1)^{\oplus 4}(*)$\\$W_{-8+\frac{8}{4}}(\mathfrak{su}(8),[4,2,1^{2}])+M(1)^{\oplus 3}(*)$\\$W_{-9+\frac{9}{3}}(\mathfrak{su}(9),[3,2^{2},1^{2}])+M(1)^{\oplus2}$\\$W_{-6+\frac{4}{4}}(D_{4},[5,1^{3}])+M(1)^{\oplus 4}(*)$\\}&\makecell{$W_{-10+3}(D_{6},[3^{2},2^{2},1^{2}])+M(1)^{\oplus 2}(*)$\\$W_{-4+1}(D_{3},[2^{2},1^{2}])+M(1)^{\oplus 3}(*)$\\$W_{-8+1}(D_{5},[7,3])]+M(1)^{\oplus5}(*)$\\$ $}\\
				\hline
				\makecell{$2,2$}&\makecell{$(\frac{19}{12},\frac{5}{3})$}&\makecell{$\mathfrak{sp}(2)_{-2}$}&\makecell{$ W_{-6+\frac{4}{2}}(D_{4},(2^{4}))$\\$W_{-6+\frac{4}{2}}(D_{4},[3,1^{5}])$\\$W_{-9+3}(B_{5},[3^{2},1^{5}])$}&\makecell{$W_{-30+\frac{24}{3}}(\mathfrak{e}_{8},A_{3}+A_{2})$\\$W_{-3+1}(B_{2},[1^{5}])$\\$W_{-6+2}(C_{5},[3^{2},1^{4}])$}&\makecell{$W_{-9+3}(B_{5},[3,2^{4}])$\\$W_{-6+2}(C_{5},[2^{5}])$\\$$}\\
				\hline
				\makecell{$\frac{12}{5},\frac{6}{5}$}&\makecell{$(\frac{163}{120},\frac{17}{12})$}&\makecell{$\mathfrak{su}(2)_{-\frac{17}{10}}$}&\makecell{$W_{-12+\frac{9}{5}}(\mathfrak{e}_{6},A_{5})$}&\makecell{$W_{-7+\frac{7}{5}}(B_{4},[4^{2},1])$}&\makecell{$W_{-4+\frac{4}{5}}(C_{3},[4,1^{2}])$}\\
				\hline
				\makecell{$\frac{5}{2},\frac{3}{2}$}&\makecell{$(\frac{7}{4},2)$}&\makecell{$\mathfrak{su}(5)_{-\frac{5}{2}}$}&\makecell{$W_{-5+\frac{5}{2}}(\mathfrak{su}(5),[1^{5}])$\\$W_{-4+\frac{3}{2}}(\mathfrak{su}(4),[1^{4}])+M(1)(*)$\\$W_{-30+\frac{30}{4}}(\mathfrak{e}_{8},A_{4})$}&\makecell{$W_{-7+\frac{7}{2}}(\mathfrak{su}(7),[2,1^{5}])$\\$W_{-4+\frac{3}{2}}(D_{3},[1^{6}])+M(1)(*)$\\$$}&\makecell{$W_{-9+\frac{9}{2}}(\mathfrak{su}(9),[2^{2},1^{5}])$\\$W_{-9+\frac{5}{2}}(B_{5},[5,1^{6}])+M(1)(*)$\\$ $}\\
				\hline
				\makecell{$\frac{5}{2},\frac{3}{2}$}&\makecell{$(\frac{19}{12},\frac{5}{3})$}&\makecell{$\mathfrak{su}(2)_{-\frac{7}{4}}\times\mathfrak{u}(1)$}&\makecell{$ W_{-4+\frac{3}{4}}(C_{3},[4,1^{2}])+M(1)$}&\makecell{$W_{-6+\frac{5}{4}}(C_{5},[4,3^{2}])+M(1)$}&\makecell{$ $}\\
				\hline
				\makecell{$\frac{8}{3},\frac{4}{3}$}&\makecell{$(\frac{13}{8},\frac{7}{4})$}&\makecell{$\mathfrak{su}(2)_{-\frac{11}{6}}\times\mathfrak{su}(2)_{-\frac{8}{3}}$}&\makecell{$W_{-8+\frac{5}{3}}(D_{5},[5,2^{2},1])+M(1)(*)$\\$W_{-5+\frac{5}{3}}(B_{3},[2^{2},1^{3}])$\\$W_{-5+\frac{5}{3}}(C_{4},[2^{3},1^{2}])$}&\makecell{$W_{-8+\frac{8}{3}}(D_{5},[3,2^{2},1^{3}])$\\$W_{-3+\frac{2}{3}}(B_{2},[2^{2},1])+M(1)(*)$\\$ $}&\makecell{$W_{-18+\frac{14}{3}}(\mathfrak{e}_{7},A_{3}+2A_{1})$\\$W_{-7+\frac{4}{3}}(B_{4},[4^{2},1])+M(1)(*)$\\$ $}\\
				\hline
				\makecell{$3,\frac{3}{2}$}&\makecell{$(\frac{47}{24},\frac{13}{6})$}&\makecell{$\mathfrak{su}(2)_{-2}\times\mathfrak{su}(3)_{-3}\times\mathfrak{u}(1)$}&\makecell{$W_{-8+\frac{5}{2}}(D_{5},[3,2^{2},1^{3}])+M(1)(*)$\\$W_{-5+\frac{3}{2}}(B_{3},[2^{2},1^{3}])+M(1)(*) $}&\makecell{$W_{-6+\frac{6}{4}}(D_{4},[3,2^{2},1])+M(1)^{\oplus 2}(*)$\\$W_{-9+\frac{5}{2}}(B_{5},[3^{2},2^{2},1])+M(1)(*)$}&\makecell{$W_{-12+\frac{9}{2}}(\mathfrak{e}_{6},3A_{1})$\\$ $}\\
				\hline
				\makecell{$3,2$}&\makecell{$(\frac{29}{12},\frac{17}{6})$}&\makecell{$\mathfrak{su}(6)_{-3}\times\mathfrak{u}(1)$}&\makecell{$ W_{-6+\frac{6}{2}}(\mathfrak{su}(6),[1^{6}])+M(1)$\\$W_{-5+\frac{4}{2}}(\mathfrak{su}(5),[1^{5}])+M(1)^{\oplus 2}(*)$}&\makecell{$W_{-8+\frac{8}{2}}(\mathfrak{su}(8),[2,1^{6}])+M(1)$\\$ $}&\makecell{$W_{-18+\frac{14}{2}}(\mathfrak{e}_{7},A_{2})+M(1)$\\$ $}\\
				\hline
				\makecell{$3,2$}&\makecell{$(2,2)$}&\makecell{$\mathfrak{su}(2)_{-4}$}&\makecell{$W_{-12+\frac{12}{4}}(\mathfrak{e}_{6},2A_{2}+A_{1})$}&\makecell{$W_{-4+1}(\mathfrak{g}_{2},A_{1})$}&\makecell{$ $}\\
				\hline
				\makecell{$3,\frac{5}{2}$}&\makecell{$(\frac{61}{24},\frac{17}{6})$}&\makecell{$\mathfrak{sp}(3)_{-\frac{5}{2}}\times\mathfrak{u}(1)$}&\makecell{$W_{-18+\frac{14}{4}}(\mathfrak{e}_{7},D_{4})+M(1)$}&\makecell{$W_{-4+\frac{3}{2}}(C_{3},[1^{6}])+M(1)$}&\makecell{$W_{-6+\frac{5}{2}}(C_{5},[2^{2},1^{6}])+M(1)$}\\
				\hline
				\makecell{$\frac{10}{3},\frac{4}{3}$}&\makecell{$(2,\frac{13}{6})$}&\makecell{$\mathfrak{sp}(2)_{-\frac{13}{6}}$}&\makecell{$W_{-10+\frac{10}{3}}(D_{6},[3,2^{4},1])$}&\makecell{$W_{-7+\frac{7}{3}}(B_{4},[2^{4},1])$}&\makecell{$W_{-4+\frac{4}{3}}(C_{3},[2,1^{4}])$}\\
				\hline
				\makecell{$4,2$}&\makecell{$(\frac{37}{12},\frac{11}{3})$}&\makecell{$\mathfrak{so}(12)_{-4}$}&\makecell{$ W_{-10+6}(D_{6},[1^{12}])$\\$W_{-30+\frac{30}{3}}(\mathfrak{e}_{8},A_{3})$}&\makecell{$W_{-8+\frac{8}{2}}(D_{5},[1^{10}])+M(1)(*)$\\$$}&\makecell{$W_{-9+5}(B_{5},[1^{11}])$\\$$}\\
				\hline
				\makecell{$4,2$}&\makecell{$(\frac{17}{6},\frac{19}{6})$}&\makecell{$\mathfrak{sp}(2)_{-\frac{5}{2}}\times\mathfrak{u}(1)$}&\makecell{$ W_{-10+3}(D_{6},[3,2^{4},1])+M(1)^{\oplus 2}$}&\makecell{$W_{-7+2}(B_{4},[2^{4},1])+M(1)^{\oplus}(*)$}&\makecell{$W_{-9+5}(B_{5},[1^{11}])$}\\
				\hline
				\makecell{$5,3$}&\makecell{$(\frac{53}{12},\frac{16}{3})$}&\makecell{$(\mathfrak{so}_{13})_{-\frac{5}{2}}$}&\makecell{$ W_{-18+\frac{18}{2}}(\mathfrak{e}_{7},A_{1})+M(1)(*)$}&\makecell{$W_{-30+\frac{30}{2}}(\mathfrak{e}_{8},2A_{1})$}&\makecell{$ $}\\
				\hline
				\makecell{$5,4$}&\makecell{$(\frac{14}{3},\frac{16}{3})$}&\makecell{$(\mathfrak{f}_{4})_{-\frac{5}{2}}\times\mathfrak{u}(1)$}&\makecell{$ W_{-18+\frac{14}{2}}(\mathfrak{e}_{7},(3A_{1})''+M(1)$}&\makecell{$W_{-9+4}(\mathfrak{f}_{4},0)+M(1)$}&\makecell{$$}\\
				\hline
				
		\end{tabular}}
		\vspace{-2em}
		\label{table:isomRank2}
\end{table}}

{\linespread{2.5}
	\begin{table}[H]\tiny
			\caption{The VOA isomorphism from rank three  theories}
		\centering
		\resizebox{\linewidth}{!}{
			\begin{tabular}{|c|c|c|c|c|c|}\hline
				{$\Delta_{\text{Coulomb}}$}&{$(a,c)$}&$\text{AKM subalgebra} $& {$W_{\tilde{k}}(\mathfrak{g},f)$}&{$W_{\tilde{k}}(\mathfrak{g},f)$}&{$W_{\tilde{k}}(\mathfrak{g},f)$}\\
				\hline
				{$\frac{6}{5},\frac{6}{5},\frac{6}{5}$}&\makecell{$(\frac{43}{30},\frac{11}{10})$}&\makecell{$-$}&\makecell{$W_{-4+\frac{4}{5}}(\mathfrak{g}_{2},G_{2}(a_{1}))$\\$W_{-18+\frac{14}{5}}(\mathfrak{e}_{7},E_{7}(a_{5}))$}&\makecell{$W_{-4+\frac{2}{5}}(\mathfrak{g}_{2},G_{2})$\\$$}&\makecell{$W_{-18+\frac{18}{15}}(\mathfrak{e}_{7},E_{7}(a_{1}))$\\$ $}\\
				\hline
				{$\frac{4}{3},\frac{4}{3},\frac{4}{3}$}&\makecell{$(\frac{11}{8},\frac{3}{2})$}&\makecell{$\mathfrak{su}(2)^{3}_{-\frac{4}{3}}$}&\makecell{$W_{-12+\frac{12}{9}}(\mathfrak{e}_{6},D_{5})+M(1)^{\oplus2}$}&\makecell{$W_{-12+\frac{8}{3}}(\mathfrak{e}_{6},D_{4}(a_{1}))+M(1)$}&\makecell{$W_{-18+\frac{14}{3}}(\mathfrak{e}_{7},D_{4}(a_{1}))$}\\
				\hline
				{$\frac{14}{9},\frac{4}{3},\frac{10}{9}$}&$(\frac{91}{72},\frac{23}{18})$&\makecell{$-$}&\makecell{$W^{-2+\frac{2}{7}}(\mathfrak{su}(2),[2])$\\$W_{-30+\frac{20}{9}}(\mathfrak{e}_{8},E_{8}(b_{4}))$}&\makecell{$W_{-7+\frac{7}{9}}(\mathfrak{su}(7),[7])$\\$W_{-5+\frac{5}{9}}(B_{3},[7])$}&\makecell{$W_{-10+\frac{10}{9}}(D_{6},[9,3])$\\$W_{-4+\frac{4}{9}}(C_{3},[6])$}\\
				\hline
				{$\frac{8}{5},\frac{6}{5},\frac{6}{5}$}&\makecell{$(\frac{157}{120},\frac{41}{30})$}&\makecell{$\mathfrak{su}(2)_{-\frac{8}{5}}$}&\makecell{$W_{-7+\frac{7}{5}}(B_{4},[5,2^{2}])$}&\makecell{$W_{-4+\frac{4}{5}}(C_{3},[3^{2}])$}&\makecell{$ $}\\
				\hline	
				{$\frac{8}{5},\frac{7}{5},\frac{6}{5}$}&\makecell{$(\frac{167}{120},\frac{43}{30})$}&\makecell{$\mathfrak{u}(1)$}&\makecell{$W_{-2+\frac{2}{10}}(\mathfrak{su}(2),[2])+M(1)$\\$W_{-9+\frac{9}{5}}(\mathfrak{su}(9),[5,3,1])$\\$W_{-3+\frac{3}{5}}(B_{2},[3,1^{2}])$\\$W_{-6+\frac{6}{5}}(\mathfrak{su}(6),[4,2])$}&\makecell{$W_{-4+\frac{4}{5}}(\mathfrak{su}(4),[3,1])$\\$W_{-4+\frac{4}{5}}(D_{3},[3^{2}])$\\$W_{-9+\frac{9}{5}}(B_{5},[5,3^{2}])$\\$ $}&\makecell{$W_{-8+\frac{8}{10}}(\mathfrak{su}(8),[8])+M(1)$\\$W_{-8+\frac{8}{5}}(D_{5},[5,3,1^{2}])$\\$W_{-6+\frac{6}{5}}(C_{5},[4^{2},2])$\\$ $}\\
				\hline
				{$\frac{5}{3},\frac{4}{3},\frac{4}{3}$}&\makecell{$(\frac{37}{24},\frac{5}{3})$}&\makecell{$\mathfrak{su}(2)_{-\frac{5}{3}}\times\mathfrak{su}(2)_{-\frac{4}{3}}\times\mathfrak{u}(1)_{f_{0}}$}&\makecell{$W_{-18+\frac{14}{6}}(\mathfrak{e}_{7},D_{5})+M(1)$}&\makecell{$W_{-8+\frac{5}{3}}(D_{5},[4^{2},1^{2}])+M(1)(*)$}&\makecell{$ $}\\
				\hline
				{$\frac{12}{7},\frac{10}{7},\frac{8}{7}$}&$(\frac{81}{56},\frac{3}{2})$&\makecell{$\mathfrak{su}(2)_{-\frac{12}{7}}$}&\makecell{$W_{-2+\frac{2}{7}}(\mathfrak{su}(2),[1^{2}])$\\$W-{-8+\frac{8}{7}}(D_{5},[7,1^{3}])$\\$W_{-30+\frac{30}{7}}(\mathfrak{e}_{8},A_{6})$}&\makecell{$W_{-9+\frac{9}{7}}(\mathfrak{su}(9),[7,1^{2}])$\\$W_{-10+\frac{10}{7}}(D_{6},[6^{2}])$\\$W_{-9+\frac{9}{7}}(\mathfrak{f}_{4},B_{3})$}&\makecell{$W_{-6+\frac{5}{7}}(\mathfrak{su}(6),[6])+M(1)(*)$\\$W_{-12+\frac{8}{7}}(\mathfrak{e}_{6},E_{6}(a_{1}))+M(1)(*)$\\$W_{-9+\frac{9}{7}}(\mathfrak{f}_{4},C_{3})$}\\
				\hline
				{$\frac{12}{7},\frac{9}{7},\frac{8}{7}$}&$(\frac{75}{56},\frac{19}{14})$&\makecell{$-$}&\makecell{$W_{-3+\frac{3}{7}}(\mathfrak{su}(3),[3])$\\$W_{-12+\frac{12}{7}}(\mathfrak{e}_{6},E_{6}(a_{3}))$\\$W_{-5+\frac{5}{7}}(C_{4},[6,2])$}&\makecell{$W_{-4+\frac{4}{7}}(\mathfrak{su}(4),[4])$\\$W_{-12+\frac{12}{14}}(\mathfrak{e}_{6},E_{6})$\\$W_{-30+\frac{24}{14}}(\mathfrak{e}_{8},E_{8}(a_{3}))$}&\makecell{$W_{-4+\frac{4}{7}}(D_{3},[5,1])$\\$W_{-30+\frac{24}{7}}(\mathfrak{e}_{8},E_{8}(b_{6}))$\\$ $}\\
				\hline
				{$\frac{7}{4},\frac{3}{2},\frac{5}{4}$}&\makecell{$(\frac{19}{12},\frac{5}{3})$}&\makecell{$\mathfrak{su}(2)_{-\frac{7}{4}}\times\mathfrak{u}(1)$}&\makecell{$W_{-2+\frac{2}{8}}(\mathfrak{su}(2),[1^{2}])+M(1)$\\$W_{-9+\frac{9}{4}}(\mathfrak{su}(9),[4,3,1^{2}])$\\$W_{-4+\frac{3}{4}}(D_{3},[3^{2}])+M(1)(*)$\\$W_{-4+\frac{3}{4}}(C_{3},[3^{2}])+M(1)$}&\makecell{$W_{-5+\frac{5}{4}}(\mathfrak{su}(5),[3,1^{2}])$\\$W_{-4+\frac{3}{4}}(\mathfrak{su}(4),[3,1])+M(1)(*)$\\$W_{-8+\frac{5}{4}}(D_{5},[5^{2}])+M(1)(*)$\\$W_{-6+\frac{5}{4}}(C_{5},[4^{2},1^{2}])+M(1)$}&\makecell{$W_{-7+\frac{7}{4}}(\mathfrak{su}(7),[3^{2},1])$\\$W_{-7+\frac{6}{8}}(\mathfrak{su}(7),[7])+M(1)^{\oplus 2}(*)$\\$W_{-12+\frac{9}{4}}(\mathfrak{e}_{6},A_{4})$\\$ $}\\
				\hline
				{$\frac{7}{4},\frac{3}{2},\frac{5}{4}$}&\makecell{$(\frac{13}{8},\frac{7}{4})$}&\makecell{$\mathfrak{su}(2)_{-\frac{7}{4}}\times\mathfrak{u}(1)^{2}$}&\makecell{$W_{-6+\frac{5}{4}}(\mathfrak{su}(6),[4,1^{2}])+M(1)$}&\makecell{$W_{-8+\frac{7}{4}}(\mathfrak{su}(8),[4,3,1])+M(1)(*)$}&\makecell{$ $}\\
				\hline
				{$\frac{9}{5},\frac{7}{5},\frac{6}{5}$}&$(\frac{3}{2},\frac{31}{20})$&\makecell{$\mathfrak{u}(1)$}&\makecell{$W_{-3+\frac{3}{5}}(\mathfrak{su}(3),[2,1])$\\$W_{-18+\frac{18}{5}}(\mathfrak{e}_{7},A_{4}+A_{1})$}&\makecell{$W_{-7+\frac{7}{5}}(\mathfrak{su}(7),[4,3])$\\$W_{-18+\frac{18}{20}}(\mathfrak{e}_{7},E_{7})+M(1)$}&\makecell{$W_{-3+\frac{2}{5}}(\mathfrak{su}(3),[3])+M(1)$\\$W_{-8+\frac{8}{5}}(\mathfrak{su}(8),[5,2,1])$}\\
				\hline
				{$2,\frac{3}{2},\frac{3}{2}$}&\makecell{$(\frac{15}{8},2)$}&\makecell{$\mathfrak{u}(1)^{3}$}&\makecell{$W_{-3+\frac{3}{6}}(\mathfrak{su}(3),[2,1])+M(1)^{\oplus 2}$\\$W_{-6+\frac{6}{4}}(\mathfrak{su}(6),[3,2,1])+M(1)$\\$W_{-3+\frac{1}{2}}(B_{2},[3,1^{2}])+M(1)^{\oplus 2}$\\$W_{-5+\frac{1}{2}}(B_{3},[7])+M(1)^{\oplus 3}$}&\makecell{$W_{-4+\frac{4}{8}}(\mathfrak{su}(4),[4])+M(1)^{\oplus 3}$\\$W_{-9+\frac{9}{6}}(\mathfrak{su}(9),[6,2,1])+M(1)^{\oplus 2}$\\$W_{-10+\frac{6}{4}}(D_{6},[7,3,1^{2}])+M(1)^{\oplus 2}$\\$ $}&\makecell{$W_{-9+\frac{9}{6}}(\mathfrak{su}(9),[5,4])+M(1)^{\oplus2}$\\$W_{-4+\frac{4}{8}}(D_{3},[5,1])+M(1)^{\oplus3}$\\$W_{-18+\frac{14}{4}}(\mathfrak{e}_{7},A_{4}+A_{1})+M(1)$\\$ $}\\
				\hline
				{$2,2,2$}&\makecell{$(\frac{61}{24},\frac{17}{6})$}&\makecell{$\mathfrak{su}(2)_{-2}\times\mathfrak{su}(2)_{-2}\times\mathfrak{u}(1)_{-2}\times\mathfrak{u}(1)_{f_{0}}$}&\makecell{$W_{-5+\frac{5}{5}}(\mathfrak{su}(5),[3,1^{2}])+M(1)^{\oplus4}(*)$\\$W_{-8+1}(D_{5},[7,1^{3}])+M(1)^{\oplus5}(*)$}&\makecell{$W_{-6+\frac{6}{6}}(\mathfrak{su}(6),[4,2])+M(1)^{\oplus5}(*)$\\$\mathcal{T}[W_{-10+1}(D_{6},[9,3])]+M(1)^{\oplus6}(*)$}&\makecell{$W_{-8+\frac{8}{4}}(\mathfrak{su}(8),[3^{2},1^{2}])+M(1)^{\oplus3}$\\$ $}\\
				\hline
				{$2,2,2$}&\makecell{$(\frac{19}{8},\frac{5}{2})$}&\makecell{$\mathfrak{su}(2)_{-2}\times\mathfrak{su}(2)_{-2}\times\mathfrak{su}(2)_{-2}$}&\makecell{$W_{-30+\frac{24}{3}}(\mathfrak{e}_{8},D_{4}(a_{1})+A_{1})$\\$W_{-12+\frac{12}{6}}(\mathfrak{e}_{6},(A_{4}+A_{1})+M(1)^{\oplus 2}(*)$}&\makecell{$W_{-9+3}(B_{5},[3,2^{2},1^{4}])$\\$\mathcal{T}[W_{-8+\frac{8}{4}}(D_{5},[3^{2},2^{2}])]+M(1)^{\oplus}(*)$}&\makecell{$W_{-6+2}(C_{5},[2^{4},1^{2}])$\\$W_{-6+\frac{4}{2}}(D_{4},[2^{2},1^{4}])$}\\
				\hline
				{$2,2,2$}&\makecell{$(\frac{53}{24},\frac{13}{6})$}&\makecell{$\mathfrak{u}(1)$}&\makecell{$W_{-10+\frac{6}{3}}(D_{6},[5,3^{2},1])$}&\makecell{$W_{-30+\frac{24}{6}}(\mathfrak{e}_{8},D_{5}+A_{2})$}&\makecell{$W_{-5+1}(B_{3},[3^{2},1])$}\\
				\hline
				{$\frac{9}{4},\frac{3}{2},\frac{5}{4}$}&$(\frac{15}{8},2)$&\makecell{$\mathfrak{su}(3)_{-\frac{9}{4}}$}&\makecell{$W^{-3+\frac{3}{4}}(\mathfrak{su}(3),[1^{3}])$\\$W_{-18+\frac{18}{8}}(\mathfrak{e}_{7},A_{6})+M(1)(*)$}&\makecell{$W_{-7+\frac{7}{4}}(\mathfrak{su}(7),[4,1^{3}])$\\$W_{-30+\frac{30}{8}}(\mathfrak{e}_{8},E_{6}(a_{1}))$}&\makecell{$W_{-9+\frac{9}{4}}(\mathfrak{su}(9),[3^{3}])$\\$W_{-6+\frac{5}{4}}(\mathfrak{su}(6),[3^{2}])+M(1)(*)$}\\
				\hline
				{$\frac{7}{3},\frac{5}{3},\frac{4}{3}$}&\makecell{$(\frac{25}{12},\frac{9}{4})$}&\makecell{$\mathfrak{su}(3)_{-\frac{7}{3}}\times\mathfrak{u}(1)$}&\makecell{$W_{-5+\frac{5}{3}}(\mathfrak{su}(5),[2,1^{3}])$\\$W_{-3+\frac{2}{3}}(\mathfrak{su}(3),[1^{3}])+M(1)$}&\makecell{$W_{-7+\frac{7}{3}}(\mathfrak{su}(7),[2^{3},1])$\\$W_{-5+\frac{4}{3}}(\mathfrak{su}(5),[2^{2},1])+M(1)(*)$}&\makecell{$W_{-8+\frac{8}{3}}(\mathfrak{su}(8),[3,2,1^{3}])$\\$W_{-30+\frac{20}{3}}(\mathfrak{e}_{8},A_{4}+A_{1})$}\\
				\hline
				{$\frac{7}{3},\frac{5}{3},\frac{4}{3}$}&\makecell{$(\frac{17}{8},\frac{7}{3})$}&\makecell{$\mathfrak{su}(3)_{-\frac{7}{3}}\times\mathfrak{u}(1)^{2}$}&\makecell{$W_{-6+\frac{5}{3}}(\mathfrak{su}(6),[3,1^{3}])+M(1)$}&\makecell{$W_{-8+\frac{7}{3}}(\mathfrak{su}(8),[3,2^{2},1])+M(1)(*)$}&\makecell{$ $}\\
				\hline
				{$\frac{12}{5},\frac{6}{5},\frac{6}{5}$}&\makecell{$(\frac{103}{60},\frac{107}{60})$}&\makecell{$\mathfrak{su}(2)_{-\frac{17}{10}}$}&\makecell{$W_{-18+\frac{14}{5}}(\mathfrak{e}_{7},D_{6}(a_{2}))$}&\makecell{$W_{-4+\frac{4}{5}}(\mathfrak{g}_{2},\tilde{A}_{1})$}&\makecell{$ $}\\
				\hline
				{$\frac{18}{7},\frac{12}{7},\frac{8}{7}$}&\makecell{$(\frac{85}{42},\frac{25}{12})$}&\makecell{$\mathfrak{su}(2)_{-\frac{25}{14}}$}&\makecell{$W_{-12+\frac{12}{7}}(\mathfrak{e}_{6},A_{5})$}&\makecell{$W_{-30+\frac{24}{7}}(\mathfrak{e}_{8},A_{7})$}&\makecell{$W_{-5+\frac{5}{7}}(C_{4},(6,1^{2})$}\\
				\hline
				{$\frac{8}{3},\frac{4}{3},\frac{4}{3}$}&\makecell{$(\frac{25}{12},\frac{9}{4})$}&\makecell{$\mathfrak{su}(2)_{-\frac{11}{6}}\times\mathfrak{su}(2)_{-\frac{8}{3}}\times\mathfrak{su}(2)_{-\frac{4}{3}}$}&\makecell{$W_{-12+\frac{8}{3}}(\mathfrak{e}_{6},A_{3}+A_{1})+M(1)$}&\makecell{$W_{-18+\frac{14}{3}}(\mathfrak{e}_{7},(A_{3}+A_{1})')$}&\makecell{$ $}\\
				\hline
				{$\frac{8}{3},\frac{5}{3},\frac{4}{3}$}&\makecell{$(\frac{55}{24},\frac{5}{2})$}&\makecell{$\mathfrak{su}(4)_{-\frac{8}{3}}$}&\makecell{$W_{-4+\frac{4}{3}}(\mathfrak{su}(4),[1^{4}])$\\$W_{-6+\frac{5}{3}}(\mathfrak{su}(6),[2^{3}])+M(1)$\\$W_{-7+\frac{7}{3}}(B_{4},[3,1^{6}])$}&\makecell{$W_{-7+\frac{7}{3}}(\mathfrak{su}(7),[3,1^{4}])$\\$W_{-4+\frac{4}{3}}(D_{3},[1^{6}])$\\$ $}&\makecell{$W_{-8+\frac{8}{3}}(\mathfrak{su}(8),[2^{4}])$\\$W_{-10+\frac{10}{3}}(D_{6},[3^{2},1^{6}])$\\$ $}\\
				\hline
				{$\frac{8}{3},\frac{7}{3},\frac{4}{3}$}&\makecell{$(\frac{21}{8},\frac{17}{6})$}&\makecell{$\mathfrak{sp}(2)_{-\frac{7}{3}}\times\mathfrak{u}(1)_{f_{0}}$}&\makecell{$W_{-8+\frac{8}{3}}(D_{5},[2^{4},1^{2}])$\\$W_{-5+\frac{5}{3}}(C_{4},[2^{2},1^{4}])$}&\makecell{$W_{-8+\frac{5}{3}}(D_{5},[5,1^{5}])+M(1)$\\$ $}&\makecell{$W_{-3+\frac{2}{3}}(B_{2},[1^{5}])+M(1)$\\$ $}\\
				\hline
				{$\frac{14}{5},\frac{8}{5},\frac{6}{5}$}&\makecell{$(\frac{32}{15},\frac{133}{60})$}&\makecell{$\mathfrak{su}(2)_{-\frac{19}{10}}$}&\makecell{$W_{-8+\frac{8}{5}}(D_{5},[5,2^{2},1])$\\$W_{-6+\frac{6}{5}}(C_{5},[4,3^{2}])$}&\makecell{$W_{-3+\frac{5}{3}}(B_{2},[2^{2},1])$\\$ $}&\makecell{$W_{-9+\frac{9}{5}}(B_{5},[4^{2},3])$\\$ $}\\
				\hline
				
		\end{tabular}}
		\vspace{-1.5em}
		\label{table:isomRank3}
\end{table}}

{\linespread{2.5}
	\begin{table}[H]\tiny
			\caption{The VOA isomorphism from rank three AD theories-continued.}
		\centering
		\resizebox{\linewidth}{!}{
			\begin{tabular}
				{|c|c|c|c|c|c|}\hline
				{$\Delta_{\text{Coulomb}}$}&{$(a,c)$}&$\text{AKM subalgebra} $& {$W_{\tilde{k}}(\mathfrak{g},f)$}&{$W_{\tilde{k}}(\mathfrak{g},f)$}&{$W_{\tilde{k}}(\mathfrak{g},f)$}\\
				\hline
				{$3,\frac{5}{2},\frac{3}{2}$}&\makecell{$(\frac{73}{24},\frac{10}{3})$}&\makecell{$\mathfrak{sp}(2)_{-\frac{5}{2}}\times\mathfrak{u}(1)^{2}$}&\makecell{$W_{-8+\frac{5}{2}}(D_{5},[2^{4},1^{2}])+M(1)$\\$W_{-6+\frac{6}{4}}(D_{4},[2^{4}])+M(1)^{\oplus2}$}&\makecell{$W_{-6+\frac{6}{4}}(D_{4},[3,1^{5}])+M(1)^{\oplus 2}$\\$W_{-30+\frac{30}{4}}(\mathfrak{e}_{8},A_{3}+A_{2})$}&\makecell{$W_{-10+\frac{10}{4}}(D_{6},[3^{3},1^{3}])+M(1)^{\oplus2}$\\$W_{-9+
						\frac{5}{2}}(B_{5},[3^{2},1^{5}])+M(1)$}\\
				\hline
				{$3,2,\frac{3}{2}$}&\makecell{$(\frac{3}{8},\frac{17}{6})$}&\makecell{$\mathfrak{su}(2)_{-2}\times\mathfrak{u}(1)^{2}$}&\makecell{$W_{-10+\frac{6}{4}}[D_{6},(7,2^{2},1])+M(1)^{\oplus 2}$}&\makecell{$W_{-3+\frac{1}{2}}(B_{2},[2^{2},1])+M(1)^{\oplus 2}$}&\makecell{$ $}\\
				\hline
				{$3,2,2$}&\makecell{$(\frac{25}{8},\frac{7}{2})$}&\makecell{$\mathfrak{su}(4)_{-3}\times\mathfrak{u}(1)^{3}$}&\makecell{$\mathcal{T}[W_{-4+\frac{4}{4}}(\mathfrak{su}(4),[1^{4}])]+M(1)^{\oplus3}$\\$W_{-9+\frac{9}{3}}(\mathfrak{su}(9),[3,2,1^{4}])+M(1)^{\oplus2}$\\$W_{-9+\frac{8}{4}}(\mathfrak{su}(9),[3^{3}])+M(1)^{\oplus4}(*)$}&\makecell{$W_{-6+\frac{6}{3}}(\mathfrak{su}(6),[2,1^{4}])+M(1)^{\oplus2}$\\$W_{-9+\frac{9}{3}}(\mathfrak{su}(9),[2^{4},1])+M(1)^{\oplus2}$\\$W_{-4+\frac{3}{3}}(D_{3},[1^{6}])+M(1)^{\oplus3}$}&\makecell{$W_{-8+\frac{8}{4}}(\mathfrak{su}(8),[4,1^{4}])+M(1)^{\oplus3}$\\$W_{-7+\frac{6}{3}}(\mathfrak{su}(7),[2^{3},1])+M(1)^{\oplus3}(*)$\\$W_{-10+\frac{6}{2}}(D_{6},[3^{2},1^{6}])+M(1)^{\oplus2}$}\\
				\hline
				{$3,3,2$}&\makecell{$(\frac{27}{8},\frac{7}{2})$}&\makecell{$(\mathfrak{g}_{2})_{3}$}&\makecell{$W_{-12+3}(\mathfrak{e}_{6},2A_{2})$}&\makecell{$W_{-4+1}(\mathfrak{g}_{2},0)$}&\makecell{$ $}\\
				\hline
				{$\frac{10}{3},\frac{8}{3},\frac{4}{3}$}&\makecell{$(\frac{19}{6},\frac{41}{12})$}&\makecell{$\mathfrak{sp}(2)_{-\frac{13}{6}}\times\mathfrak{su}(2)_{-\frac{8}{3}}$}&\makecell{$W_{-10+\frac{10}{3}}(D_{6},[3,2^{2},1^{5}])$}&\makecell{$W_{-7+\frac{7}{3}}(B_{4},[2^{2},1^{5}])$}&\makecell{$ $}\\
				\hline
				{$\frac{10}{3},\frac{8}{3},\frac{4}{3}$}&\makecell{$(\frac{77}{24},\frac{7}{2})$}&\makecell{$\mathfrak{so}(7)_{-\frac{10}{3}}$}&\makecell{$W_{-8+\frac{8}{3}}(D_{5},[3,1^{7}])$}&\makecell{$W_{-18+\frac{14}{3}}(\mathfrak{e}_{7},(A_{3}+A_{1})'')$}&\makecell{$W_{-5+\frac{5}{3}}(B_{3},[1^{7}])$}\\
				\hline
				{$\frac{10}{3},\frac{8}{3},\frac{4}{3}$}&\makecell{$(\frac{77}{24},\frac{7}{2})$}&\makecell{$\mathfrak{sp}(3)_{-\frac{8}{3}}$}&\makecell{$W_{-10+\frac{10}{3}}(D_{6},[2^{6}])$}&\makecell{$W_{-4+\frac{4}{3}}(C_{3},[1^{6}])$}&\makecell{$ $}\\
				\hline
				{$\frac{7}{2},\frac{5}{2},\frac{3}{2}$}&\makecell{$(\frac{7}{2},4)$}&\makecell{$\mathfrak{su}(7)_{-\frac{7}{2}}$}&\makecell{$W_{-7+\frac{7}{2}}(\mathfrak{su}(7),[1^{7}])$}&\makecell{$W_{-9+\frac{9}{2}}(\mathfrak{su}(9),[2,1^{7}])$}&\makecell{$W_{-6+\frac{5}{2}}(\mathfrak{su}(6),[1^{6}])+M(1)(*)$}\\
				\hline
				{$\frac{7}{2},3,\frac{3}{2}$}&\makecell{$(\frac{29}{8},4)$}&\makecell{$\mathfrak{so}(7)_{-\frac{7}{2}}\times\mathfrak{u}(1)$}&\makecell{$W_{-12+\frac{9}{2}}(\mathfrak{e}_{6},2A_{1})$}&\makecell{$W_{-5+\frac{3}{2}}(B_{3},[1^{7}])+M(1)$}&\makecell{$W_{-8+\frac{5}{2}}(D_{5},[3,1^{7}])+M(1)$}\\
				\hline
				{$\frac{18}{5},\frac{12}{5},\frac{6}{5}$}&\makecell{$(\frac{71}{24},\frac{46}{15})$}&\makecell{$\mathfrak{su}(2)_{-\frac{23}{5}}$}&\makecell{$W_{-18+\frac{14}{5}}(\mathfrak{e}_{7},A_{5}+A_{1})$}&\makecell{$W_{-4+\frac{4}{5}}(\mathfrak{g}_{2},A_{1})$}&\makecell{$$}\\
				\hline
				{$4,\frac{8}{3},\frac{4}{3}$}&\makecell{$(\frac{83}{24},\frac{11}{3})$}&\makecell{$\mathfrak{su}(2)_{-5}\times\mathfrak{su}(2)_{-4}$}&\makecell{$W_{-12+\frac{8}{3}}(\mathfrak{e}_{6},2A_{2}+A_{1})+M(1)(*)$}&\makecell{$W_{-18+\frac{14}{3}}(\mathfrak{e}_{7},2A_{2}+A_{1})$}&\makecell{$$}\\
				\hline	
				{$4,2,2$}&\makecell{$(\frac{83}{24},\frac{11}{3})$}&\makecell{$\mathfrak{sp}(2)_{-\frac{5}{2}}\times\mathfrak{su}(2)_{-4}$}&\makecell{$W_{-30+\frac{24}{3}}(\mathfrak{e}_{8},A_{3}+2A_{1})$}&\makecell{$W_{-9+3}(B_{5},[2^{4},1^{3}])$}&\makecell{$W_{-6+2}(C_{5},[2^{3},1^{4}])$}\\
				\hline
				{$4,3,2$}&\makecell{$(\frac{107}{24},\frac{31}{6})$}&\makecell{$\mathfrak{su}(8)_{-4}\times\mathfrak{u}(1)$}&\makecell{$W_{-8+\frac{8}{2}}(\mathfrak{su}(8),[1^{8}])+M(1)$}&\makecell{$W_{-7+\frac{6}{2}}(\mathfrak{su}(7),[1^{7}])+M(1)^{\oplus2}(*)$}&\makecell{$$}\\
				\hline
				{$4,3,2$}&\makecell{$(\frac{25}{6},\frac{55}{12})$}&\makecell{$\mathfrak{sp}(3)_{-3}\times\mathfrak{su}(2)_{\frac{5}{2}}\times\mathfrak{u}(1)^{2}$}&\makecell{$W_{-7+2}(B_{4},[2^{2},1^{5}])+M(1)^{\oplus 2}$}&\makecell{$W_{-10+\frac{6}{2}}(D_{6},[3,2^{2},1^{5}])+M(1)^{\oplus2}$}&\makecell{$$}\\
				\hline
				{$4,4,2$}&\makecell{$(\frac{39}{8},\frac{11}{2})$}&\makecell{$\mathfrak{so}(7)_{-4}\times\mathfrak{su}(2)_{-12}$}&\makecell{$W_{-12+\frac{8}{2}}(\mathfrak{e}_{6},2A_{1})+M(1)^{\oplus2}$}&\makecell{$W_{-30+\frac{30}{3}}(\mathfrak{e}_{8},A_{2}+2A_{1})$}&\makecell{$$}\\
				\hline
				{$6,2,2$}&\makecell{$(\frac{19}{4},\frac{21}{4})$}&\makecell{$\mathfrak{sp}(3)_{-\frac{7}{2}}$}&\makecell{$W_{-30+\frac{24}{4}}(\mathfrak{e}_{8},D_{4}+A_{1})$}&\makecell{$W_{-9+3}(\mathfrak{f}_{4},A_{1})$}&\makecell{$W_{-18+\frac{18}{3}(\mathfrak{e}_{7},4A_{1})}$}\\
				\hline{$6,3,2$}&\makecell{$(\frac{45}{8},\frac{13}{2})$}&\makecell{$(\mathfrak{e}(6))_{-6}$}&\makecell{$W_{-12+\frac{12}{2}}(\mathfrak{e}_{6},0)$}&\makecell{$W_{-30+\frac{24}{2}}(\mathfrak{e}_{8},A_{2})$}&\makecell{$ $}\\
				\hline
				
		\end{tabular}}
		\vspace{-1.5em}
		\label{table:isomRank3more}
\end{table}}

\section{Some results on rank four theories}
\label{sec:higherRank}
It is possible to scan the rank four theories, although it takes significant more time. In this section, a different approach is taken, i.e. we 
look at those theories with special flavor symmetry, namely only the special type of regular singularities would be considered.

Since there are universal formulae of dimensions of trivial, minimal, subregular and principal nilpotent orbit of $\fkg$ \cite{Collingwood:1993rr}, one can have a universal treatment for AD theories whose regular singularity is of such type.
Below we list  values of $k$(or $k_t$) of rank $\dim\CB$ theories with $f_0=0$ when the regular puncture is chosen to be full, minimal, subregular and principal. Since $k$ should be an integer and $f_0=0$, there are only finite number of choices for a given rank $\dim\CB$.

\begin{table}[htb]
	\begin{center}
		\begin{tabular}{ |c|c|c|c|c|c| }
			
			\hline
			$\fkg$ & $ b $ ~& $k$ ($f=\CO_{triv}$) & $k$ ($f=\CO_{min}$) & $k$ ($f=\CO_{subreg}$) & $k$ ($f=\CO_{prin}$) \\ \hline
			
			\hline
			$A_{n-1}$ & $n$   &$-n + 1+\frac{2\,\text{dim}\mathcal{B}}{n-1}$  &$-n + 1+\frac{2\,\text{dim}\mathcal{B}+2}{n-1}$ & $-1+\frac{2\,\text{dim}\mathcal{B}}{n-1}$  &$1+\frac{2\,\text{dim}\mathcal{B}}{n-1}$\\     \hline
			
			\hline
			
			$D_n$  & $n$   & $-n +\frac{1}{2} + \frac{2\text{dim}\mathcal{B}+1}{2(n-1)}$ & $-n +\frac{1}{2} + \frac{2\text{dim}\mathcal{B}+3}{2(n-1)}$ & $-\frac{3}{2} + \frac{2\text{dim}\mathcal{B}+3}{2(n-1)}$  &$\frac{1}{2} + \frac{2\text{dim}\mathcal{B}+1}{2(n-1)}$\\     \hline
			~ & $2n-2$   &$-2n + 3+\frac{2\text{dim}\mathcal{B}}{n}$  &$-2n + 3+\frac{2\text{dim}\mathcal{B}+2}{n}$ & $-3+\frac{2(\text{dim}\mathcal{B}+3)}{n}$  &$1+\frac{2\text{dim}\mathcal{B}}{n}$\\     \hline
			
			\hline
		
			$E_6$ & $12$   &$-11+\frac{\text{dim}\mathcal{B}}{3}$ &$-11+\frac{\text{dim}\mathcal{B}+1}{3}$ & $-3 + \frac{\text{dim}\mathcal{B}+1}{3}$  &$1+\frac{\text{dim}\mathcal{B}}{3}$\\     \hline
			~ & $9$   &$-8+\frac{\text{dim}\mathcal{B}-1}{4}$ &$-8+\frac{\text{dim}\mathcal{B}}{4}$ & $-2 + \frac{\text{dim}\mathcal{B}}{4}$  &$1 + \frac{\text{dim}\mathcal{B}-1}{4}$\\     \hline
			~ & $8$   &$-7+\frac{2\text{dim}\mathcal{B}-3}{9}$ &$-7+\frac{2\text{dim}\mathcal{B}-1}{9}$ & $-2+\frac{2\text{dim}\mathcal{B}+2}{9}$  &$1+ \frac{2\text{dim}\mathcal{B}-3}{9}$\\     \hline
			
			\hline
		
			$ E_7$ & $18$   &$-17+\frac{2\text{dim}\mathcal{B}}{7}$  &$-17+\frac{2\text{dim}\mathcal{B}+2}{7}$ & $-4+ \frac{2\text{dim}\mathcal{B}+1}{7}$  &$1 + \frac{2\text{dim}\mathcal{B}}{7}$ \\     \hline
			~ & $14$   &$- 13 + \frac{2\text{dim}\mathcal{B}-2}{9}$ &$- 13 + \frac{2\text{dim}\mathcal{B}}{9}$ & $-3 +\frac{2\text{dim}\mathcal{B}}{9}$  &$1 + \frac{2\text{dim}\mathcal{B}-2}{9}$\\     \hline
		
		 \hline
			$E_8$ &$30$   &$-29+ \frac{\text{dim}\mathcal{B}}{4}$ &$-29+ \frac{\text{dim}\mathcal{B}+1}{4}$ & $-7 + \frac{\text{dim}\mathcal{B} +3 }{4}$  &$1+\frac{\text{dim}\mathcal{B}}{4}$\\     \hline
			~ &$24$   &$-23 + \frac{\text{dim}\mathcal{B}-1}{5}$ &$-23 + \frac{\text{dim}\mathcal{B}}{5}$ & $-5 + \frac{\text{dim}\mathcal{B}}{5}$  &$1 + \frac{\text{dim}\mathcal{B}-1}{5}$\\     \hline
			~ & $20$   &$-19 + \frac{\text{dim}\mathcal{B}-2}{6}$ &$-19 + \frac{\text{dim}\mathcal{B}-1}{6}$ & $-4 + \frac{\text{dim}\mathcal{B}-1}{6}$  &$ 1 + \frac{\text{dim}\mathcal{B}-2}{6}$\\     \hline
		\end{tabular}
	\end{center}
	\caption{Values of $k$ for a rank $\dim\CB$ untwisted theories with $f_0=0$. The requirements of $k\in\bbZ$ and $\mathrm{gcd}(b,k)=1$ determine  possible theories for a given $\dim\CB$.}
	\label{table:rankCBuntwisted}
\end{table}

\begin{table}[htb]
	\hspace{-2cm}
			\begin{tabular}{ |c| c |c|c|c|c|c| }
			
			\hline
			$\fkj/o$ & $\fkg$ & $b_t$ &$k_t$ ($f=\CO^{Hitchin}_{prin}$) &$k_t$ ($f=\CO^{Hitchin}_{subreg}$) & $k_t$ ($f=\CO^{Hitchin}_{min}$)  & $k_t$ ($f=\CO^{Hitchin}_{triv}$)\\ \hline
			
			\hline
			$A_{2n-1}/\mathbb{Z}_{2}$ & $B_n$ &$4n-2$   &$-4n +3 +\frac{2\text{dim}\mathcal{B}}{n}$  &$-4n +3 +\frac{2\text{dim}\mathcal{B}+2}{n}$ & $-2n-1+\frac{2(\mathrm{dim}\mathcal{B}+1)}{n}$  &$-2n +3 +\frac{2\text{dim}\mathcal{B}}{ n}$\\     \hline
			~ & ~ & $2n$   &$-2n +\frac{1}{2} + \frac{4\text{dim}\mathcal{B}+1}{2(2n-1)}$ &$-2n +\frac{1}{2} + \frac{4\text{dim}\mathcal{B}+5}{2(2n-1)}$ & $-n-1 + \frac{4\text{dim}\mathcal{B}+2}{2(2n-1)}$  &$-n+1 + \frac{4\text{dim}\mathcal{B}+2}{2(2n-1)}$\\     \hline

 \hline
			$D_{n+1}/\mathbb{Z}_{2}$ & $C^{(2)}_n$ & $2n+2$   &$-2n - 1 +\frac{2\text{dim}\mathcal{B}}{n}$ &$-2n - 1 +\frac{2\text{dim}\mathcal{B}+2}{n}$ & $-5+\frac{2(\text{dim}\mathcal{B} + 1)}{n}$  &$-1+\frac{2\text{dim}\mathcal{B}}{n}$\\     \hline
			~ & ~ & $2n$   &$ -2n+1 + \frac{2\text{dim}\mathcal{B}-1}{n+1}$ &$ -2n+1 + \frac{2\text{dim}\mathcal{B}+1}{n+1}$ & $- 5 + \frac{2\text{dim}\mathcal{B} + 7}{n+1 }$ &$-1 + \frac{2\text{dim}\mathcal{B} +1}{ n + 1}$ \\     \hline
			
	 \hline
			$D_{4}/\mathbb{Z}_{3}$ & $G_2$ & $12$   &$-11+\text{dim}\mathcal{B}$ &$-10+\text{dim}\mathcal{B}$ & $-10+\text{dim}\mathcal{B}$ &$-5+\text{dim}\mathcal{B}$ \\     \hline
			~ & ~ & $6$   &$-5+\frac{\text{dim}\mathcal{B}-1}{2}$  &$-5 + \frac{\text{dim}\mathcal{B}}{2}$ & $-5 + \frac{\text{dim}\mathcal{B}}{2}$ &$\frac{-5+\text{dim}\mathcal{B}}{2}$ \\     \hline

\hline
			$E_{6}/\mathbb{Z}_{2}$ & $F_4$ & $18$   &$-17 + \frac{\text{dim}\mathcal{B}}{2}$  &$-17 + \frac{\text{dim}\mathcal{B}+1}{2}$ & $-11+ \frac{\text{dim}\mathcal{B} +1}{2}$ &$-5 + \frac{\text{dim}\mathcal{B}}{2}$ \\     \hline
			~ & ~ & $12$   &$-11 + \frac{\text{dim}\mathcal{B}-1}{3}$ &$-11 + \frac{\text{dim}\mathcal{B}}{3}$ & $-7 + \frac{\text{dim}\mathcal{B}}{3}$ &$-3 + \frac{\text{dim}\mathcal{B}-1}{3}$ \\     \hline
			~ & ~ &$8$   &$-7 + \frac{2\text{dim}\mathcal{B}-5}{9}$ &$-7 + \frac{2\text{dim}\mathcal{B}-3}{9}$ & $-5+ \frac{2\text{dim}\mathcal{B} +3}{9}$  &$-2 + \frac{2\text{dim}\mathcal{B}-2}{9}$\\     \hline
		\end{tabular}
	\caption{Values of $k_t$ for a rank $\dim\CB$ twisted theories with $f_0=0$. The requirements of $k_t\in\bbZ$ and $\mathrm{gcd}(b_t,k_t)=1$ determine  possible theories for a given $\dim\CB$.}
	\label{table:rankCBtwisted}
\end{table}

{\bf Example: } For $(A_{n-1},n,k,f=triv)$ theories, the dimension formula requires that $k=-n+1+\frac{2\text{dim}\mathcal{B}}{n-1}$. Since $k$ should be an integer and $f_0=0$, $n-1$ must be a divisor of $2\dim\CB$ such that $n$ and $k$ are coprime. Let the integer $q=\frac{2\dim\CB}{n-1}+1$ satisfy $\mathrm{gcd}(n,q)=1$, then $k=-n+q$. The CB spectrum is the following set
\begin{equation}
	\{\Delta_{Coulomb}\} =\left\{ \Delta_{l,j}=l-j\frac{n}{q}\  |\  (l,j)\in\bbZ^2,2\leq l\leq n, 1\leq j \leq \lfloor \frac{(l-1)q}{n} \rfloor \right\}.
\end{equation}
The flavor symmetry should be $A_n$, and the corresponding VOA is $V_{-n+\frac{n}{q}}(\fsu(n))$. The 4d central charge $c_{4d}$ is
\begin{equation}
c_{4d}=\frac{1}{12}(q-1)(n^2-1).
\end{equation}

%
%
%
%
%
%
%
%

In table \ref{table:rank4sum} we summarized all rank four theories with special nilpotent orbit discussed in this section. One immediate implication is the following VOA isomorphisms
\begin{equation}
W_{-9+\frac{3}{5}}(\mathfrak{f}_{4},F_{4})\cong \mathrm{Vir}(2,5)^{\oplus 4},
\end{equation}
and
\begin{equation}
W_{-9+\frac{5}{7}}(B_{5},[11])\cong W_{-2+\frac{2}{7}}(\mathfrak{su}(2),[2])^{\oplus 2}.
\end{equation}

\begin{table}[H]\small
	\centering
	\begin{tabular}{|c|c|c|c|c|c|}
		\hline
		$(a,c)$ & $\Delta_{\mathrm{Coulomb}}$ & $\mathfrak{f}$ & VOA & $\CM_H$/Asso. Var. & Singularity\\
		\hline
		$*(\frac{43}{30},\frac{22}{15})$ & $\frac{6}{5},\frac{6}{5},\frac{6}{5},\frac{6}{5}$ & $-$ & $ W_{-9+\frac{3}{5}}(\mathfrak{f}_{4},F_{4})$ & $-$ &\\
		\hline
		$*(\frac{67}{42},\frac{34}{21})$ & $\frac{10}{7},\frac{10}{7},\frac{8}{7},\frac{8}{7}$ & $-$ & $ W_{-9+\frac{5}{7}}(B_{5},[11])$ & $-$ &\\
		\hline
		$*(\frac{115}{66},\frac{58}{33})$ & $\frac{18}{11},\frac{16}{11},\frac{14}{11},\frac{12}{11}$ & $-$ & $ W_{-2+\frac{2}{11}}(\mathfrak{su}(2),[2])$ & $-$ & $(A_1, A_8)$\\
		\hline
		$*(\frac{15}{8},\frac{23}{12})$ & $\frac{5}{3},\frac{3}{2},\frac{4}{3},\frac{7}{6}$ & $\mathfrak{u}(1)$ & $ W_{-5+\frac{5}{6}}(\mathfrak{su}(5),[4,1])$ & $\overline{\CO}_{[5]}\cap S_{[4,1]}$ & \\
		\hline
		$*(\frac{35}{18},2)$ & $\frac{16}{9},\frac{14}{9},\frac{4}{3},\frac{10}{9}$ & $\mathfrak{su}(2)_{\frac{16}{9}}$ & \textcolor{red}{$ V_{-\frac{16}{9}}(\mathfrak{su}(2))$} & $[2]$ & $(A_1, D_9)$ \\
		\hline	
		$*(\frac{91}{48},\frac{23}{12})$ & $\frac{15}{8},\frac{3}{2},\frac{5}{4},\frac{9}{8}$ & $-$ & $ W_{-3+\frac{3}{8}}(\mathfrak{su}(3),[3])$ & $-$ &\\
		\hline
		$*(2,2)$ & $2,\frac{4}{3},\frac{4}{3},\frac{4}{3}$ & $-$ & $ W_{-6+\frac{2}{3}}(D_{4},[7,1])$ & $-$ &\\
		\hline
		$*(\frac{38}{15},\frac{8}{3})$ & $\frac{12}{5},\frac{9}{5},\frac{7}{5},\frac{6}{5}$ &$\mathfrak{su}(3)_{\frac{12}{5}}$ &  \textcolor{red}{$ V_{-\frac{12}{5}}(\mathfrak{su}(3))$} & $[3]$ & $D_5(SU(3))$ \\
		\hline
		$(\frac{19}{6},\frac{10}{3})$ & $\frac{14}{5},\frac{12}{5},\frac{8}{5},\frac{6}{5}$ & $\mathfrak{so}(5)_{\frac{12}{5}}$ & \textcolor{red}{$ V_{-\frac{12}{5}}(B_{2})$} & $[5]$ & $D_5(SO(5))$ \\
		\hline
		$*(\frac{11}{3},4)$ & $\frac{10}{3},\frac{7}{3},\frac{5}{3},\frac{4}{3}$ & $\mathfrak{su}(5)_{\frac{10}{3}}$ & \textcolor{red}{$ V_{-\frac{10}{3}}(\mathfrak{su}(5))$} & $[3,2]$ & $D_3 (SU(5))$\\
		\hline
		$(\frac{25}{6},\frac{9}{2})$ & $\frac{10}{3},\frac{8}{3},\frac{7}{3},\frac{4}{3}$ & $\mathfrak{so}(6)_{\frac{10}{3}}\times\mathfrak{su}(2)_{\frac{7}{3}}$ & $ W_{-8+\frac{8}{3}}(D_{5},[2^{2},1^{6}])$ & $\overline{\CO}_{[3^3,1]}\cap S_{[2^2,1^6]}$ &\\
		\hline
		$(\frac{133}{30},\frac{14}{3})$ & $\frac{18}{5},\frac{16}{5},\frac{12}{5},\frac{6}{5}$ & $(\mathfrak{g}_{2})_{\frac{16}{5}}$ & $ V_{-\frac{16}{5}}(\mathfrak{g}_{2})$ & $G_2(a_1)$  &\\
		\hline
		$*(\frac{13}{3},\frac{14}{3})$ & $4,2,2,2$ & $\mathfrak{so}(8)_{4}$ & $ V_{-4}(\mathfrak{so}(8))$ & $[3^2,1^2]$ &\\
		\hline
		$(\frac{29}{6},\frac{31}{6})$ & $4,3,2,2$ & $\mathfrak{sp}(3)_{3}\times\mathfrak{u}(1)_{2}$ & $ W_{-6+2}(C_{5},[2^{2},1^{6}])$ & $\overline{\CO}_{[3^2,2^2]} \cap S_{[2^{2},1^{6}]}$ &\\
		\hline
		$(\frac{43}{8},\frac{23}{4})$ & $4,4,2,2$ & $\mathfrak{so}(7)_{4}\times\mathfrak{su}(2)_{\frac{5}{2}}$ & $ W_{-9+3}(B_{5},[2^{2},1^{7}])$ & $\overline{\CO}_{[3^3,1^2]}\cap S_{[2^{2},1^{7}]}$ &\\
		\hline
		$*(\frac{35}{6},\frac{20}{3})$ & $\frac{9}{2},\frac{7}{2},\frac{5}{2},\frac{3}{2}$ & $\mathfrak{su}(9)_{\frac{9}{2}}$ & \textcolor{red}{$ V_{-\frac{9}{2}}(\mathfrak{su}(9))$} & $[2^4,1]$ & $D_2(SU(9))$\\
		\hline
		$(\frac{35}{6},\frac{77}{12})$ & $\frac{9}{2},\frac{7}{2},3,\frac{3}{2}$ & $\mathfrak{su}(6)_{\frac{9}{2}}$ & $ W_{-12+\frac{9}{2}}(\mathfrak{e}_{6},A_{1})$ & $\overline{\CO}_{A_2+2A_1}\cap S_{A_1}$  &\\
		\hline
		$(\frac{11}{2},6)$ & $\frac{14}{3},\frac{10}{3},\frac{8}{3},\frac{4}{3}$ & $\mathfrak{so}(9)_{\frac{14}{3}}$ & \textcolor{red}{$ V_{-\frac{14}{3}}(B_{4})$} & $[3^3]$ & $D_3 SO(9)$  \\
		\hline
		$(\frac{13}{2},7)$ & $6,4,2,2$ & $\mathfrak{sp}(4)_{\frac{7}{2}}$ & $ W_{-6+2}(C_{5},[2,1^{8}])$  & $\overline{\CO}_{[3^2,2^2]} \cap S_{[2,1^{8}]}$ & \\
		\hline
		$(\frac{47}{6},\frac{26}{3})$ & $6,6,2,2$ & $(\mathfrak{f}_{4})_{6}$ & \textcolor{red}{$ V_{-6}(\mathfrak{f}_{4})$} & $F_4(a_3)$ &\\
		\hline
		$*(\frac{107}{6},\frac{62}{3})$ & $15,9,5,3$ & $(\mathfrak{e}_{8})_{15}$ & \textcolor{red}{$ V_{-15}(\mathfrak{e}_{8})$} & $A_2+A_1$ &\\
		\hline
	\end{tabular}
	\caption{\label{table:rank4sum}Rank four theories with regular singularity being full, minimal, subregular or regular. Central charges $(a,c)$, CB spectrum, flavor symmetry algebra $\mathfrak{f}$, one corresponding VOA and the Higgs branch are listed. Theories with $*$ saturate the bound \ref{eq:boundSCFTs}. }
\end{table}

\section{Conclusion}
\label{sec:conclusion}
In this paper, we perform a systematical search for small rank theories in the space of geometrically engineered 4d $\mathcal{N}=2$ SCFTs. Some of the rank one and rank two theories constructed 
in \cite{Argyres:2016xua, Argyres:2022lah} are found here, and new properties of these theories such as the corresponding VOA and the Higgs branch can be obtained. The main new input is the dimension formula \ref{eq:CBdimUT} and \ref{eq:CBdimT}
for the CB. It is now possible to study these theories with arbitrary rank. 

Let us now list some interesting questions for further study
\begin{enumerate}
\item The main  feature of the new theories found in \cite{Argyres:2016xua, Argyres:2022lah} is that undeformable singularities except the $I_1$ type appear in generic deformations.
These theories also show up in our twisted constructions. While these special undeformable singularities are related to 
discrete gauging \cite{Argyres:2016yzz}, they are related to the outer automophsim twist in our construction. It would be interesting to further clarify the relation between the discrete gauging and the outer automorphism twist.
\item For generic rank it is difficult to classify all possible theories, therefore one should try to look for bounds on the rank of flavor symmetry, the maximal scaling dimension, and etc at a given rank.
\item  If the mass parameter of the irregular singularity is not zero, the corresponding VOA is not completely determined (might be an extension between the W algebra and Heisenberg algebra). it is interesting to find the full corresponding 
VOA. The isomorphism of  VOAs found in this paper might be a clue for the general correspondence.
\end{enumerate}

\acknowledgments

BL, DX and WY are supported by Yau Mathematical Science Center at Tsinghua
University. DX and WY are supported by  national key research
and development program of China (NO. 2020YFA0713000), and NNSF of China with
Grant NO: 11847301 and 12047502. BL and WY are supported by Dushi program of Tsinghua.

\newpage
\appendix

\section{List of $d(n)$}
\label{sec:dnlists}

In this section we list $d(n)$ used to compute the growth function for theories engineered using one irregular and one regular singularities.

\begin{table}[h]
	\begin{center}
			\begin{tabular}{|l|l|}
				\hline
				$\mathfrak{g}$ & $d_{\mathfrak{g}}(n)$\\ \hline
				$A_{2}$ &$ d(2)=3,~d(1)=0,$ \\ \hline
				$A_{3}$&  $d(3)=8,~d(2)=3,~d(1)=0$ \\ \hline
				$A_{4}$ &$ d(4)=15,~d(2)=6,~d(3)=8,~d(1)=0$ \\ \hline
				$A_{5}$&  $d(5)=24,~d(4)=15,~d(2)=6,~d(1)=0$ \\ \hline
				$A_{6}$&  $d(6)=35,~d(5)=24,~d(3)=16,~d(2)=9,~d(1)=0$ \\ \hline
				$A_{7}$&  $d(7)=48,~d(6)=35,~d(3)=16,~d(2)=9$ \\ \hline
				$A_{8}$&  $d(8)=63,~d(7)=48,~d(4)=30,~d(2)=12$ \\ \hline
				$A_{9}$&  $d(9)=80,~d(8)=63,~d(4)=30,~d(3)=24$ \\ \hline
			\end{tabular}
			\begin{tabular}{|l|l|}
				\hline
				$\mathfrak{g}$ & $d_{\mathfrak{g}}(n)$\\ \hline
				$B_{1}$ &$ d(1)=3 \,\text{or} \,0$ \\ \hline
				$B_{2}$ &$ d(3)=10,~d(2)=3,~d(1)=6\, \text{or}\,0$ \\ \hline
				$B_{3}$ &$ d(5)=21,~d(3)=11\, \text{or}\,8,~d(1)=9\, \text{or}\, 0$ \\ \hline
				$B_{4}$ &$ d(7)=36,~d(4)=15,~d(2)=6,~d(1)=12$ \\ \hline
				$B_{5}$ &$ d(9)=55,~d(5)=27\, \text{or}\, 24,~d(3)=21$ \\ \hline
			\end{tabular}
			\begin{tabular}{|l|l|}
				\hline
				$\mathfrak{g}$ & $d_{\mathfrak{g}}(n)$\\ \hline
				$C_{1}$ &$ d(2)=3,~d(1)=0$ \\ \hline
				$C_{2}$ &$ d(3)=10,~d(1)=0$ \\ \hline
				$C_{3}$ &$ d(4)=21,~d(3)=16$ \\ \hline
				$C_{4}$ &$ d(5)=36$ \\ \hline
				$C_{5}$ &$ d(6)=55,~d(5)=48,d(2)=21$ \\ \hline
			\end{tabular}
			\begin{tabular}{|l|l|}
				\hline
				$\mathfrak{g}$ & $d_{\mathfrak{g}}(n)$\\ \hline
				$D_{3}$ &$ d(4)=15,~d(3)=8,~d(2)=6,~d(1)=0$ \\ \hline
				$D_{4}$ &$ d(6)=28,~d(4)=18,~d(3)=8,~d(2)=12,~d(1)=0$ \\ \hline
				$D_{5}$ &$ d(8)=45,~d(5)=24.~d(4)=18,~d(2)=12,~d(1)=0$ \\ \hline
				$D_{6}$ &$ d(10)=66,~d(6)=38,~d(5)=24,~d(3)=16,~d(2)=18,~d(1)=0$ \\ \hline
			\end{tabular}
			\begin{tabular}{|l|l|}
\hline
$\mathfrak{g}$ & $d_{\mathfrak{g}}(n)$\\ \hline
$\mathfrak{e}_6$ &$ d(12)=78,~d(9)=56,~d(8)=45,~d(6)=36,~d(4)=18,~d(3)=24,~d(2)=12$ \\ \hline
$\mathfrak{e}_7$&  $d(18)=133,~d(14)=99,~d(9)=56,~d(7)=48,~d(6)=39,~d(3)=24,~d(2)=21$ 
\\ \hline
$\mathfrak{e}_8$&$d(30)=248,~d(24)=190,~d(20)=156,~d(15)=112,~d(12)=92,~d(10)=72$ \\ 
~& $~d(8)=66,~d(6)=40,~d(5)=48,~d(4)=36,~d(3)=32,~d(2)=24$ \\ \hline
$\mathfrak{g}_2$ &  $d(4)=14,~d(2)=12 (u=3k),~d(2)=4 (u\neq 3k),~d(1)=8$ \\ \hline
$\mathfrak{f}_4$ &  $d(9)=52,~d(6)=30,~d(3)=24 (u=3k-1),~d(3)=12(u=3k+1),~d(4)=21$ \\ 
~& $~d(2)=18, ~d(1)=12$ \\ \hline
\end{tabular}
	\end{center}
	\caption{\label{table:dnsA}Data used to compute the growth function of Kac-Moody algebra $V_{-h^\vee+{n\over u}} (\mathfrak{g})$, see formula. \ref{growth}.}
	
\end{table}

\newpage

\section{ List of AD theories}

One can choose $k$, $b$ and regular puncture such that the CB dimension formular satisfied. In the following, we list all $\text{rank}1$, $\text{rank}2$, $\text{rank}3$ non-twisted AD theories with type ADE from the CB dimension formular. We list the folowing data:
\begin{itemize}
	\item The parameters which describe the Higgs field of the Hitchin system near the irregular singularity $b$ and $k$. 
	\item The Hitchin label (HL) and Nahm label (NL) of nilpotent orbits, for $A_{n-1}$ type, we only list the Nahm label denoted by $\mathcal{O}_{f}$.
	\item The number of mass deformation $f_{0}$ and the number of exact marginal deformation $f_{1}$\cite{Xie:2017aqx}. For $E_{8}$ type, $f_{0}=0$ for all cases, we only list the $f_{1}$.
	\item The dimension of nilpotent orbits correspond to Hichin label and Nahm label denoted by HD and ND respectively
	\item The CB operator scaling dimension $\Delta_{\text{Coulomb}}$.
	\item The flavor symmetry $\mathfrak{f}$ correspond to regular puncture and flavor central charge $k_{2d}$ for each simple factor.
	\item Conformal central charge $(a_{4d},c_{4d})$.
	\item The notation "Theory" label AD theories.
	\item $\fsp(r)$ means $C_r$.
\end{itemize}

\subsection{Rank 1}

\label{sec:completList1}

{\linespread{2.0}
\begin{table}[H]\tiny
\caption{$A_{n-1}$, $b=n$}
	\centering
	\resizebox{\linewidth}{!}{
\begin{tabular}{|c|c|c|c|c|c|c|c|c|c|}\hline
		{$n$}& {$k$}&{$\mathcal{O}_{f}$}&{$f_{0}$}& {$f_{1}$}&{ND}&{$\Delta_{\text{Coulomb}}$}&{$\mathfrak{f}$}&{$(a,c)$}&{Theory}\\ \hline
		$2$&\makecell{$1$\\$2$\\$3$\\$4$} &\makecell{$(1^{2})$\\$(1^{2})$\\$(2)$\\$(2)$}&\makecell{$0$\\$1$\\$0$\\$1$}&\makecell{$0$\\$1$\\$0$\\$1$} &\makecell{$0$\\$0$\\$2$\\$2$}&\makecell{$\frac{4}{3}$\\$\frac{3}{2}$\\$\frac{6}{5}$\\$\frac{4}{3}$}&\makecell{$\mathfrak{su}(2)_{\frac{4}{3}}$\\$\mathfrak{su}(2)_{\frac{3}{2}}\times\mathfrak{u}(1)_{f_{0}}$\\$-$\\$\mathfrak{u}(1)_{f_{0}}$}&\makecell{$(\frac{11}{24},\frac{1}{2})$\\$(\frac{7}{12},\frac{2}{3})$\\$(\frac{43}{120},\frac{11}{30})$\\$(\frac{11}{24},\frac{1}{2})$}&\makecell{$\mathcal{T}[W_{-\frac{4}{3}}(\mathfrak{su}(2),(1^{2}))]$\\$\mathcal{T}[W_{-\frac{3}{2}}(\mathfrak{su}(2)),(1^{2})]+\mathfrak{u}(1)$\\$\mathcal{T}[W_{-\frac{8}{5}}(\mathfrak{su}(2),(2))]$\\$\mathcal{T}[W_{-\frac{5}{3}}(\mathfrak{su}(2),(2))]+\mathfrak{u}(1)$}\\
		\hline
		$3$&\makecell{$-1$\\$0$\\$2$\\$3$} &\makecell{$(1^{3})$\\$(1^{3})$\\$(3)$\\$(3)$}&\makecell{$0$\\$2$\\$0$\\$2$}&\makecell{$0$\\$2$\\$0$\\$2$}&\makecell{$0$\\$0$\\$6$\\$6$}&\makecell{$\frac{3}{2}$\\$2$\\$\frac{6}{5}$\\$\frac{3}{2}$}&\makecell{$\mathfrak{su}(3)_{\frac{3}{2}}$\\$\mathfrak{su}(3)_{2}\times\mathfrak{u}(1)^{2}_{f_{0}}$\\$-$\\$\mathfrak{u}(1)^{2}_{f_{0}}$}&\makecell{$(\frac{7}{12},\frac{2}{3})$\\$(\frac{23}{24},\frac{7}{6})$\\$(\frac{43}{120},\frac{11}{30})$\\$(\frac{7}{12},\frac{2}{3})$}&\makecell{$\mathcal{T}[W_{-3+\frac{3}{2}}(\mathfrak{su}(3),(1^{3}))]$\\$\mathcal{T}[W_{-3+\frac{3}{3}}(\mathfrak{su}(3),(1^{3}))]+2\mathfrak{u}(1)$\\$\mathcal{T}[W_{-3+\frac{3}{5}}(\mathfrak{su}(3),(3))]$\\$\mathcal{T}[W_{-3+\frac{3}{6}}(\mathfrak{su}(3),(3))]+2\mathfrak{u}(1)$}\\
\hline
	\end{tabular}}
\vspace{-1.5em}
\end{table}

}

{\linespread{2.0}
	\begin{table}[H]\tiny
		\caption{$A_{n-1}$, $b=n-1$}
	\centering
	\resizebox{\linewidth}{!}{
	\begin{tabular}{|c|c|c|c|c|c|c|c|c|c|}\hline
		{$n$}& {$k$}&{$\mathcal{O}_{f}$}&{$f_{0}$}& {$f_{1}$}&{ND}&{$\mathfrak{f}$}&{$\Delta_{\text{Coulomb}}$}&{$(a,c)$}&{Theory}\\ \hline
		$4$&\makecell{$-1$} &\makecell{$(2,1^{2})$}&\makecell{$1$}&\makecell{$0$}&\makecell{$6$}&\makecell{$\mathfrak{su}(2)_{\frac{3}{2}}\times\mathfrak{u}(1)_{1}\times\mathfrak{u}(1)_{f_{0}}$}&\makecell{$\frac{3}{2}$}&\makecell{$(\frac{5}{8},\frac{3}{4})$}&\makecell{$\mathcal{T}[W_{-4+\frac{3}{2}}(\mathfrak{su}(4),(1^{4}))]+\mathfrak{u}(1)$}\\
		\hline
		$5$&\makecell{$-2$\\$-1$} &\makecell{$(2,1^{3})$\\$(3,2)$}&\makecell{$2$\\$1$}&\makecell{$1$\\$0$}&\makecell{$8$\\$16$}&\makecell{$\mathfrak{su}(3)_{2}\times\mathfrak{u}(1)_{1}\times\mathfrak{u}(1)^{2}_{f_{0}}$\\$\mathfrak{u}(1)_{1}\times\mathfrak{u}(1)_{f_{0}}$}&\makecell{$2$\\$\frac{4}{3}$}&\makecell{$(1,\frac{5}{4})$\\$(\frac{1}{2},\frac{7}{12})$}&\makecell{$\mathcal{T}[W_{-5+\frac{4}{2}}(\mathfrak{su}(5),(2,1^{3}))]+2\mathfrak{u}(1)$\\$\mathcal{T}[W_{-5+\frac{4}{3}}(\mathfrak{su}(5),(3,2))]+\mathfrak{u}(1)$}\\
		\hline
		$6$&\makecell{$-3$} &\makecell{$(2^{2},1^{2})$}&\makecell{$1$}&\makecell{$0$}&\makecell{$16$}&\makecell{$\mathfrak{su}(2)_{\frac{3}{2}}\times\mathfrak{su}(2)_{1}\times\mathfrak{u}(1)_{1}\times\mathfrak{u}(1)_{f_{0}}$}&\makecell{$\frac{3}{2}$}&\makecell{$(\frac{2}{3},\frac{5}{6})$}&\makecell{$\mathcal{T}[W_{-6+\frac{5}{2}}(\mathfrak{su}(6),(2^{2},1^{2}))]+\mathfrak{u}(1)$}\\
\hline
	\end{tabular}}
\vspace{-3em}
\end{table}

}

{\linespread{2.0}\begin{table}[H]\tiny
	\caption{$D_{n}$, $b=n$}
	\centering
	\resizebox{\linewidth}{!}{
	\begin{tabular}{|c|c|c|c|c|c|c|c|c|c|c|c|}\hline
		{$n$}& {$k$}&{HL}&{NL}&{HD}&{ND}&{$f_{0}$}& {$f_{1}$}&{$\mathfrak{f}$}&{$\Delta_{\text{Coulomb}}$}&{$(a,c)$}&{Theory}\\ \hline
		$3$&\makecell{$-1$\\$0$} &\makecell{$(3,3)$\\$(3,1^{3})$}&\makecell{$(2^{2},1^{2})$\\$(3,1^{3})$}&\makecell{$10$\\$8$}&\makecell{$6$\\$8$}&\makecell{$1$\\$3$}&\makecell{$0$\\$2$}&\makecell{$\mathfrak{su}(2)_{\frac{3}{2}}\times\mathfrak{u}(1)_{\frac{1}{2}}\times\mathfrak{u}(1)_{f_{0}}$\\$\mathfrak{su}(2)_{1}\times\mathfrak{u}(1)^{3}_{f_{0}}$}&\makecell{$\frac{3}{2}$\\$2$}&\makecell{$(\frac{5}{8},\frac{3}{4})$\\$(\frac{23}{24},\frac{7}{6})$}&\makecell{$\mathcal{T}[W_{-4+\frac{3}{2}}(D_{3},(2^{2},1^{2}))]+\mathfrak{u}(1)$\\$\mathcal{T}[W_{-4+\frac{3}{3}}(D_{3},(3,1^{3}))]+3\mathfrak{u}(1)$}\\
\hline
	\end{tabular}}
\vspace{-1.5em}
\end{table}

}

{\linespread{2.0}\begin{table}[H]\tiny
	\caption{$D_{n}$, $b=2n-2$}
	\centering
	\resizebox{\linewidth}{!}{
	\begin{tabular}{|c|c|c|c|c|c|c|c|c|c|c|c|}\hline
		{$n$}& {$k$}&{HL}&{NL}&{HD}&{ND}&{$f_{0}$}& {$f_{1}$}&{$\mathfrak{f}$}&{$\Delta_{\text{Coulomb}}$}&{$(a,c)$}&{Theory}\\ \hline
		$3$&\makecell{$-2$\\$-1$\\$2$} &\makecell{$(5,1)$\\$(3,1^{3})$\\$(1^{6})$}&\makecell{$(1^{6})$\\$(3,1^{3})$\\$(5,1)$}&\makecell{$12$\\$8$\\$0$}&\makecell{$0$\\$8$\\$12$}&\makecell{$1$\\$0$\\$1$}&\makecell{$1$\\$0$\\$1$}&\makecell{$\mathfrak{su}(4)_{2}\times\mathfrak{u}(1)_{f_{0}}$\\$\mathfrak{so}(3)_{\frac{2}{3}}$\\$\mathfrak{u}(1)_{f_{0}}$}&\makecell{$2$\\$\frac{4}{3}$\\$\frac{4}{3}$}&\makecell{$(\frac{23}{24},\frac{7}{6})$\\$(\frac{11}{24},\frac{1}{2})$\\$(\frac{11}{24},\frac{1}{2})$}&\makecell{$\mathcal{T}[W_{-4+\frac{4}{2}}(D_{3},(1^{6}))]+\mathfrak{u}(1)$\\$\mathcal{T}[W_{-4+\frac{4}{3}}(D_{3},(3,1^{3}))]$\\$\mathcal{T}[W_{-4+\frac{4}{6}}(D_{3},(5,1))]+\mathfrak{u}(1)$}\\
		\hline
		$4$&\makecell{$-4$\\$-1$\\$2$} &\makecell{$(7,1)$\\$(2^{2},1^{4})$\\$(1^{8})$}&\makecell{$(1^{8})$\\$(5,3)$\\$(7,1)$}&\makecell{$24$\\$10$\\$0$}&\makecell{$0$\\$22$\\$24$}&\makecell{$2$\\$0$\\$2$}&\makecell{$0$\\$0$\\$0$}&\makecell{$\mathfrak{spin}(8)_{3}\times\mathfrak{u}(1)^{2}_{f_{0}}$\\$-$\\$\mathfrak{u}(1)^{2}_{f_{0}}$}&\makecell{$3$\\$\frac{6}{5}$\\$\frac{3}{2}$}&\makecell{$(\frac{41}{24},\frac{13}{6})$\\$(\frac{43}{120},\frac{11}{30})$\\$(\frac{7}{12},\frac{2}{3})$}&\makecell{$\mathcal{T}[W_{-6+\frac{6}{2}}(D_{4},(1^{8}))]+2\mathfrak{u}(1)$\\$\mathcal{T}[W_{-6+\frac{6}{5}}(D_{4},(5,3))]$\\$\mathcal{T}[W_{-6+\frac{6}{8}}(D_{4},(7,1))]+2\mathfrak{u}(1)$}\\
\hline
	\end{tabular}}
\vspace{-1.5em}
\end{table}

}

{\linespread{2.0}
\begin{table}[H]\tiny
	\centering
	\caption{$E_{6}$}
	\resizebox{\linewidth}{!}{
	\begin{tabular}{|c|c|c|c|c|c|c|c|c|c|c|c|}\hline
		{$b$}& {$k$}&{HL}&{NL}&{HD}&{ND}&{$f_{0}$}& {$f_{1}$}&{$\mathfrak{f}$}&{$\Delta_{\text{Coulomb}}$}&{$(a,c)$}&{Theory}\\ \hline
		$12$&\makecell{$-7$} &\makecell{$A_{2}+2A_{1}$}&\makecell{$A_{4}+A_{1}$}&\makecell{$50$} &\makecell{$62$}&\makecell{$0$}&\makecell{$0$}&\makecell{$\mathfrak{u}(1)_{0}$}&\makecell{$\frac{6}{5}$}&\makecell{$(\frac{43}{120},\frac{11}{30})$}&\makecell{$\mathcal{T}[W_{-12+\frac{12}{5}}(\mathfrak{e}_{6},A_{4}+A_{1})]$}\\
		\hline
		$12$&\makecell{$-10$} &\makecell{$D_{5}$}&$2A_{1}$&\makecell{$68$}&\makecell{$32$}&\makecell{$0$}&\makecell{$1$}&\makecell{$\mathfrak{so}(7)_{2}\times\mathfrak{u}(1)_{0}$}&\makecell{$2$}&\makecell{$(\frac{23}{24},\frac{7}{6})$}&\makecell{$\mathcal{T}[W_{-12+\frac{12}{2}}(\mathfrak{e}_{6},2A_{1})]$}\\
		\hline
		$12$&\makecell{$-9$} &\makecell{$D_{5}(a_{1})$}&$A_{2}+A_{1}$&\makecell{$64$}&\makecell{$46$}&\makecell{$2$}&\makecell{$1$}&\makecell{$\mathfrak{su}(3)_{2}\times\mathfrak{u}(1)_{3}\times\mathfrak{u}(1)^{2}_{f_{0}}$}&\makecell{$2$}&\makecell{$(\frac{13}{12},\frac{17}{12})$}&\makecell{$\mathcal{T}[W_{-12+\frac{12}{3}}(\mathfrak{e}_{6},A_{2}+A_{1})]+2\mathfrak{u}(1)$}\\
		\hline
		$12$&\makecell{$-4$} &\makecell{$2A_{1}$}&\makecell{$D_{5}$}&\makecell{$32$}&\makecell{$68$}&\makecell{$0$}&\makecell{$2$}&\makecell{$\mathfrak{u}(1)_{0}$}&\makecell{$1$}&\makecell{$(\frac{5}{24},\frac{1}{6})$}&\makecell{$\mathcal{T}[W_{-12+\frac{12}{8}}(\mathfrak{e}_{6},D_{5})]$}\\
		\hline
		$12$&\makecell{$-6$} &\makecell{$A_{2}+A_{1}$}&$D_{5}(a_{1})$&\makecell{$46$}&\makecell{$64$}&\makecell{$2$}&\makecell{$3$}&\makecell{$\mathfrak{u}(1)_{3}\times\mathfrak{u}(1)^{2}_{f_{0}}$}&\makecell{$1$}&\makecell{$(\frac{1}{3},\frac{5}{12})$}&\makecell{$\mathcal{T}[W_{-12+\frac{12}{6}}(\mathfrak{e}_{6},D_{5}(a_{1}))]+2\mathfrak{u}(1)$}\\
\hline
	\end{tabular}}
\vspace{-1.5em}
	
\end{table}

}

{\linespread{2.0}
\begin{table}[H]\tiny
		\caption{$E_{7}$}
	\centering
	\resizebox{\linewidth}{!}{
	\begin{tabular}{|c|c|c|c|c|c|c|c|c|c|c|}\hline
		{$b$}& {$k$}&{HL}&{NL}&{HD}&{ND}& {$(f_{0},f_{1})$}&{$\mathfrak{f}$}&{$\Delta_{\text{Coulomb}}$}&{$(a,c)$}&{$\text{Theory}$}\\ \hline
		$18$&\makecell{$-13$} &\makecell{$A_{3}+A_{2}+A_{1}$}&\makecell{$A_{4}+A_{2}$}&\makecell{$100$}&\makecell{$106$} &\makecell{$(0,0)$}&\makecell{$\mathfrak{su}(2)_{0}$}&\makecell{$\frac{6}{5}$}&\makecell{$(\frac{43}{120},\frac{11}{30})$}&\makecell{$\mathcal{T}[W_{-18+\frac{18}{5}}(\mathfrak{e}_{7},A_{4}+A_{2})]$}\\
		\hline
		$18$&\makecell{$-16$\\$-14$} &\makecell{$E_{7}(a_{2})$\\$D_{5}(a_{1})+A_{1}$}&\makecell{$2A_{1}$\\$A_{3}+A_{2}$}&\makecell{$122$\\$108$}&\makecell{$52$\\$98$}&\makecell{$(1,0)$}&\makecell{$\mathfrak{so}(9)_{3}\times\mathfrak{su}(2)_{1}\times\mathfrak{u}(1)_{f_{0}}$\\$\mathfrak{su}(2)_{\frac{3}{2}}\times\mathfrak{u}(1)_{\frac{1}{6}}\times\mathfrak{u}(1)_{f_{0}}$}&\makecell{$3$\\$\frac{3}{2}$}&\makecell{$(\frac{43}{24},\frac{7}{3})$\\$(\frac{2}{3},\frac{5}{6})$}&\makecell{$\mathcal{T}[W_{-18+\frac{18}{2}}(\mathfrak{e}_{7},2A_{1})]+\mathfrak{u}(1)$\\$\mathcal{T}[W_{-18+\frac{18}{4}}(\mathfrak{e}_{7},A_{3}+A_{2})]+\mathfrak{u}(1)$}\\
		\hline
		$18$&\makecell{$-15$\\$-15$\\$-15$} &\makecell{$A_{6}$\\$D_{5}+A_{1}$\\$D_{6}(a_{1})$}&\makecell{$A_{2}+3A_{1}$\\$2A_{2}$\\$A_{3}$}&\makecell{$114$\\$114$\\$114$}&\makecell{$84$\\$84$\\$84$}&\makecell{$(0,2)$}&\makecell{$(\mathfrak{g}_{2})_{2}$\\$(\mathfrak{g}(2))_{2}\times\mathfrak{su}(2)_{0}$\\$\mathfrak{so}(7)_{2}\times\mathfrak{su}(2)_{0}$}&\makecell{$2$\\$2$\\$2$}&\makecell{$(\frac{23}{24},\frac{7}{6})$\\$(\frac{23}{24},\frac{7}{6})$\\$(\frac{23}{24},\frac{7}{6})$}&\makecell{$\mathcal{T}[W_{-18+\frac{18}{3}}(\mathfrak{e}_{7},A_{2}+3A_{1})]$\\$\mathcal{T}[W_{-18+\frac{18}{3}}(\mathfrak{e}_{7},2A_{2})]$\\$\mathcal{T}[W_{-18+\frac{18}{3}}(\mathfrak{e}_{7},A_{3})]$}\\
\hline
	\end{tabular}}
    \vspace{-1.5em}
\end{table}

}

{\linespread{2.0}
\begin{table}[H]\tiny
	\caption{$E_{8}$}
	\centering
	\resizebox{\linewidth}{!}{
	\begin{tabular}{|c|c|c|c|c|c|c|c|c|c|c|}\hline
		{$b$}& {$k$}&{Hitchin label}&{Nahm label}&{HD}&{$f_{0}$}&{$f_{1}$}&{$\mathfrak{f}$}&{$\Delta_{\text{Coulomb}}$}&{$(a,c)$}&{Theory}\\ \hline
		$30$&\makecell{$-28$\\$-28$\\$-26$} &\makecell{$E_{8}(a_{3})$\\$(E_{8}(a_{3}),\mathbb{Z}_{2})$\\$E_{6}(a_{1})+A_{1}$}&\makecell{$A_{2}$\\$3A_{1}$\\$A_{4}+A_{1}$}&\makecell{$234$\\$234$\\$218$}&\makecell{$0$}&\makecell{$0$}&\makecell{$(\mathfrak{e}_{6})_{3}$\\$(\mathfrak{f}_{4})_{3}\times\mathfrak{su}(2)_{\frac{1}{2}}$\\$\mathfrak{su}(3)_{\frac{3}{2}}\times\mathfrak{u}(1)_{0}$}&\makecell{$3$\\$3$\\$\frac{3}{2}$}&\makecell{$(\frac{41}{24},\frac{13}{6})$\\$(\frac{7}{4},\frac{9}{4})$\\$(\frac{7}{12},\frac{2}{3})$}&\makecell{$\mathcal{T}[W_{-30+\frac{30}{2}}(\mathfrak{e}_{8},A_{2})]$\\$\mathcal{T}[W_{-30+\frac{30}{2}}(\mathfrak{e}_{8},3A_{1})]$\\$\mathcal{T}[W_{-30+\frac{30}{4}}(\mathfrak{e}_{8},A_{4}+A_{1})]$}\\
		\hline
		$30$&\makecell{$-27$\\$-27$\\$-27$\\$-21$} &\makecell{$E_{8}(b_{5})$\\$(E_{8}(b_{5},S_{3}))$\\$(E_{8}(b_{5},\mathbb{Z}_{2}))$\\$A_{3}+A_{2}$}&\makecell{$D_{4}(a_{1})$\\$2A_{2}+A_{1}$\\$A_{3}+A_{1}$\\$D_{7}(a_{1})$}&\makecell{$226$\\$226$\\$226$\\$178$}&\makecell{$0$}&\makecell{$1$}&\makecell{$\mathfrak{so}(8)_{2}$\\$(\mathfrak{g}_{2})_{2}\times\mathfrak{su}(2)_{1}$\\$\mathfrak{so}(7)_{2}\times\mathfrak{su}(2)_{\frac{1}{2}}$\\$\mathfrak{u}(1)_{\frac{2}{3}}$}&\makecell{$2$\\$2$\\$2$\\$\frac{4}{3}$}&\makecell{$(\frac{23}{24},\frac{7}{6})$\\$(\frac{25}{24},\frac{4}{3})$\\$(1,\frac{5}{4})$\\$(\frac{11}{24},\frac{1}{2})$}&\makecell{$\mathcal{T}[W_{-30+\frac{30}{3}}(\mathfrak{e}_{8},D_{4}(a_{1}))]$\\$\mathcal{T}[W_{-30+\frac{30}{3}}(\mathfrak{e}_{8},2A_{2}+A_{1})]$\\$\mathcal{T}[W_{-30+\frac{30}{3}}(\mathfrak{e}_{8},A_{3}+A_{1})]$\\$\mathcal{T}[W_{-30+\frac{30}{9}}(\mathfrak{e}_{8},D_{7}(a_{1}))]$}\\
\hline
	\end{tabular}}
\vspace{-1.5em}
\end{table}

}
\newpage

{\linespread{2.0}
\begin{table}[H]\tiny
	\caption{$A_{2n-1}/\mathbb{Z}_{2}$, $b_{t}=4n-2$}
	\centering
		\resizebox{\linewidth}{!}{
	\begin{tabular}{|c|c|c|c|c|c|c|c|c|c|c|c|}\hline
		{$n$}& {$k_{t}$}&{ C-partition}&{B-partition}&{HD}&{ND}&{$f_{0}$}& {$f_{1}$}&{$\mathfrak{f}$}&{$\Delta_{\text{Coulomb}}$}&{$(a,c)$}&{$\text{Theory}$}\\ \hline
        $1$&\makecell{$1$\\$3$} &\makecell{$(2)$\\$(1^{2})$}&\makecell{$(1^{3})$\\$(3)$}&\makecell{$2$\\$0$}&\makecell{$0$\\$2$}&\makecell{$0$\\$0$} &\makecell{$0$\\$0$}&\makecell{$\mathfrak{so}(3)_{\frac{2}{3}}$\\$-$}&\makecell{$\frac{4}{3}$\\$\frac{6}{5}$}&\makecell{$(\frac{11}{24},\frac{1}{2})$\\$(\frac{43}{120},\frac{11}{30})$}&\makecell{$W_{-1+\frac{1}{3}}(B_{1},(1^{3}))$\\$W_{-1+\frac{1}{5}}(B_{1},(3))$}\\
		\hline
		$2$&\makecell{$-3$\\$-3$} &\makecell{$(2^{2})$\\$((2^{2}),\mathbb{Z}_{2})$}&\makecell{$(3,1^{2})$\\$(2^{2},1)$}&\makecell{$6$\\$6$}&\makecell{$6$\\$4$}&\makecell{$0$\\$0$} &\makecell{$2$\\$2$}&\makecell{$\mathfrak{u}(1)_{0}$\\$\mathfrak{su}(2)_{\frac{3}{2}}$}&\makecell{$1$\\$2$}&\makecell{$(\frac{5}{24},\frac{1}{6})$\\$(\frac{3}{4},\frac{3}{4})$}&\makecell{$\mathcal{T}[W_{-3+1}(B_{2},(3,1^{2}))]$\\$\mathcal{T}[W_{-3+1}(B_{2},(2^{2},1))]$}\\
		\hline
		$3$&\makecell{$-7$} &\makecell{$(3^{2})$}&\makecell{$(3,2^{2})$}&\makecell{$14$}&\makecell{$12$}&\makecell{$0$}&\makecell{$0$}&\makecell{$\mathfrak{su}(2)_{\frac{4}{3}}$}&\makecell{$\frac{4}{3}$}&\makecell{$(\frac{11}{24},\frac{1}{2})$}&\makecell{$\mathcal{T}[W_{-5+\frac{5}{3}}(B_{3},(3,2^{2}))]$}\\
		\hline
		$4$&\makecell{$-11$} &\makecell{$(4,2^{2})$}&\makecell{$(3^{2},1^{3})$}&\makecell{$26$}&\makecell{$22$}&\makecell{$0$}&\makecell{$0$}&\makecell{$\mathfrak{su}(2)_{\frac{4}{3}}\times\mathfrak{u}(1)_{0}$}&\makecell{$\frac{4}{3}$}&\makecell{$(\frac{11}{24},\frac{1}{2})$}&\makecell{$\mathcal{T}[W_{-7+\frac{7}{3}}(B_{4},(3^{2},1^{3}))]$}\\
\hline
	\end{tabular}}
\vspace{-1.5em}
\end{table}

}

{\linespread{2.0}
\begin{table}[H]\tiny
	\caption{$A_{2n-1}/\mathbb{Z}_{2}$, $b_{t}=2n$}
	\centering
	\resizebox{\linewidth}{!}{
	\begin{tabular}{|c|c|c|c|c|c|c|c|c|c|c|c|}\hline
		{$n$}& {$k_{t}$}&{HL}&{NL}&{HD}&{ND}&{$f_{0}$}& {$f_{1}$}&{$\mathfrak{f}$}&{$\Delta_{\text{Coulomb}}$}&{$(a,c)$}&{$\text{Theory}$}\\ \hline
        $1$&\makecell{$2$\\$4$} &\makecell{$(2)$\\$(1^{2})$}&\makecell{$(1^{3})$\\$(3)$}&\makecell{$2$\\$0$}&\makecell{$0$\\$2$}&\makecell{$1$\\$1$} &\makecell{$0$\\$0$}&\makecell{$\mathfrak{so}(3)_{\frac{3}{4}}\times\mathfrak{u}(1)_{f_{0}}$\\$\mathfrak{u}(1)_{f_{0}}$}&\makecell{$\frac{3}{2}$\\$\frac{4}{3}$}&\makecell{$(\frac{7}{12},\frac{2}{3})$\\$(\frac{11}{24},\frac{1}{2})$}&\makecell{$W_{-1+\frac{1}{4}}(B_{1},(1^{3}))$\\$W_{-1+\frac{1}{6}}(B_{1},(3))$}\\
		\hline
		$3$&\makecell{$-5$\\$-4$} &\makecell{$(6)$\\$(3^{2})$}&\makecell{$(1^{7})$\\$(3,2^{2})$}&\makecell{$18$\\$14$}&\makecell{$0$\\$12$}&\makecell{$0$\\$1$}&\makecell{$1$\\$0$}&\makecell{$\mathfrak{so}(7)_{2}$\\$\mathfrak{su}(2)_{\frac{3}{2}}\times\mathfrak{u}(1)_{f_{0}}$}&\makecell{$2$\\$\frac{3}{2}$}&\makecell{$(\frac{23}{24},\frac{7}{6})$\\$(\frac{5}{8},\frac{3}{4})$}&\makecell{$\mathcal{T}[W_{-5+3}(B_{3},(1^{7}))]$\\$\mathcal{T}[W_{-5+\frac{3}{2}}(B_{3},(3,2^{2}))]+\mathfrak{u}(1)$}\\
		\hline
		$4$&\makecell{$-7$\\$-6$} &\makecell{$(8)$\\$(4,2^{2})$}&\makecell{$(1^{9})$\\$(3^{2},1^{3})$}&\makecell{$32$\\$26$}&\makecell{$0$\\$22$}&\makecell{$1$\\$2$}&\makecell{$0$\\$1$}&\makecell{$\mathfrak{so}(9)_{3}\times\mathfrak{u}(1)_{f_{0}}$\\$\mathfrak{so}(3)_{1}\times\mathfrak{u}(1)_{1}\times\mathfrak{u}(1)^{2}_{f_{0}}$}&\makecell{$3$\\$2$}&\makecell{$(\frac{41}{24},\frac{13}{6})$\\$(\frac{23}{24},\frac{7}{6})$}&\makecell{$\mathcal{T}[W_{-7+4}(B_{4},(1^{9}))]+\mathfrak{u}(1)$\\$\mathcal{T}[W_{-7+2}(B_{4},(3^{2},1^{3}))]+2\mathfrak{u}(1)$}\\
\hline
	\end{tabular}}
\vspace{-1.5em}
\end{table}

}
{\linespread{2.0}
\begin{table}[H]\tiny
	\caption{$D_{n+1}/\mathbb{Z}_{2}$, $b_{t}=2n+2$}
	\centering
	\resizebox{\linewidth}{!}{
	\begin{tabular}{|c|c|c|c|c|c|c|c|c|c|c|c|}\hline
		{$n$}& {$k_{t}$}&{HL}&{NL}&{HD}&{ND}&{$f_{0}$}& {$f_{1}$}&{$\mathfrak{f}$}&{$\Delta_{\text{Coulomb}}$}&{$(a,c)$}&{$\text{Theory}$}\\ \hline
        $1$&\makecell{$-1$\\$1$} &\makecell{$(3)$\\$(1^{3})$}&\makecell{$(1^{2})$\\$(2)$}&\makecell{$2$\\$0$}&\makecell{$0$\\$2$}&\makecell{$0$\\$0$} &\makecell{$0$\\$0$}&\makecell{$\mathfrak{su}(2)_{\frac{4}{3}}$\\$-$}&\makecell{$\frac{4}{3}$\\$\frac{6}{5}$}&\makecell{$(\frac{11}{24},\frac{1}{2})$\\$(\frac{43}{120},\frac{11}{30})$}&\makecell{$W_{-2+\frac{2}{3}}(C_{1},(1^{2}))$\\$W_{-2+\frac{2}{5}}(C_{1},(2))$}\\
		\hline
		$3$&\makecell{$-5$} &\makecell{$(3^{2},1)$}&\makecell{$(2^{3})$}&\makecell{$14$}&\makecell{$12$}&\makecell{$0$} &\makecell{$0$}&\makecell{$\mathfrak{so}(3)_{\frac{2}{3}}$}&\makecell{$\frac{4}{3}$}&\makecell{$(\frac{11}{24},\frac{1}{2})$}&\makecell{$\mathcal{T}[W_{-4+\frac{4}{3}}(C_{3},(2^{3})]$}\\
		\hline
		$4$&\makecell{$-7$} &\makecell{$(5,2^{2})$}&\makecell{$(3^{2},1^{2})$}&\makecell{$26$}&\makecell{$22$}&\makecell{$0$} &\makecell{$0$}&\makecell{$\mathfrak{su}(2)_{\frac{4}{3}}\times\mathfrak{su}(2)_{0}$}&\makecell{$\frac{4}{3}$}&\makecell{$(\frac{11}{24},\frac{1}{2})$}&\makecell{$\mathcal{T}[W_{-5+\frac{5}{3}}(C_{4},(3^{2},1^{2})]$}\\
		\hline
		$5$&\makecell{$-9$\\$-9$} &\makecell{$(5,3^{2})$\\$((5,3^{2}),\mathbb{Z}_{2})$}&\makecell{$(3^{2},2^{2})$\\$(3^{2},2,1^{2})$}&\makecell{$42$\\$42$}&\makecell{$36$\\$34$}&\makecell{$0$\\$0$} &\makecell{$2$\\$2$}&\makecell{$\mathfrak{su}(2)_{0}\times\mathfrak{u}(1)$\\$\mathfrak{su}(2)_{\frac{3}{2}}\times\mathfrak{su}(2)_{0}$}&\makecell{$1$\\$2$}&\makecell{$(\frac{5}{24},\frac{1}{6})$\\$(\frac{3}{4},\frac{3}{4})$}&\makecell{$\mathcal{T}[W_{-6+2}(C_{5},(3^{2},2^{2})]$\\$\mathcal{T}[W_{-6+2}(C_{5},(3^{2},2,1^{2})]$}\\
		\hline
	\end{tabular}}
\vspace{-1.5em}
\end{table}

}
{\linespread{2.0}
\begin{table}[H]\tiny
	\caption{$D_{n+1}/\mathbb{Z}_{2}$, $b_{t}=2n$}
	\centering
	\resizebox{\linewidth}{!}{
	\begin{tabular}{|c|c|c|c|c|c|c|c|c|c|c|c|}\hline
		{$n$}& {$k_{t}$}&{HL}&{NL}&{HD}&{ND}&{$f_{0}$}& {$f_{1}$}&{$\mathfrak{f}$}&{$\Delta_{\text{Coulomb}}$}&{$(a,c)$}&{$\text{Theory}$}\\ \hline
        $1$&\makecell{$0$} &\makecell{$(3)$}&\makecell{$(1^{2})$}&\makecell{$2$}&\makecell{$0$}&\makecell{$1$} &\makecell{$0$}&\makecell{$\mathfrak{su}(2)_{\frac{3}{2}}\times\mathfrak{u}(1)_{f_{0}}$}&\makecell{$\frac{3}{2}$}&\makecell{$(\frac{7}{12},\frac{2}{3})$}&\makecell{$W_{-2+\frac{1}{2}}(C_{1},(1^{2}))+M(1)$}\\
		\hline
		$3$&\makecell{$-4$\\$-4$} &\makecell{$(5,1^{2})$\\$((5,1^{2}),\mathbb{Z}_{2})$}&\makecell{$(2^{2},1^{2})$\\$(2,1^{4})$}&\makecell{$16$\\$16$}&\makecell{$10$\\$6$}&\makecell{$1$\\$1$} &\makecell{$0$\\$0$}&\makecell{$\mathfrak{su}(2)_{\frac{3}{2}}\times\mathfrak{u}(1)_{0}\times\mathfrak{u}(1)_{f_{0}}$\\$\mathfrak{sp}(2)_{2}\times\mathfrak{u}(1)_{f_{0}}$}&\makecell{$\frac{3}{2}$\\$3$}&\makecell{$(\frac{7}{12},\frac{2}{3})$\\$(\frac{17}{12},\frac{19}{12})$}&\makecell{$\mathcal{T}[W_{-4+\frac{3}{2}}(C_{3},(2^{2},1^{2})]+\mathfrak{u}(1)$\\$\mathcal{T}[W_{-4+\frac{3}{2}}(C_{3},(2,1^{4})]+\mathfrak{u}(1)$}\\
		\hline
		$5$&\makecell{$-8$\\$-8$} &\makecell{$(7,3,1)$\\$((7,3,1),\mathbb{Z}_{2})$}&\makecell{$(2^{4},1^{2})$\\$(2^{3},1^{4})$}&\makecell{$46$\\$46$}&\makecell{$28$\\$24$}&\makecell{$1$\\$1$} &\makecell{$0$\\$0$}&\makecell{$\mathfrak{so}(4)_{0}\times\mathfrak{su}(2)_{\frac{3}{2}}\times\mathfrak{u}(1)_{f_{0}}$\\$\mathfrak{sp}(2)_{2}\times\mathfrak{so}(3)_{0}\times\mathfrak{u}(1)_{f_{0}}$}&\makecell{$\frac{3}{2}$\\$3$}&\makecell{$(\frac{7}{12},\frac{2}{3})$\\$(\frac{17}{12},\frac{19}{12})$}&\makecell{$\mathcal{T}[W_{-6+\frac{5}{2}}(C_{5},(2^{4},1^{2})]+\mathfrak{u}(1)$\\$\mathcal{T}[W_{-6+\frac{5}{2}}(C_{5},(2^{3},1^{4})]+\mathfrak{u}(1)$}\\
		\hline
	\end{tabular}}
\vspace{-1.5em}
\end{table}

}
{\linespread{2.0}
\begin{table}[H]\tiny
	\caption{$D_{4}/\mathbb{Z}_{3}$, $b_{t}=12$,  $b_{t}=6$}
	\centering
	\begin{tabular}{|c|c|c|c|c|c|c|c|c|c|c|}\hline
		{$b_{t}$}& {$k_{t}$}&{HL}&{NL}&{HD}&{$f_{0}$}& {$f_{1}$}&{$\mathfrak{f}$}&{$\Delta_{\text{Coulomb}}$}&{$(a,c)$}&{$\text{Theory}$}\\ \hline
		$12$&\makecell{$-10$\\$-4$} &\makecell{$G_{2}$\\$0$}&\makecell{$0$\\$G_{2}$}&\makecell{$12$\\$0$}&\makecell{$0$\\$0$} &\makecell{$1$\\$1$}&\makecell{$(\mathfrak{g}_{2})_{2}$\\$-$}&\makecell{$2$\\$1$}&\makecell{$(\frac{23}{24},\frac{7}{6})$\\$(\frac{5}{24},\frac{1}{6})$}&\makecell{$\mathcal{T}[W_{-4+2}(\mathfrak{g}_{2},0)]$\\$\mathcal{T}[W_{-4+\frac{1}{2}}(\mathfrak{g}_{2},G_{2})]$}\\
		\hline
	\end{tabular}
\end{table}

}
{\linespread{2.0}
\begin{table}[H]\tiny
	\caption{$E_{6}/\mathbb{Z}_{2}$}
	\centering
	\resizebox{\linewidth}{!}{
	\begin{tabular}{|c|c|c|c|c|c|c|c|c|c|c|c|}\hline
		{$b_{t}$}& {$k_{t}$}&{HL}&{NL}&{HD}&{ND}&{$f_{0}$}& {$f_{1}$}&{$\mathfrak{f}$}&{$\Delta_{\text{Coulomb}}$}&{$(a,c)$}&{$\text{Theory}$}\\ \hline
		$12$&\makecell{$-11$\\$-10$\\$-10$\\$-8$\\$-8$\\$-3$} &\makecell{$F_{4}$\\$B_{3}$\\$C_{3}$\\$A_{2}$\\$\tilde{A}_{2}$\\$0$}&\makecell{$0$\\$A_{2}$\\$\tilde{A}_{2}$\\$B_{3}$\\$C_{3}$\\$F_{4}$}&\makecell{$48$\\$42$\\$42$\\$30$\\$30$\\$0$}&\makecell{$0$\\$30$\\$30$\\$42$\\$42$\\$48$}&\makecell{$0$\\$0$\\$0$\\$0$\\$0$\\$0$} &\makecell{$0$\\$2$\\$2$\\$1$\\$1$\\$1$}&\makecell{$(\mathfrak{f}_{4})_{3}$\\$\mathfrak{su}(3)_{2}$\\$(\mathfrak{g}_{2})_{2}$\\$\mathfrak{su}(2)_{0}$\\$\mathfrak{su}(2)_{\frac{3}{2}}$\\$-$}&\makecell{$3$\\$2$\\$2$\\$1$\\$\frac{3}{2}$\\$1$}&\makecell{$(\frac{41}{24},\frac{13}{6})$\\$(\frac{23}{24},\frac{7}{6})$\\$(\frac{23}{24},\frac{7}{6})$\\$(\frac{5}{24},\frac{1}{6})$\\$(\frac{23}{24},\frac{11}{12})$\\$(\frac{5}{24},\frac{1}{6})$}&\makecell{$\mathcal{T}[W_{-9+6}(\mathfrak{f}_{4},0)]$\\$\mathcal{T}[W_{-9+3}(\mathfrak{f}_{4},A_{2})]$\\$\mathcal{T}[W_{-9+3}(\mathfrak{f}_{4},\tilde{A}_{2})]$\\$\mathcal{T}[W_{-9+\frac{3}{2}}(\mathfrak{f}_{4},B_{3})]$\\$\mathcal{T}[W_{-9+\frac{3}{2}}(\mathfrak{f}_{4},C_{3})]$\\$\mathcal{T}[W_{-9+\frac{2}{3}}(\mathfrak{f}_{4},F_{4})]$}\\
\hline
	\end{tabular}}
\vspace{-2em}
\end{table}
}

\newpage

\subsection{Rank 2}
\label{sec:completList2}

{\linespread{2.0}

\begin{table}[H]\tiny
	\caption{$A_{n-1}$, $b=n$}
	\centering
	\resizebox{\linewidth}{!}{
	\begin{tabular}{|c|c|c|c|c|c|c|c|c|c|}
		\hline
		{$n$}& {$k$}&{$\mathcal{O}_{f}$}&{$f_{0}$}& {$f_{1}$}&{ND}&{$\mathfrak{f}$}&{$\Delta_{\text{Coulomb}}$}&{$(a,c)$}&{Theory}\\
		\hline
		$2$&\makecell{$3$\\$4$\\$5$\\$6$} &\makecell{$(1^{2})$\\$(1^{2})$\\$(2)$\\$(2)$}&\makecell{$0$\\$1$\\$0$\\$1$}&\makecell{$0$\\$1$\\$0$\\$1$} &\makecell{$0$\\$0$\\$2$\\$2$}&\makecell{$\mathfrak{su}(2)_{\frac{8}{5}}$\\$\mathfrak{su}(2)_{\frac{5}{3}}\times\mathfrak{u}(1)_{f_{0}}$\\$-$\\$\mathfrak{u}(1)_{f_{0}}$}&\makecell{$\frac{8}{5},\frac{6}{5}$\\$\frac{5}{3},\frac{4}{3}$\\$\frac{10}{7},\frac{8}{7}$\\$\frac{3}{2},\frac{5}{4}$}&\makecell{$(\frac{19}{20},1)$\\$(\frac{13}{12},\frac{7}{6})$\\$(\frac{67}{84},\frac{17}{21})$\\$(\frac{11}{12},\frac{23}{24})$}&\makecell{$\mathcal{T}[W_{-2+\frac{2}{5}}(\mathfrak{su}(2),(1^{2}))]$\\$\mathcal{T}[W_{-2+\frac{2}{6}}(\mathfrak{su}(2),(1^{2}))]+\mathfrak{u}(1)$\\$\mathcal{T}[W_{-2+\frac{2}{7}}(\mathfrak{su}(2),(2))]$\\$\mathcal{T}[W_{-2+\frac{2}{8}}(\mathfrak{su}(2),(2))]+\mathfrak{u}(1)$}\\
		\hline
		$3$&\makecell{$1$} &\makecell{$(2,1)$}&\makecell{$0$}&\makecell{$0$}&\makecell{$4$}&\makecell{$\mathfrak{u}(1)_{\frac{3}{2}}$}&\makecell{$\frac{3}{2},\frac{5}{4}$}&\makecell{$(\frac{11}{12},\frac{23}{24})$}&\makecell{$\mathcal{T}[W_{-3+\frac{3}{4}}(\mathfrak{su}(3),(2,1))]$}\\
		\hline
		$4$&\makecell{$-1$\\$0$} &\makecell{$(2,1^{2})$\\$(2,1^{2})$}&\makecell{$0$\\$3$}&\makecell{$0$\\$3$}&\makecell{$6$\\$6$}&\makecell{$\mathfrak{su}(2)_{\frac{5}{3}}\times\mathfrak{u}(1)_{\frac{4}{3}}$\\$\mathfrak{su}(2)_{2}\times\mathfrak{u}(1)_{2}\times\mathfrak{u}(1)^{3}_{f_{0}}$}&\makecell{$\frac{4}{3},\frac{5}{3}$\\$2,2$}&\makecell{$(\frac{13}{12},\frac{7}{6})$\\$(\frac{7}{4},2)$}&\makecell{$\mathcal{T}[W_{-4+\frac{4}{3}}(\mathfrak{su}(4),(2,1^{2}))]$\\$\mathcal{T}[W_{-4+\frac{4}{4}}(\mathfrak{su}(4),(2,1^{2}))]+3\mathfrak{u}(1)$}\\
\hline
	\end{tabular}}
\vspace{-1.5em}
\end{table}
}

{\linespread{2.0}

\begin{table}[H]\tiny
	\caption{$A_{n-1}$, $b=n-1$}
	\centering
	\resizebox{\linewidth}{!}{
	\begin{tabular}{|c|c|c|c|c|c|c|c|c|c|}\hline
		{$n$}& {$k$}&{$\mathcal{O}_{f}$}&{$f_{0}$}& {$f_{1}$}&{ND}&{$\mathfrak{f}$}&{$\Delta_{\text{Coulomb}}$}&{$(a,c)$}&{Theory}\\ \hline
		$3$&\makecell{$1$} &\makecell{$(2,1)$}&\makecell{$1$}&\makecell{$0$}&\makecell{$4$}&\makecell{$\mathfrak{u}(1)_{\frac{5}{3}}\times\mathfrak{u}(1)_{f_{0}}$}&\makecell{$\frac{5}{3},\frac{4}{3}$}&\makecell{$(\frac{13}{12},\frac{7}{6})$}&\makecell{$\mathcal{T}[W_{-3+\frac{2}{3}}(\mathfrak{su}(3),(2,1))]+\mathfrak{u}(1)$}\\
		\hline
		$4$&\makecell{$-1$\\$2$} &\makecell{$(1^{4})$\\$(4)$}&\makecell{$1$\\$1$}&\makecell{$0$\\$0$}&\makecell{$0$\\$12$}&\makecell{$\mathfrak{su}(4)_{\frac{5}{2}}\times\mathfrak{u}(1)_{f_{0}}$\\$\mathfrak{u}(1)_{f_{0}}$}&\makecell{$\frac{5}{2},\frac{3}{2}$\\$\frac{8}{5},\frac{6}{5}$}&\makecell{$(\frac{7}{4},2)$\\$(\frac{19}{20},1)$}&\makecell{$\mathcal{T}[W_{-4+\frac{3}{2}}(\mathfrak{su}(4),(1^{4}))]+\mathfrak{u}(1)$\\$\mathcal{T}[W_{-4+\frac{3}{5}}(\mathfrak{su}(4),(4))]+\mathfrak{u}(1)$}\\
		\hline
		$5$&\makecell{$-2$\\$-1$\\$2$} &\makecell{$(1^{5})$\\$(3,1^{2})$\\$(5)$}&\makecell{$2$\\$1$\\$2$}&\makecell{$1$\\$0$\\$1$}&\makecell{$0$\\$14$\\$20$}&\makecell{$\mathfrak{su}(5)_{3}\times\mathfrak{u}(1)^{2}_{f_{0}}$\\$\mathfrak{su}(2)_{\frac{5}{3}}\times\mathfrak{u}(1)_{1}\times\mathfrak{u}(1)_{f_{0}}$\\$\mathfrak{u}(1)^{2}_{f_{0}}$}&\makecell{$3,2$\\$\frac{4}{3},\frac{5}{3}$\\$\frac{5}{3},\frac{4}{3}$}&\makecell{$(\frac{29}{12},\frac{17}{6})$\\$(\frac{9}{8},\frac{5}{4})$\\$(\frac{13}{12},\frac{7}{6})$}&\makecell{$\mathcal{T}[W_{-5+\frac{4}{2}}(\mathfrak{su}(5),(1^{5}))]+2\mathfrak{u}(1)$\\$\mathcal{T}[W_{-5+\frac{4}{3}}(\mathfrak{su}(5),(3,1^{2}))]+\mathfrak{u}(1)$\\$\mathcal{T}[W_{-5+\frac{4}{6}}(\mathfrak{su}(5),(5))]+2\mathfrak{u}(1)$}\\
\hline
	\end{tabular}}
\vspace{-3em}
\end{table}
}
\newpage

{\linespread{2.0}

\begin{table}[H]\tiny
	\caption{$D_{n}$, $b=n$}
	\centering
	\resizebox{\linewidth}{!}{
	\begin{tabular}{|c|c|c|c|c|c|c|c|c|c|c|c|}\hline
		{$n$}& {$k$}&{HL}&{NL}&{HD}&{ND}&{$f_{0}$}& {$f_{1}$}&{$\makecell{f}$}&{$\Delta_{\text{Coulomb}}$}&{$(a,c)$}&{Theory}\\ \hline
		$3$&\makecell{$-1$\\$0$\\$2$} &\makecell{$(5,1)$\\$(3,3)$\\$(1^{6})$}&\makecell{$(1^{6})$\\$(2^{2},1^{2})$\\$(5,1)$}&\makecell{$12$\\$10$\\$0$}&\makecell{$0$\\$6$\\$12$}&\makecell{$1$\\$3$\\$1$}&\makecell{$0$\\$2$\\$0$}&\makecell{$\mathfrak{su}(4)_{\frac{5}{2}}\times\mathfrak{u}(1)_{f_{0}}$\\$\mathfrak{su}(2)_{2}\times\mathfrak{u}(1)_{1}\times\mathfrak{u}(1)^{3}_{f_{0}}$\\$\mathfrak{u}(1)_{f_{0}}$}&\makecell{$\frac{5}{2},\frac{3}{2}$\\$2,2$\\$\frac{8}{5},\frac{6}{5}$}&\makecell{$(\frac{7}{4},2)$\\$(\frac{7}{4},2)$\\$(\frac{19}{20},1)$}&\makecell{$\mathcal{T}[W_{-4+\frac{3}{2}}(D_{3},(1^{6}))]+\mathfrak{u}(1)$\\$\mathcal{T}[W_{-4+\frac{3}{3}}(D_{3},(2^{2},1^{2}))]+3\mathfrak{u}(1)$\\$\mathcal{T}[W_{-4+\frac{3}{5}}(D_{3},(5,1))]+\mathfrak{u}(1)$}\\
		\hline
		$4$&\makecell{$-2$\\$-2$\\$0$\\$0$} &\makecell{$(4,4)$\\$(5,1^{3})$\\$(2^{4})$\\$(3,1^{5})$}&\makecell{$(2^{4})$\\$(3,1^{5})$\\$(4^{2})$\\$(5,1^{3})$}&\makecell{$20$\\$20$\\$12$\\$12$}&\makecell{$12$\\$12$\\$20$\\$20$}&\makecell{$0$\\$0$\\$4$\\$4$}&\makecell{$3$\\$3$\\$3$\\$3$}&\makecell{$\mathfrak{sp}(2)_{2}$\\$\mathfrak{sp}(2)_{2}$\\$\mathfrak{su}(2)_{2}\times\mathfrak{u}(1)^{4}_{f_{0}}$\\$\mathfrak{su}(2)_{2}\times\mathfrak{u}(1)^{4}_{f_{0}}$}&\makecell{$2,2$\\$2,2$\\$ 2,2$\\$2,2$}&\makecell{$(\frac{19}{12},\frac{5}{3})$\\$(\frac{19}{12},\frac{5}{3})$\\$(\frac{7}{4},2)$\\$(\frac{7}{4},2)$}&\makecell{$\mathcal{T}[W_{-6+\frac{4}{2}}(D_{4},(2^{4}))]$\\$\mathcal{T}[W_{-6+\frac{4}{2}}(D_{4},(3,1^{5}))]$\\$\mathcal{T}[W_{-6+\frac{4}{4}}(D_{4},(4^{2}))]+4\mathfrak{u}(1)$\\$\mathcal{T}[W_{-6+\frac{4}{4}}(D_{4},(5,1^{3}))]+4\mathfrak{u}(1)$}\\
\hline
	\end{tabular}}
\vspace{-1.5em}
\end{table}
}
\newpage

{\linespread{2.0}
\begin{table}[H]\tiny
	\caption{$D_{n}$, $b=2n-2$}
	\centering
	\resizebox{\linewidth}{!}{
	\begin{tabular}{|c|c|c|c|c|c|c|c|c|c|c|c|}\hline
		{$n$}& {$k$}&{HL}&{NL}&{HD}&{ND}&{$f_{0}$}& {$f_{1}$}&{$\mathfrak{f}$}&{$\Delta_{\text{Coulomb}}$}&{$(a,c)$}&{Theory}\\ \hline
		$3$&\makecell{$-1$} &\makecell{$(3,3)$}&\makecell{$(2^{2},1^{2})$}&\makecell{$10$}&\makecell{$6$}&\makecell{$0$}&\makecell{$0$}&\makecell{$\mathfrak{su}(2)_{\frac{5}{3}}\times\mathfrak{u}(1)_{\frac{2}{3}}$}&\makecell{$\frac{4}{3},\frac{5}{3}$}&\makecell{$(\frac{13}{12},\frac{7}{6})$}&\makecell{$\mathcal{T}[W_{-4+\frac{4}{3}}(D_{3},(2^{2},1^{2}))]$}\\
		\hline
		$4$&\makecell{$-2$\\$-2$\\$-1$\\$-1$} &\makecell{$(3^{2},1^{2})$\\$(3^{2},1^{2})$\\$(2^{4})$\\$(3,1^{5})$}&\makecell{$(3^{2},1^{2})$\\$(3,2^{2},1)$\\$(4^{2})$\\$(5,1^{3})$}&\makecell{$18$\\$18$\\$12$\\$12$}&\makecell{$18$\\$16$\\$20$\\$20$}&\makecell{$2$\\$2$\\$0$\\$0$}&\makecell{$0$\\$0$\\$0$\\$0$}&\makecell{$\mathfrak{u}(1)\times\mathfrak{u}(1)\times\mathfrak{u}(1)^{2}_{f_{0}}$\\$\mathfrak{su}(2)_{2}\times\mathfrak{u}(1)^{2}_{f_{0}}$\\$\mathfrak{su}(2)_{\frac{8}{5}}$\\$\mathfrak{so}(3)_{\frac{4}{5}}$}&\makecell{$\frac{3}{2},\frac{3}{2}$\\$3,\frac{3}{2}$\\$\frac{6}{5},\frac{8}{5}$\\$\frac{8}{5},\frac{6}{5}$}&\makecell{$(\frac{7}{6},\frac{4}{3})$\\$(\frac{47}{24},\frac{13}{6})$\\$(\frac{19}{20},1)$\\$(\frac{19}{20},1)$}&\makecell{$\mathcal{T}[W_{-6+\frac{6}{4}}(D_{4},(3^{2},1^{2}))]+2\mathfrak{u}(1)$\\$\mathcal{T}[W_{-6+\frac{6}{4}}(D_{4},(3,2^{2},1))]+2\mathfrak{u}(1)$\\$\mathcal{T}[W_{-6+\frac{6}{5}}(D_{4},(4^{2}))]$\\$\mathcal{T}[W_{-6+\frac{6}{5}}(D_{4},(5,1^{3}))]$}\\
\hline
	\end{tabular}}
\vspace{-2em}
\end{table}
}

{\linespread{2.0}

\begin{table}[H]\tiny
\caption{$E_{6}$}
\centering
\resizebox{\linewidth}{!}{
\begin{tabular}{|c|c|c|c|c|c|c|c|c|c|c|}\hline
	{$b$}& {$k$}&{HL}&{NL}&{HD}&{ND}& {$(f_{0},f_{1})$}&{$\mathfrak{f}$}&{$\Delta_{\text{Coulomb}}$}&{$(a,c)$}&{Theory}\\ \hline
	$12$&\makecell{$-7$} &\makecell{$A_{3}$}&\makecell{$A_{4}$}&\makecell{$52$} &\makecell{$60$}&\makecell{$(0,0)$}&\makecell{$\mathfrak{su}(2)_{\frac{8}{5}}\times\mathfrak{u}(1)_{0}$}&\makecell{$\frac{6}{5},\frac{8}{5}$}&\makecell{$(\frac{19}{20},1)$}&\makecell{$\mathcal{T}[W_{-12+\frac{12}{5}}(\mathfrak{e}_{6},A_{4})]$}\\
	\hline
	$12$&\makecell{$-10$\\$-2$} &\makecell{$E_{6}(a_{1})$\\$A_{1}$}&\makecell{$A_{1}$\\$E_{6}(a_{1})$}&\makecell{$70$\\$22$}&\makecell{$22$\\$70$}&\makecell{$(0,1)$}&\makecell{$\mathfrak{su}(6)_{3}$\\$-$}&\makecell{$3,2$\\$\frac{6}{5},\frac{6}{5}$}&\makecell{$(\frac{29}{12},\frac{17}{6})$\\$(\frac{43}{60},\frac{11}{15})$}&\makecell{$\mathcal{T}[W_{-12+\frac{12}{2}}(\mathfrak{e}_{6},A_{1})]$\\$\mathcal{T}[W_{-12+\frac{12}{10}}(\mathfrak{e}_{6},E_{6}(a_{1}))]$}\\
	\hline
	$12$&\makecell{$-9$\\$-9$} &\makecell{$E_{6}(a_{3})$\\$(E_{6}(a_{3}),\mathbb{Z}_{2})$}&\makecell{$A_{2}$\\$3A_{1}$}&\makecell{$66$\\$66$}&\makecell{$42$\\$40$}&\makecell{$(2,1)$}&\makecell{$\mathfrak{su}(3)_{2}\times\mathfrak{su}(3)_{2}\times\mathfrak{u}(1)^{2}_{f_{0}}$\\$\mathfrak{su}(3)_{4}\times\mathfrak{su}(2)_{\frac{5}{2}}\times\mathfrak{u}(1)^{2}_{f_{0}}$}&\makecell{$2,2$\\$4,2$}&\makecell{$(\frac{23}{12},\frac{7}{3})$\\$(\frac{71}{24},\frac{41}{12})$}&\makecell{$\mathcal{T}[W_{-12+\frac{12}{3}}(\mathfrak{e}_{6},A_{2})]+2\mathfrak{u}(1)$\\$\mathcal{T}[W_{-12+\frac{12}{3}}(\mathfrak{e}_{6},3A_{1})]+2\mathfrak{u}(1)$}\\
\hline
\end{tabular}}
\vspace{-1.5em}
\end{table}
}

{\linespread{2.0}

\begin{table}[H]\tiny
	\caption{$E_{7}$}
	\centering
	\resizebox{\linewidth}{!}{
	\begin{tabular}{|c|c|c|c|c|c|c|c|c|c|c|}\hline
		{$b$}& {$k$}&{HL}&{NL}&{HD}&{ND}& {$f_{0},f_{1}$}&{$\mathfrak{f}$}&{$\Delta_{\text{Coulomb}}$}&{$(a,c)$}&{$\text{Theory}$}\\ \hline
		$18$&\makecell{$-16$\\$-14$\\$-14$\\$-10$} &\makecell{$E_{7}(a_{1})$\\$E_{6}(a_{3})$\\$(E_{6}(a_{3}),\mathbb{Z}_{2})$\\$A_{2}+2A_{1}$}&\makecell{$A_{1}$\\$D_{4}(a_{1})+A_{1}$\\$A_{3}+2A_{1}$\\$E_{7}(a_{4})$}&\makecell{$124$\\$110$\\$110$\\$82$}&\makecell{$34$\\$96$\\$94$\\$116$}&\makecell{$(1,0)$}&\makecell{$\mathfrak{so}(12)_{5}\times\mathfrak{u}(1)_{f_{0}}$\\$\mathfrak{su}(2)_{\frac{3}{2}}\times\mathfrak{su}(2)_{\frac{3}{2}}\times\mathfrak{u}(1)_{f_{0}}$\\$\mathfrak{su}(2)_{2}\times\mathfrak{su}(2)_{3}\times\mathfrak{u}(1)_{f_{0}}$\\$\mathfrak{u}(1)_{f_{0}}$}&\makecell{$5,3$\\$\frac{3}{2},\frac{3}{2}$\\$3,\frac{3}{2}$\\$\frac{5}{4},\frac{3}{2}$}&\makecell{$(\frac{53}{12},\frac{16}{3})$\\$(\frac{29}{24},\frac{17}{12})$\\$(2,\frac{9}{4})$\\$(\frac{23}{24},\frac{25}{24})$}&\makecell{$\mathcal{T}[W_{-18+\frac{18}{2}}(\mathfrak{e}_{7},A_{1})]+\mathfrak{u}(1)$\\$\mathcal{T}[W_{-18+\frac{9}{2}}(\mathfrak{e}_{7},D_{4}(a_{1})+A_{1})]+\mathfrak{u}(1)$\\$\mathcal{T}[W_{-18+\frac{9}{2}}(\mathfrak{e}_{7},A_{3}+2A_{1})]+\mathfrak{u}(1)$\\$\mathcal{T}[W_{-18+\frac{18}{8}}(\mathfrak{e}_{7},E_{7}(a_{4}))]+\mathfrak{u}(1)$}\\
		\hline
		$18$&\makecell{$-15$} &\makecell{$E_{7}(a_{4})$}&\makecell{$A_{2}+2A_{1}$}&\makecell{$116$}&\makecell{$82$}&\makecell{$(0,2)$}&\makecell{$\mathfrak{su}(2)_{6}\times\mathfrak{su}(2)_{2}^{2}$}&\makecell{$2,2$}&\makecell{$(\frac{7}{4},2)$}&\makecell{$\mathcal{T}[W_{-18+\frac{18}{3}}(\mathfrak{e}_{7},A_{2}+2A_{1})]$}\\
\hline
	\end{tabular}}
    \vspace{-1.5em}
\end{table}
}

{\linespread{2.0}

\begin{table}[H]\tiny
	\caption{$E_{8}$}
	\centering
	\resizebox{\linewidth}{!}{
	\begin{tabular}{|c|c|c|c|c|c|c|c|c|c|c|}\hline
		{$b$}& {$k$}&{Hitchin label}&{Nahm label}&{HD}&{$f_{0}$}&{$f_{1}$}&{$\mathfrak{f}$}&{$\Delta_{\text{Coulomb}}$}&{$(a,c)$}&{$\text{Theory}$}\\ \hline
		$30$&\makecell{$-23$} &\makecell{$A_{4}+A_{2}+A_{1}$}&\makecell{$A_{6}+A_{1}$}&\makecell{$196$}&\makecell{$0$}&\makecell{$0$}&\makecell{$\mathfrak{su}(2)_{0}$}&\makecell{$\frac{8}{7},\frac{10}{7}$}&\makecell{$(\frac{67}{84},\frac{17}{21})$}&\makecell{$\mathcal{T}[W_{-30+\frac{30}{7}}(\mathfrak{e}_{8},A_{6}+A_{1})]$}\\
		\hline
		$30$&\makecell{$-28$\\$-26$\\$-26$\\$-26$\\$-22$} &\makecell{$E_{8}(a_{2})$\\$E_{7}(a_{3})$\\$E_{8}(b_{6})$\\$(E_{8}(b_{6}),\mathbb{Z}_{2})$\\$A_{4}+A_{1}$}&\makecell{$2A_{1}$\\$A_{4}$\\$D_{4}(a_{1})+A_{2}$\\$A_{3}+A_{2}+A_{1}$\\$E_{6}(a_{1})+A_{1}$}&\makecell{$236$\\$220$\\$220$\\$220$\\$188$}&\makecell{$0$}&\makecell{$0$}&\makecell{$\mathfrak{so}(13)_{5}$\\$\mathfrak{su(5)}_{\frac{5}{2}}$\\$\mathfrak{su}(3)_{3}$\\$\mathfrak{su}(2)^{2}_{12,2}$\\$\mathfrak{u}(1)_{\frac{3}{2}}$}&\makecell{$5,3$\\$\frac{3}{2},\frac{5}{2}$\\$\frac{3}{2},\frac{3}{2}$\\$\frac{3}{2},3$\\$\frac{3}{2},\frac{5}{4}$}&\makecell{$(\frac{53}{12},\frac{16}{3})$\\$(\frac{7}{4},2)$\\$(\frac{7}{6},\frac{4}{3})$\\$(\frac{47}{24},\frac{13}{6})$\\$(\frac{11}{12},\frac{23}{24})$}&\makecell{$\mathcal{T}[W_{-30+\frac{30}{2}}(\mathfrak{e}_{8},2A_{1})]$\\$\mathcal{T}[W_{-30+\frac{30}{4}}(\mathfrak{e}_{8},A_{4})]$\\$\mathcal{T}[W_{-30+\frac{30}{4}}(\mathfrak{e}_{8},D_{4}(a_{1})+A_{2})]$\\$\mathcal{T}[W_{-30+\frac{30}{4}}(\mathfrak{e}_{8},A_{3}+A_{2}+A_{1})]$\\$\mathcal{T}[W_{-30+\frac{30}{8}}(\mathfrak{e}_{8},E_{6}(a_{1})+A_{1})]$}\\
\hline
	\end{tabular}}
\vspace{-1.5em}
\end{table}
}

{\linespread{2.0}

\begin{table}[H]\tiny
	\caption{$A_{2n-1}/\mathbb{Z}_{2}$, $b_{t}=4n-2$}
	\centering
	\resizebox{\linewidth}{!}{
	\begin{tabular}{|c|c|c|c|c|c|c|c|c|c|c|c|}\hline
		{$n$}& {$k_{t}$}&{C-partition}&{B-partition}&{HD}&{ND}&{$f_{0}$}& {$f_{1}$}&{$\mathfrak{f}$}&{$\Delta_{\text{Coulomb}}$}&{$(a,c)$}&{$\text{Theory}$}\\ \hline
        $1$&\makecell{$3$\\$5$} &\makecell{$(2)$\\$(1^{2})$}&\makecell{$(1^{3})$\\$(3)$}&\makecell{$2$\\$0$}&\makecell{$0$\\$2$}&\makecell{$0$\\$0$} &\makecell{$0$\\$0$}&\makecell{$\mathfrak{so}(3)_{\frac{4}{5}}$\\$-$}&\makecell{$\frac{8}{5},\frac{6}{5}$\\$\frac{10}{7},\frac{8}{7}$}&\makecell{$(\frac{19}{20},1)$\\$(\frac{67}{84},\frac{17}{21})$}&\makecell{$W_{-1+\frac{1}{5}}(B_{1},(1^{3}))$\\$W_{-1+\frac{1}{7}}(B_{1},(3))$}\\
		\hline
		$2$&\makecell{$-3$\\$1$} &\makecell{$(4)$\\$(1^{4})$}&\makecell{$(1^{5})$\\$(5)$}&\makecell{$8$\\$0$}&\makecell{$0$\\$8$}&\makecell{$0$\\$0$} &\makecell{$2$\\$0$}&\makecell{$\mathfrak{sp}(2)_{2}$\\$-$}&\makecell{$2,2$\\$\frac{8}{7},\frac{10}{7}$}&\makecell{$(\frac{19}{12},\frac{5}{3})$\\$(\frac{67}{84},\frac{17}{21})$}&\makecell{$\mathcal{T}[W_{-3+1}(B_{2},(1^{5}))]$\\$\mathcal{T}[W_{-3+\frac{3}{7}}(B_{2},(5))]$}\\
		\hline
		$3$&\makecell{$-7$\\$-7$\\$-5$} &\makecell{$(4,2)$\\$((4,2),\mathbb{Z}_{2})$\\$(2^{2},1^{2})$}&\makecell{$(3,1^{4})$\\$(2^{2},1^{3})$\\$(5,1^{2})$}&\makecell{$16$\\$16$\\$10$}&\makecell{$10$\\$8$\\$16$}&\makecell{$0$\\$0$\\$0$}&\makecell{$0$\\$0$\\$4$}&\makecell{$\mathfrak{su}(2)_{\frac{4}{3}}\times\mathfrak{su}(2)_{\frac{4}{3}}$\\$\mathfrak{su}(2)_{\frac{11}{6}}\times\mathfrak{so}(3)_{\frac{4}{3}}$\\$\mathfrak{u}(1)_{0}$}&\makecell{$\frac{4}{3},\frac{4}{3}$\\$\frac{4}{3},\frac{8}{3}$\\$1,1$}&\makecell{$(\frac{11}{12},1)$\\$(\frac{13}{8},\frac{7}{4})$\\$(\frac{5}{12},\frac{1}{3})$}&\makecell{$\mathcal{T}[W_{-5+\frac{5}{3}}(B_{3},(3,1^{4}))]$\\$\mathcal{T}[W_{-5+\frac{5}{3}}(B_{3},(2^{2},1^{3}))]$\\$\mathcal{T}[W_{-5+1}(B_{3},(5,1^{2}))]$}\\
\hline
	\end{tabular}}
\vspace{-1.5em}
\end{table}
}

{\linespread{2.0}

\begin{table}[H]\tiny
	\caption{$A_{2n-1}/\mathbb{Z}_{2}$, $b_{t}=2n$}
	\centering
	\resizebox{\linewidth}{!}{
	\begin{tabular}{|c|c|c|c|c|c|c|c|c|c|c|c|}\hline
		{$n$}& {$k_{t}$}&{HL}&{NL}&{HD}&{ND}&{$f_{0}$}& {$f_{1}$}&{$\mathfrak{f}$}&{$\Delta_{\text{Coulomb}}$}&{$(a,c)$}&{$\text{Theory}$}\\ \hline
        $1$&\makecell{$4$} &\makecell{$(2)$}&\makecell{$(1^{3})$}&\makecell{$2$}&\makecell{$0$}&\makecell{$1$} &\makecell{$0$}&\makecell{$\mathfrak{so}(3)_{\frac{5}{6}}\times\mathfrak{u}(1)_{f_{0}}$}&\makecell{$\frac{5}{3},\frac{4}{3}$}&\makecell{$(\frac{13}{12},\frac{7}{6})$}&\makecell{$W_{-1+\frac{1}{6}}(B_{1},(1^{3}))$}\\
		\hline
		$2$&\makecell{$-1$\\$-1$\\$1$} &\makecell{$(2^{2})$\\$((2^{2}),\mathbb{Z}_{2})$\\$(1^{4})$}&\makecell{$(3,1^{2})$\\$(2^{2},1)$\\$(5)$}&\makecell{$6$\\$6$\\$0$}&\makecell{$6$\\$4$\\$8$}&\makecell{$1$\\$1$\\$1$} &\makecell{$0$\\$0$\\$0$}&\makecell{$\mathfrak{u}(1)_{\frac{1}{3}}\times\mathfrak{u}(1)_{f_{0}}$\\$\mathfrak{su}(2)_{\frac{11}{6}}\times\mathfrak{u}(1)_{f_{0}}$\\$\mathfrak{u}(1)_{f_{0}}$}&\makecell{$\frac{4}{3},\frac{4}{3}$\\$\frac{8}{3},\frac{4}{3}$\\$\frac{6}{5},\frac{8}{5}$}&\makecell{$(\frac{11}{12},1)$\\$(\frac{13}{8},\frac{7}{4})$\\$(\frac{19}{20},1)$}&\makecell{$\mathcal{T}[W_{-3+\frac{2}{3}}(B_{2},(3,1^{2}))]+\mathfrak{u}(1)$\\$\mathcal{T}[W_{-3+\frac{2}{3}}(B_{2},(2^{2},1))]+\mathfrak{u}(1)$\\$\mathcal{T}[W_{-3+\frac{2}{5}}(B_{2},(5))]+\mathfrak{u}(1)$}\\
\hline
	\end{tabular}}
\vspace{-1.5em}
\end{table}
}

{\linespread{2.0}

\begin{table}[H]\tiny
	\caption{$D_{n+1}/\mathbb{Z}_{2}$, $b_{t}=2n+2$}
	\centering
	\resizebox{\linewidth}{!}{
	\begin{tabular}{|c|c|c|c|c|c|c|c|c|c|c|c|}\hline
		{$n$}& {$k_{t}$}&{HL}&{NL}&{HD}&{ND}&{$f_{0}$}& {$f_{1}$}&{$\mathfrak{f}$}&{$\Delta_{\text{Coulomb}}$}&{$(a,c)$}&{$\text{Theory}$}\\ \hline
        $1$&\makecell{$1$\\$3$} &\makecell{$(3)$\\$(1^{3})$}&\makecell{$(1^{2})$\\$(2)$}&\makecell{$2$\\$0$}&\makecell{$0$\\$2$}&\makecell{$0$\\$0$} &\makecell{$0$\\$0$}&\makecell{$\mathfrak{su}(2)_{\frac{5}{8}}$\\$-$}&\makecell{$\frac{8}{5},\frac{6}{5}$\\$\frac{10}{7},\frac{8}{7}$}&\makecell{$(\frac{19}{20},1)$\\$(\frac{67}{84},\frac{17}{21})$}&\makecell{$W_{-2+\frac{2}{5}}(C_{1},(1^{2}))$\\$W_{-2+\frac{2}{7}}(C_{1},(2))$}\\
		\hline
		$2$&\makecell{$1$} &\makecell{$(1^{5})$}&\makecell{$(4)$}&\makecell{$0$}&\makecell{$8$}&\makecell{$0$} &\makecell{$0$}&\makecell{$-$}&\makecell{$\frac{10}{7},\frac{8}{7}$}&\makecell{$(\frac{67}{84},\frac{17}{21})$}&\makecell{$W_{-3+\frac{3}{7}}(C_{2},(4))$}\\
		\hline
		$3$&\makecell{$-5$\\$-5$\\$-3$\\$-3$} &\makecell{$(5,1^{2})$\\$((5,1^{2}),\mathbb{Z}_{2})$\\$(3,1^{4})$\\$((3,1^{4}),\mathbb{Z}_{2})$}&\makecell{$(2^{2},1^{2})$\\$(2,1^{4})$\\$(4,2)$\\$(4,1^{2})$}&\makecell{$16$\\$16$\\$10$\\$10$}&\makecell{$10$\\$6$\\$16$\\$14$}&\makecell{$0$\\$0$\\$0$\\$0$} &\makecell{$0$\\$0$\\$0$\\$0$}&\makecell{$\mathfrak{su}(2)_{\frac{5}{3}}\times\mathfrak{u}(1)_{\frac{2}{3}}$\\$\mathfrak{sp}(2)_{\frac{13}{6}}$\\$-$\\$\mathfrak{su}(2)_{\frac{17}{10}}$}&\makecell{$\frac{4}{3},\frac{5}{3}$\\$\frac{4}{3},\frac{10}{3}$\\$\frac{6}{5},\frac{6}{5}$\\$\frac{12}{5},\frac{6}{5}$}&\makecell{$(\frac{13}{12},\frac{7}{6})$\\$(2,\frac{13}{6})$\\$(\frac{43}{60},\frac{11}{15})$\\$(\frac{163}{120},\frac{17}{12})$}&\makecell{$\mathcal{T}[W_{-4+\frac{4}{3}}(C_{3},(2^{2},1^{2})]$\\$\mathcal{T}[W_{-4+\frac{4}{3}}(C_{3},(2,1^{4})]$\\$\mathcal{T}[W_{-4+\frac{4}{5}}(C_{3},(4,2)]$\\$\mathcal{T}[W_{-4+\frac{4}{5}}(C_{3},(4,1^{2})]$}\\
\hline
	\end{tabular}}
\vspace{-1.5em}
\end{table}
}
{\linespread{2.0}

\begin{table}[H]\tiny
	\caption{$D_{n+1}/\mathbb{Z}_{2}$, $b_{t}=2n$}
	\centering
	\resizebox{\linewidth}{!}{
	\begin{tabular}{|c|c|c|c|c|c|c|c|c|c|c|c|}\hline
		{$n$}& {$k_{t}$}&{HL}&{NL}&{HD}&{ND}&{$f_{0}$}& {$f_{1}$}&{$\mathfrak{f}$}&{$\Delta_{\text{Coulomb}}$}&{$(a,c)$}&{$\text{Theory}$}\\ \hline
        $1$&\makecell{$2$} &\makecell{$(1^{3})$}&\makecell{$(2)$}&\makecell{$0$}&\makecell{$2$}&\makecell{$1$} &\makecell{$0$}&\makecell{$\mathfrak{u}(1)_{f_{0}}$}&\makecell{$\frac{3}{2},\frac{5}{4}$}&\makecell{$(\frac{11}{12},\frac{23}{24})$}&\makecell{$W_{-2+\frac{1}{4}}(C_{1},(2))+M(1)$}\\
		\hline
		$3$&\makecell{$-4$\\$-2$\\$-2$} &\makecell{$(7)$\\$(3,1^{4})$\\$((3,1^{4}),\mathbb{Z}_{2})$}&\makecell{$(1^{6})$\\$(4,2)$\\$(4,1^{2})$}&\makecell{$18$\\$10$\\$10$}&\makecell{$0$\\$16$\\$14$}&\makecell{$1$\\$1$\\$1$} &\makecell{$0$\\$0$\\$0$}&\makecell{$\mathfrak{sp}(3)_{\frac{5}{2}}\times\mathfrak{u}(1)_{f_{0}}$\\$\mathfrak{u}(1)_{f_{0}}$\\$\mathfrak{su}(2)_{\frac{7}{4}}\times\mathfrak{u}(1)_{f_{0}}$}&\makecell{$\frac{5}{2},3$\\$\frac{3}{2},\frac{5}{4}$\\$\frac{5}{2},\frac{3}{2}$}&\makecell{$(\frac{61}{24},\frac{17}{6})$\\$(\frac{11}{12},\frac{23}{24})$\\$(\frac{19}{12},\frac{5}{3})$}&\makecell{$\mathcal{T}[W_{-4+\frac{3}{2}}(C_{3},(1^{6})]+\mathfrak{u}(1)$\\$\mathcal{T}[W_{-4+\frac{3}{4}}(C_{3},(4,2)]+\mathfrak{u}(1)$\\$\mathcal{T}[W_{-4+\frac{3}{4}}(C_{3},(4,1^{2})]+\mathfrak{u}(1)$}\\
\hline
	\end{tabular}}
\vspace{-1.5em}
\end{table}
}
{\linespread{2.0}

\begin{table}[H]\tiny
	\caption{$D_{4}/\mathbb{Z}_{3}$, $b_{t}=12$, $b_{t}=6$}
	\centering
	\begin{tabular}{|c|c|c|c|c|c|c|c|c|c|c|}\hline
		{$b_{t}$}& {$k_{t}$}&{HL}&{NL}&{HD}&{$f_{0}$}& {$f_{1}$}&{$\mathfrak{f}$}&{$\Delta_{\text{Coulomb}}$}&{$(a,c)$}&{$\text{Theory}$}\\ \hline
		$12$&\makecell{$-8$\\$-8$} &\makecell{$G_{2}(a_{1})$\\$(G_{2}(a_{1}),S_{3})$}&\makecell{$G_{2}(a_{1})$\\$A_{1}$}&\makecell{$10$\\$10$}&\makecell{$0$\\$0$} &\makecell{$1$\\$1$}&\makecell{$-$\\$\mathfrak{su}(2)_{4}$}&\makecell{$1,1$\\$3,2$}&\makecell{$(\frac{5}{12},\frac{1}{3})$\\$(2,2)$}&\makecell{$\mathcal{T}[W_{-4+1}(\mathfrak{g}_{2},G_{2}(a_{1}))]$\\$\mathcal{T}[W_{-4+1}(\mathfrak{g}_{2},A_{1})]$}\\
		\hline
	\end{tabular}
\end{table}
}
{\linespread{2.0}

\begin{table}\tiny
	\centering
		\caption{$E_{6}/\mathbb{Z}_{2}$}
	\resizebox{\linewidth}{!}{
	\begin{tabular}{|c|c|c|c|c|c|c|c|c|c|c|c|}\hline
		{$b_{t}$}& {$k_{t}$}&{HL}&{NL}&{HD}&{ND}&{$f_{0}$}& {$f_{1}$}&{$\mathfrak{f}$}&{$\Delta_{\text{Coulomb}}$}&{$(c,a)$}&{$\text{Theory}$}\\ \hline
		$18$&\makecell{$-15$\\$-11$} &\makecell{$F_{4}(a_{2})$\\$\tilde{A}_{1}+A_{1}$}&\makecell{$A_{1}+\tilde{A}_{1}$\\$F_{4}(a_{2})$}&\makecell{$44$\\$28$}&\makecell{$28$\\$44$}&\makecell{$0$\\$0$} &\makecell{$2$\\$0$}&\makecell{$\mathfrak{su}(2)_{2}\times\mathfrak{su}(2)_{8}$\\$-$}&\makecell{$2,2$\\$\frac{8}{7},\frac{10}{7}$}&\makecell{$(\frac{7}{4},2)$\\$(\frac{67}{84},\frac{17}{21})$}&\makecell{$\mathcal{T}[W_{-9+3}(\mathfrak{f}_{4},A_{1}+\tilde{A}_{1})]$\\$\mathcal{T}[W_{-9+\frac{9}{7}}(\mathfrak{f}_{4},F_{4}(a_{2}))]$}\\
		\hline
		$8$&\makecell{$-7$\\$-5$\\$-5$} &\makecell{$F_{4}$\\$A_{2}$\\$\tilde{A}_{2}$}&\makecell{$0$\\$B_{3}$\\$C_{3}$}&\makecell{$48$\\$30$\\$30$}&\makecell{$0$\\$42$\\$42$}&\makecell{$1$\\$1$\\$1$} &\makecell{$0$\\$0$\\$0$}&\makecell{$(\mathfrak{f}_{4})_{5}\times\mathfrak{u}(1)_{f_{0}}$\\$\mathfrak{su}(2)_{\frac{4}{3}}\times\mathfrak{u}(1)_{f_{0}}$\\$\mathfrak{su}(2)_{\frac{5}{3}}\times\mathfrak{u}(1)_{f_{0}}$}&\makecell{$5,4$\\$\frac{4}{3},\frac{4}{3}$\\$\frac{5}{3},\frac{4}{3}$}&\makecell{$(\frac{14}{3},\frac{16}{3})$\\$(\frac{11}{12},1)$\\$(\frac{13}{12},\frac{7}{6})$}&\makecell{$\mathcal{T}[W_{-9+4}(\mathfrak{f}_{4},0)]+\mathfrak{u}(1)$\\$\mathcal{T}[W_{-9+\frac{4}{3}}(\mathfrak{f}_{4},B_{3})]+\mathfrak{u}(1)$\\$\mathcal{T}[W_{-9+\frac{4}{3}}(\mathfrak{f}_{4},C_{3})]+\mathfrak{u}(1)$}\\
		\hline
	\end{tabular}}
\vspace{-2em}
\end{table}}

\subsection{Rank 3}
\label{sec:completList3}

{\linespread{2.0}

\begin{table}[H]\tiny
	\caption{$A_{n-1}$, $b=n$}
	\centering
	\resizebox{\linewidth}{!}{
	\begin{tabular}{|c|c|c|c|c|c|c|c|c|c|}\hline{$n$}& {$k$}&{$\mathcal{O}_{f}$}&{$f_{0}$}& {$f_{1}$}&{ND}&{$\mathfrak{f}$}&{$\Delta_{\text{Coulomb}}$}&{$(a,c)$}&{Theory}\\ 
	\hline
	$2$&\makecell{$5$\\$6$\\$7$\\$8$} &\makecell{$(1^{2})$\\$(1^{2})$\\$(2)$\\$(2)$}&\makecell{$0$\\$1$\\$0$\\$1$}&\makecell{$0$\\$1$\\$0$\\$1$}&\makecell{$0$\\$0$\\$2$\\$2$} &\makecell{$\mathfrak{su}(2)_{\frac{12}{7}}$\\$\mathfrak{su}(2)_{\frac{7}{4}}\times\mathfrak{u}(1)_{f_{0}}$\\$-$\\$\mathfrak{u}(1)_{f_{0}}$}&\makecell{$\frac{12}{7},\frac{10}{7},\frac{8}{7}$\\$\frac{7}{4},\frac{3}{2},\frac{5}{4}$\\$\frac{14}{9},\frac{4}{3},\frac{10}{9}$\\$\frac{8}{5},\frac{7}{5},\frac{6}{5}$}&\makecell{$(\frac{81}{56},\frac{3}{2})$\\$(\frac{19}{12},\frac{5}{3})$\\$(\frac{91}{72},\frac{23}{18})$\\$(\frac{167}{120},\frac{43}{30})$}&\makecell{$\mathcal{T}[W_{-2+\frac{2}{7}}(\mathfrak{su}(2),(1^{2}))]$\\$\mathcal{T}[W_{-2+\frac{2}{8}}(\mathfrak{su}(2),(1^{2}))]+\mathfrak{u}(1)$\\$\mathcal{T}[W_{-2+\frac{2}{9}}(\mathfrak{su}(2),(2))]$\\$\mathcal{T}[W_{-2+\frac{2}{10}}(\mathfrak{su}(2),(2))]+\mathfrak{u}(1)$}\\
		\hline
		$3$&\makecell{$1$\\$2$\\$3$\\$4$} &\makecell{$(1^{3})$\\$(2,1)$\\$(2,1)$\\$(3)$}&\makecell{$0$\\$0$\\$2$\\$0$}&\makecell{$0$\\$0$\\$2$\\$0$}&\makecell{$0$\\$4$\\$4$\\$6$}&\makecell{$\mathfrak{su}(3)_{\frac{9}{4}}$\\$\mathfrak{u}(1)_{\frac{9}{5}}$\\$\mathfrak{u}(1)_{2}\times\mathfrak{u}(1)^{2}_{f_{0}}$\\$-$}&\makecell{$\frac{9}{4},\frac{3}{2},\frac{5}{4}$\\$\frac{9}{5},\frac{6}{5},\frac{7}{5}$\\$2,\frac{3}{2},\frac{3}{2}$\\$\frac{12}{7},\frac{9}{7},\frac{8}{7}$}&\makecell{$(\frac{15}{8},2)$\\$(\frac{3}{2},\frac{31}{20})$\\$(\frac{15}{8},2)$\\$(\frac{75}{56},\frac{19}{14})$}&\makecell{$\mathcal{T}[W_{-3+\frac{3}{4}}(\mathfrak{su}(3),(1^{3}))]$\\$\mathcal{T}[W_{-3+\frac{3}{5}}(\mathfrak{su}(3),(2,1))]$\\$\mathcal{T}[W_{-3+\frac{3}{6}}(\mathfrak{su}(3),(2,1))]+2\mathfrak{u}(1)$\\$\mathcal{T}[W_{-3+\frac{3}{7}}(\mathfrak{su}(3),(3))]$}\\
\hline
	\end{tabular}}
\vspace{-1.5em}
\end{table}}

{\linespread{2.0}

\begin{table}[H]\tiny
	\caption{$A_{n-1}$, $b=n-1$}
	\centering
	\resizebox{\linewidth}{!}{
	\begin{tabular}{|c|c|c|c|c|c|c|c|c|c|}\hline
		{$n$}& {$k$}&{$\mathcal{O}_{f}$}&{$f_{0}$}& {$f_{1}$}&{ND}&{$\mathfrak{f}$}&{$\Delta_{\text{Coulomb}}$}&{$(a,c)$}&{Theory}\\ \hline
		$3$&\makecell{$1$\\$3$} &\makecell{$(1^{3})$\\$(3)$}&\makecell{$1$\\$1$}&\makecell{$0$\\$0$}&\makecell{$0$\\$6$}&\makecell{$\mathfrak{su}(3)_{\frac{7}{3}}\times\mathfrak{u}(1)_{f_{0}}$\\$\mathfrak{u}(1)_{f_{0}}$}&\makecell{$\frac{7}{3},\frac{5}{3},\frac{4}{3}$\\$\frac{9}{5},\frac{7}{5},\frac{6}{5}$}&\makecell{$(\frac{25}{12},\frac{9}{4})$\\$(\frac{3}{2},\frac{31}{20})$}&\makecell{$\mathcal{T}[W_{-3+\frac{2}{3}}(\mathfrak{su}(3),(1^{3}))]+\mathfrak{u}(1)$\\$\mathcal{T}[W_{-3+\frac{2}{5}}(\mathfrak{su}(3),(3))]+\mathfrak{u}(1)$}\\
		\hline
		$4$&\makecell{$1$} &\makecell{$(3,1)$}&\makecell{$1$}&\makecell{$0$}&\makecell{$10$}&\makecell{$\mathfrak{u}(1)_{\frac{7}{4}}\times\mathfrak{u}(1)_{f_{0}}$}&\makecell{$\frac{7}{4},\frac{3}{2},\frac{5}{4}$}&\makecell{$(\frac{19}{12},\frac{5}{3})$}&\makecell{$\mathcal{T}[W_{-4+\frac{3}{4}}(\mathfrak{su}(4),(3,1))]+\mathfrak{u}(1)$}\\
		\hline
		$5$&\makecell{$-1$} &\makecell{$(2^{2},1)$}&\makecell{$1$}&\makecell{$0$}&\makecell{$12$}&\makecell{$\mathfrak{su}(2)_{\frac{7}{3}}\times\mathfrak{u}(1)_{\frac{7}{3}}\times\mathfrak{u}(1)_{f_{0}}$}&\makecell{$\frac{7}{3},\frac{4}{3},\frac{5}{3}$}&\makecell{$(\frac{25}{12},\frac{9}{4})$}&\makecell{$\mathcal{T}[W_{-5+\frac{4}{3}}(\mathfrak{su}(5),(2^{2},1))]+\mathfrak{u}(1)$}\\
\hline
	\end{tabular}}
\vspace{-1.5em}
\end{table}}

{\linespread{2.0}

\begin{table}[H]\tiny
	\caption{$D_{n}$, $b=n$}
	\centering
	\resizebox{\linewidth}{!}{
	\begin{tabular}{|c|c|c|c|c|c|c|c|c|c|c|c|}\hline
		{$n$}& {$k$}&{Hitchin label}&{Nahm label}&{HD}&{ND}&{$f_{0}$}& {$f_{1}$}&{$\mathfrak{f}$}&{$\Delta_{\text{Coulomb}}$}&{$(a,c)$}&{Theory}\\ \hline
		$3$&\makecell{$0$\\$1$\\$3$} &\makecell{$(5,1)$\\$(2^{2},1^{2})$\\$(1^{6})$}&\makecell{$(1^{6})$\\$(3^{2})$\\$(5,1)$}&\makecell{$12$\\$6$\\$0$}&\makecell{$0$\\$10$\\$12$}&\makecell{$3$\\$1$\\$3$}&\makecell{$2$\\$0$\\$2$}&\makecell{$\mathfrak{su}(4)_{3}\times\mathfrak{u}(1)^{3}_{f_{0}}$\\$\mathfrak{u}(1)\times\mathfrak{u}(1)_{f_{0}}$\\$\mathfrak{u}(1)^{3}_{f_{0}}$}&\makecell{$3,2,2$\\$\frac{5}{4},\frac{7}{4},\frac{3}{2}$\\$2,\frac{3}{2},\frac{3}{2}$}&\makecell{$(\frac{25}{8},\frac{7}{2})$\\$(\frac{19}{12},\frac{5}{3})$\\$(\frac{15}{8},2)$}&\makecell{$\mathcal{T}[W_{-4+\frac{3}{3}}(D_{3},(1^{6}))]+3\mathfrak{u}(1)$\\$\mathcal{T}[W_{-4+\frac{3}{4}}(D_{3},(3^{2}))]+\mathfrak{u}(1)$\\$\mathcal{T}[W_{-4+\frac{3}{6}}(D_{3},(5,1))]+3\mathfrak{u}(1)$}\\
		\hline
		$4$&\makecell{$-2$} &\makecell{$(5,3)$}&\makecell{$(2^{2},1^{4})$}&\makecell{$22$}&\makecell{$10$}&\makecell{$0$}&\makecell{$3$}&\makecell{$\mathfrak{su}(2)_{2}\times\mathfrak{su}(2)_{2}\times\mathfrak{su}(2)_{2}$}&\makecell{$2,2,2$}&\makecell{$(\frac{19}{8},\frac{5}{2})$}&\makecell{$\mathcal{T}[W_{-6+\frac{4}{2}}(D_{4},(2^{2},1^{4}))]$}\\
\hline
	\end{tabular}}
\vspace{-1.5em}
\end{table}}

{\linespread{2.0}

\begin{table}[H]\tiny
	\caption{$D_{n}$, $b=2n-2$}
	\centering
		\resizebox{\linewidth}{!}{
	\begin{tabular}{|c|c|c|c|c|c|c|c|c|c|c|c|}\hline
		{$n$}& {$k$}&{HL}&{NL}&{HD}&{ND}&{$f_{0}$}& {$f_{1}$}&{$\mathfrak{f}$}&{$\Delta_{\text{Coulomb}}$}&{$(a,c)$}&{Theory}\\ \hline
		$3$&\makecell{$-1$\\$1$\\$3$} &\makecell{$(5,1)$\\$(2^{2},1^{2})$\\$(1^{6})$}&\makecell{$(1^{6})$\\$(3^{2})$\\$(5,1)$}&\makecell{$12$\\$6$\\$0$}&\makecell{$0$\\$10$\\$12$}&\makecell{$0$\\$0$\\$0$}&\makecell{$0$\\$0$\\$0$}&\makecell{$\mathfrak{su}(4)_{\frac{8}{3}}$\\$\mathfrak{u}(1)$\\$-$}&\makecell{$\frac{8}{3},\frac{4}{3},\frac{5}{3}$\\$\frac{6}{5},\frac{8}{5},\frac{7}{5}$\\$\frac{12}{7},\frac{8}{7},\frac{9}{7}$}&\makecell{$(\frac{55}{24},\frac{5}{2})$\\$(\frac{167}{120},\frac{43}{30})$\\$(\frac{75}{56},\frac{19}{14})$}&\makecell{$\mathcal{T}[W_{-4+\frac{4}{3}}(D_{3},(1^{6}))]$\\$\mathcal{T}[W_{-4+\frac{4}{5}}(D_{3},(3^{2}))]$\\$\mathcal{T}[W_{-4+\frac{4}{7}}(D_{3},(5,1))]$}\\
		\hline
		$4$&\makecell{$-2$\\$-2$} &\makecell{$(4^{2})$\\$(5,1^{3})$}&\makecell{$(2^{4})$\\$(3,1^{5})$}&\makecell{$20$\\$20$}&\makecell{$12$\\$12$}&\makecell{$2$\\$2$}&\makecell{$0$\\$0$}&\makecell{$\mathfrak{sp}(2)_{\frac{5}{2}}\times\mathfrak{u}(1)^{2}_{f_{0}}$\\$\mathfrak{sp}(2)_{\frac{5}{2}}\times\mathfrak{u}(1)^{2}_{f_{0}}$}&\makecell{$3,\frac{3}{2},\frac{5}{2}$\\$\frac{5}{2},3,\frac{3}{2}$}&\makecell{$(\frac{73}{24},\frac{10}{3})$\\$(\frac{73}{24},\frac{10}{3})$}&\makecell{$\mathcal{T}[W_{-6+\frac{6}{4}}(D_{4},2^{4}))]+2\mathfrak{u}(1)$\\$\mathcal{T}[W_{-6+\frac{6}{4}}(D_{4},(3,1^{5}))]+2\mathfrak{u}(1)$}\\
\hline
	\end{tabular}}
\vspace{-2em}
\end{table}}

{\linespread{2.0}

\begin{table}[H]\tiny
\caption{$E_{6}$}
\centering
\resizebox{\linewidth}{!}{
\begin{tabular}{|c|c|c|c|c|c|c|c|c|c|c|}\hline
	{$b$}& {$k$}&{HL}&{NL}&{HD}&{ND}& {$(f_{0},f_{1})$}&{$\mathfrak{f}$}&{$\Delta_{\text{Coulomb}}$}&{$(a,c)$}&{Theory}\\ \hline
	$12$&\makecell{$-5$\\$-5$} &\makecell{$A_{2}$\\$(A_{2},\mathbb{Z}_{2})$}&\makecell{$E_{6}(a_{3})$\\$A_{5}$}&\makecell{$42$\\$42$} &\makecell{$66$\\$64$}&\makecell{$(0,0)$}&\makecell{$-$\\$\mathfrak{su}(2)_{\frac{25}{14}}$}&\makecell{$\frac{12}{7},\frac{8}{7},\frac{9}{7}$\\$\frac{12}{7},\frac{8}{7},\frac{18}{7}$}&\makecell{$(\frac{75}{56},\frac{19}{14})$\\$(\frac{85}{42},\frac{25}{12})$}&\makecell{$\mathcal{T}[W_{-12+\frac{12}{7}}(\mathfrak{e}_{6},E_{6}(a_{3}))]$\\$\mathcal{T}[W_{-12+\frac{12}{7}}(\mathfrak{e}_{6},A_{5})]$}\\
	\hline
	$12$&\makecell{$-10$\\$2$} &\makecell{$E_{6}$\\$0$}&\makecell{$0$\\$E_{6}$}&\makecell{$72$\\$0$}&\makecell{$0$\\$72$}&\makecell{$(0,1)$}&\makecell{$(\mathfrak{e}_{6})_{6}$\\$-$}&\makecell{$6,3,2$\\$\frac{8}{7},\frac{9}{7},\frac{12}{7}$}&\makecell{$(\frac{45}{8},\frac{13}{2})$\\$(\frac{75}{56},\frac{19}{14})$}&\makecell{$\mathcal{T}[W_{-12+\frac{12}{2}}(\mathfrak{e}_{6},0)]$\\$\mathcal{T}[W_{-12+\frac{12}{14}}(\mathfrak{e}_{6},E_{6})]$}\\
	\hline
	$12$&\makecell{$-9$\\$-3$} &\makecell{$D_{5}$\\$2A_{1}$}&\makecell{$2A_{1}$\\$D_{5}$}&\makecell{$68$\\$32$}&\makecell{$32$\\$68$}&\makecell{$(2,1)$}&\makecell{$\mathfrak{so}(7)_{4}\times\mathfrak{u}(1)_{6}\times\mathfrak{u}(1)^{2}_{f_{0}}$\\$\mathfrak{u}(1)_{\frac{1}{6}}\times\mathfrak{u}(1)^{2}_{f_{0}}$}&\makecell{$4,4,2$\\$\frac{4}{3},\frac{4}{3},\frac{4}{3}$}&\makecell{$(\frac{39}{8},\frac{11}{2})$\\$(\frac{11}{8},\frac{3}{2})$}&\makecell{$\mathcal{T}[W_{-12+\frac{12}{3}}(\mathfrak{e}_{6},2A_{1})]+2\mathfrak{u}(1)$\\$ \mathcal{T}[W_{-12+\frac{12}{9}}(\mathfrak{e}_{6},D_{5})]+2\mathfrak{u}(1)$}\\
\hline
\end{tabular}}
\vspace{-1.5em}
\end{table}}

{\linespread{2.0}

\begin{table}[H]\tiny
	\caption{$E_{7}$}
	\centering
	\resizebox{\linewidth}{!}{
	\begin{tabular}{|c|c|c|c|c|c|c|c|c|c|c|}\hline
		{$b$}& {$k$}&{HL}&{NL}&{HD}&{ND}& {$f_{0},f_{1}$}&{$\mathfrak{f}$}&{$\Delta_{\text{Coulomb}}$}&{$(a,c)$}&{$\text{Theory}$}\\ \hline
		$18$&\makecell{$-13$} &\makecell{$A_{4}+A_{1}$}&\makecell{$A_{4}+A_{1}$}&\makecell{$104$} &\makecell{$104$}&\makecell{$(0,0)$}&\makecell{$\mathfrak{u}(1)^{2}$}&\makecell{$\frac{6}{5},\frac{7}{5},\frac{9}{5}$}&\makecell{$(\frac{3}{2},\frac{31}{20})$}&\makecell{$\mathcal{T}[W_{-18+\frac{18}{5}}(\mathfrak{e}_{7},A_{4}+A_{1})]$}\\
\hline
	\end{tabular}}
    \vspace{-1.5em}
\end{table}}

{\linespread{2.0}

\begin{table}[H]\tiny
	\caption{$E_{8}$}
	\centering
	\resizebox{\linewidth}{!}{
	%
}
\vspace{-1.5em}
\end{table}}

{\linespread{2.0}

\begin{table}[H]\tiny
	\caption{$A_{2n-1}/\mathbb{Z}_{2}$, $b_{t}=4n-2$}
	\centering
	\resizebox{\linewidth}{!}{
	%
}
\vspace{-1.5em}
\end{table}}

{\linespread{2.0}

\begin{table}[H]\tiny
	\caption{$A_{2n-1}/\mathbb{Z}_{2}$, $b_{t}=2n$}
	\centering
	\resizebox{\linewidth}{!}{
	%
}
\vspace{-1.5em}
\end{table}}

{\linespread{2.0}

\begin{table}[H]\tiny
	\caption{$D_{n+1}/\mathbb{Z}_{2}$, $b_{t}=2n+2$}
	\centering
	\resizebox{\linewidth}{!}{
	%
}
\vspace{-1.5em}
\end{table}}

{\linespread{2.0}

\begin{table}[H]\tiny
	\caption{$D_{n+1}/\mathbb{Z}_{2}$, $b_{t}=2n$}
	\centering
	\resizebox{\linewidth}{!}{
	%
}
\vspace{-1.5em}
\end{table}}

{\linespread{2.0}

\begin{table}[H]\tiny
	\caption{$D_{4}/\mathbb{Z}_{3}$, $b_{t}=12$, $b_{t}=6$}
	\centering
	%
\end{table}}

{\linespread{1.8}
\begin{table}[H]\tiny
	\caption{$E_{6}/\mathbb{Z}_{2}$}
	\centering
	\resizebox{\linewidth}{!}{
	%
}
\vspace{-2em}
\end{table}}

\newpage

\bibliographystyle{jhep}
\bibliography{ref.bib}
\end{document}